\documentclass[10pt,twocolumn]{article}
\usepackage[top=1.9cm, bottom=1.9cm, left=1.8cm, right=1.8cm]{geometry}

\usepackage{amsmath,amssymb,amsfonts}
\usepackage{listings}                       %
\usepackage[noend]{algpseudocode}
\usepackage{multirow}
\usepackage{graphicx}
\usepackage{textcomp}
\usepackage{xcolor}
\usepackage{flushend}
\usepackage[lined,boxed]{algorithm2e}

\usepackage{graphicx,subfigure}
\usepackage[small,bf]{caption}
\usepackage[square,sort,comma,numbers]{natbib}

\SetCommentSty{mycommfont}

\usepackage{xcolor}
\usepackage{booktabs}
\usepackage{graphicx}
\usepackage{paralist}

\usepackage{times}

\usepackage[compact]{titlesec}
\titlespacing*{\section}{0pt}{*4}{4pt}
\titlespacing{\subsection}{0pt}{*3}{3pt}
\titlespacing{\subsubsection}{0pt}{*3}{3pt}

\definecolor{linkcol}{rgb}{0,0,0.5}
\definecolor{citecol}{rgb}{0,0.5,0.3}
\definecolor{urlcol}{rgb}{0.3,0,0}

\usepackage{xspace}

\renewenvironment{thebibliography}[1]{
  \begin{oldthebibliography}{#1}
    \setlength{\itemsep}{0.0em}
    \setlength{\parskip}{0.5em}
}
{
  \end{oldthebibliography}
}

\usepackage[hang,flushmargin]{footmisc}

\renewcommand{\footnoterule}{%
  \kern -3pt
  \hrule width 1in
  \kern 2pt
}

\usepackage{url}
\makeatletter
\def\url@leostyle{%
  \@ifundefined{selectfont}{\def\UrlFont{}}%
  {\def\UrlFont{}}%
}
\makeatother
\usepackage[hyphenbreaks]{breakurl}

\definecolor{darkred}{RGB}{153,0,0}
\definecolor{darkblue}{RGB}{0,0,99}
\usepackage[colorlinks=true, linkcolor = darkred,   citecolor = darkred, urlcolor = darkblue]{hyperref}

\newtheorem{ldp-definition}{Definition}
\newtheorem{dp-definition}{Definition}
\newcommand{\descr}[1]{\smallskip\noindent\textbf{#1}}

\begin{document}

\sloppy 

\title{\bf Adherence to Misinformation on Social Media Through Socio-Cognitive and Group-Based Processes\thanks{To Appear at the 25th ACM Conference on Computer Supported Cooperative Work and Social Computing (CSCW 2022).}}

\author{Alexandros Efstratiou and Emiliano De Cristofaro\\[1ex]
\fontsize{11}{12}\selectfont University College London\\ 
\fontsize{11}{12}\selectfont alexandros.efstratiou.20@ucl.ac.uk, e.decristofaro@ucl.ac.uk}
\date{}

\maketitle

\begin{abstract}
Previous work suggests that people's preference for different kinds of information depends on more than just accuracy. This could happen because the messages contained within different pieces of information may either be well-liked or repulsive. Whereas factual information must often convey uncomfortable truths, misinformation can have little regard for veracity and leverage psychological processes which increase its attractiveness and proliferation on social media. In this review, we argue that when misinformation proliferates, this happens because the social media environment enables adherence to misinformation by reducing, rather than increasing, the psychological cost of doing so. We cover how attention may often be shifted away from accuracy and towards other goals, how social and individual cognition is affected by misinformation and the cases under which debunking it is most effective, and how the formation of online groups affects information consumption patterns, often leading to more polarization and radicalization. Throughout, we make the case that polarization and misinformation adherence are closely tied. We identify ways in which the psychological cost of adhering to misinformation can be increased when designing anti-misinformation interventions or resilient affordances, and we outline open research questions that the CSCW community can take up in further understanding this cost.
\end{abstract}

\section{Introduction}

The era of social media, where content consumers are simultaneously content creators, has caused a surge in the availability of information. 
This is strengthening the ``attention economy,'' where human attention is among the scarcest resources for which we compete~\cite{davenport_attention_2002}.
As a consequence, reduced veracity or quality of the information that people consume is a real risk.
For example, false news articles have a \textit{lower} potential for reach (i.e., they are typically disseminated by less popular users or channels) than fact-checker verified articles, yet they spread faster, broader, deeper into networks, and are more viral~\cite{vosoughi_spread_2018}.
All of these four properties are due to this false information simply being shared by more users than verified information (i.e., its cascade size)~\cite{juul_comparing_2021}.

Nonetheless, studies of representative US samples find that the vast majority of users never share misinformation-laden links~\cite{guess_less_2019}, and that, when considering holistic media diets (including offline sources such as TV, as well as non-news information such as entertainment), misinformation only constitutes about 0.15\% of an average user's media diet~\cite{allen_evaluating_2020}.
Furthermore, social media posts promoting content which has been fact-checked occasionally garner disapproval from other users, indicating that corrections to misinformation are often taken up by the wider community~\cite{jiang_linguistic_2018}.
Given such findings, misinformation may be a more contextual problem rather than an omnipresent one.
To that end, understanding the contexts and situations under which misinformation is likely to be a prominent problem is an important endeavor.

In this paper, we attempt to enhance this understanding.
We focus on the ways in which misinformation is enabled by genuine user belief, user biases, and ideological passions, following the typology of Zannettou et al.~\cite{zannettou_web_2019}.
In other words, we fixate on the {\em receivers} of misinformation, and not necessarily on deliberate spreaders such as information operation actors, disinformation campaigns, or trolls (although these are closely related in often producing much of the misinformation consumed by the receivers; see~\citet{starbird_disinformation_2019}).
We use the term ``adherence to misinformation'' in order to probe insights from existing studies and to identify directions for future research.
Adherence to misinformation refers to the likelihood or degree to which a user will believe in, share, or otherwise endorse false information.

\paragraph{Problem Statement and Scope.} 
Proliferation of misinformation is best understood when considering how users, along with their cognitive limitations and implicit motivations, interact with technological affordances. 
Some work~\cite{caulfield_why_2019} has highlighted how the public can be defended against false information via social means, such as recognizing the biases and limited attention of recipients.
Nevertheless, we are still lacking a clear understanding of the motivations that users have when parsing information, and the situations under which accuracy can be at the forefront as one such motivation. 

Here, we set out to draw together insights from various disciplines, aiming to better understand the role of social and cognitive processes in shaping users' adherence to misinformation, as well as the degree to which balancing the network environment with corrections and accurate information can solve the problem.
We primarily focus on the areas of social, cognitive, and political psychology, while also discussing work conducted within the domains of computational social science, human-computer interaction, and web data mining.
We believe that unpacking insights provided from these disciplines can offer CSCW scholars important considerations for creating, modifying, or probing platform affordances and anti-misinformation interventions.
Furthermore, through this review we aim to bridge together several related fields and to bring them in more explicit conversation with existing work in CSCW, particularly by identifying open research questions that the CSCW community can pursue.

\paragraph{Motivation.} Misinformation can be made to draw a disproportionate amount of users' already limited attention because it is not bound by accuracy restrictions. 
Whereas facts may be difficult to comprehend due to their usually arduous complexity, misinformation can arguably be constructed in any way that makes it more attractive.
Furthermore, polarity and group-based conflicts can implicitly make information that maintains group image very valuable, even when it is not accurate~\cite{gillespie_disruption_2020, slater_reinforcing_2007}. 
Many of these cognitive and social amplifiers of misinformation adherence, which are largely inherent in human nature, may be further exacerbated in online environments.

Indeed, people online interact differently than how they do offline.
Homophily -- i.e., the human disposition to surround oneself with like-minded people~\cite{rogers_homophily-heterophily_1970} -- seems more readily enabled online due to the increased connectivity and availability of people similar to us. 
Confirmation bias, a cognitive pattern whereby we seek information in agreement with pre-existing beliefs while avoiding discordant information~\cite{wason_failure_1960}, is also more pronounced online~\cite{pearson_is_2019}. 
However, people with ideologically diverse offline circles also tend to have more diverse networks online~\cite{vaccari_echo_2016}.
Evidently, the interplay between social media affordances and the inherent psychological processes which affect users' information parsing is quite complex.

These processes, which often occur at an implicit (subconscious) level~\cite{evans_two_2003}, can create problems for the veracity of information disseminated on social media.
As mentioned, it is necessary to understand when misinformation is likely to be a prominent problem and how this problem can be thwarted.
To do so, we must first understand how psychological processes affect adherence to different types of information, how these may interact with social media environments and affordances, and how these interactions can affect the goals and motivations of a user who is parsing information. 

\paragraph{State of the Art Limitations.} Previous studies have attempted to algorithmically solve the problem of information diversity, either by directly maximizing network diversity~\cite{aslay_maximizing_2018} or through visualization tools helping users see the true (non)diversity of their networks and nudge them towards alternative viewpoints~\cite{gillani_me_2018}.
These interventions assume that if a user can be exposed to more diverse kinds of information, they will update their views accordingly. 
In other words, users are viewed as rational Bayesian agents, where all information is assimilated and considered equally. 
Consequently, the primary aim of such research is disrupting echo chambers, i.e., homogeneous online social groups where members' opinions go unchallenged in the like-minded social environment. 

Unfortunately, implicit cognitive and social biases may hamper the effectiveness of such approaches. 
As mentioned, people may not take information at face value.
Therefore, while such approaches could be very promising under specific circumstances, they could have little effect in others where echo chambers play a less important role.

\paragraph{Contributions.} 
Our main contribution in this work is drawing together several related literatures that have thus far remained somewhat disjointed, in order to paint a more comprehensive picture of the current state of scholarship on the issue of misinformation.
Through this synthesis, we draw out several recommendations for CSCW researchers and practitioners working in the intervention and affordance design space.
These mainly concern understanding and considering users' dynamic social identities and how this affects their information parsing motivations, their limitations in processing information under heavy cognitive load, and the informational environments within which they interact with other users and new information.
We summarize these recommendations in the discussion.

We also highlight several avenues for future research on the topic, and call for scholars to adopt new viewpoints in studying misinformation-adjacent issues.
Primarily, we identify the need to map out combative, \emph{as well as deliberative} spaces online.
Throughout the review, we highlight how highly polarized spaces make information-sharing combative, thus reducing the focus on the accuracy of information.
We observe that a vast majority of the literature has focused on these combative spaces, however, understanding deliberative information-sharing remains a very important open challenge in order to form a more balanced view.
Furthermore, we identify a need to consider echo chambers along a continuous spectrum; that is, \emph{the degree to which} users may be involved in them, and not just \emph{whether} they are involved in them or not.
This would allow researchers to quantify the impact of seclusion within particular communities on allowing misinformation to proliferate.
Once again, we summarize the directions for future research that we identify in the discussion section.

\paragraph{Roadmap and Paper Organization.} In this article, we use the idea of cognitive dissonance as a conceptual binder for the insights we discuss. 
Cognitive dissonance is a process through which individuals resolve their conflicting beliefs, by explaining some of them away (so as to remove the conflict)~\cite{festinger_cognitive_1959, festinger_theory_1957}. 
The belief that is retained is the least psychologically costly.
For the purposes of this paper, we conceptually assume that users will encounter varying degrees of dissonance when encountering misinformation.
We treat successful anti-misinformation efforts as those which make adherence to misinformation more psychologically costly (i.e., less comfortable), and we treat enablers of misinformation as whatever makes adhering to misinformation less psychologically costly (i.e., more comfortable).
We return to this idea at the ``take-aways'' summary of each major section, in order to synthesize the research we discuss into the actions that are likely to either enable or thwart adherence to misinformation.

\begin{figure*}[t]
    \centering
    \includegraphics[width=0.6\textwidth]{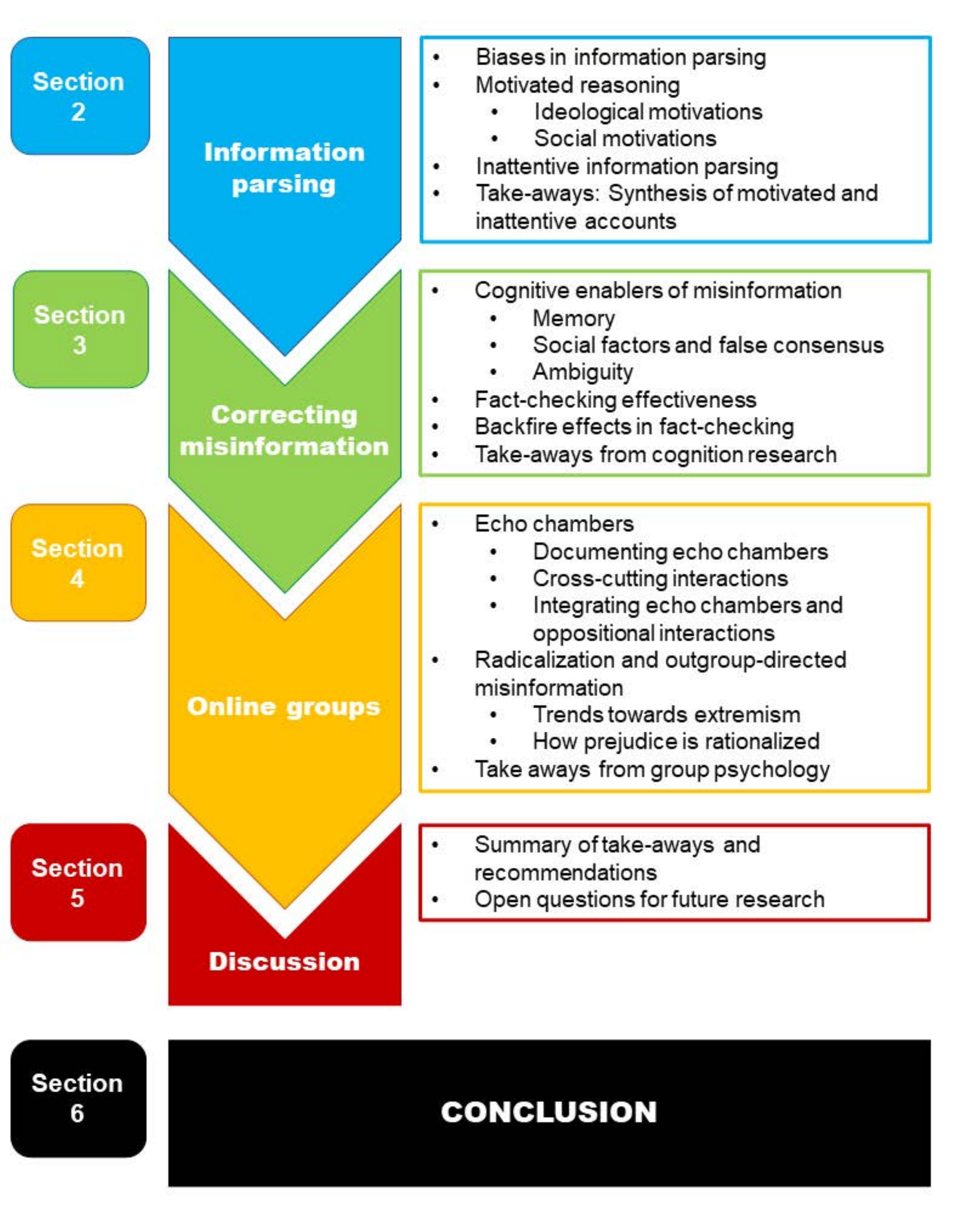}
    \caption{A roadmap of the review which outlines, in order, the topics covered in each section.}
    \label{fig:roadmap}
\end{figure*}

We present a visual roadmap of the entire paper in Figure~\ref{fig:roadmap}.
In Section~\ref{sec:confirming}, we examine biased information parsing from the perspective of two distinct theoretical accounts, namely, motivated reasoning and ``lazy'' inattention.
Then, we synthesize insights from both of these accounts.
Having discussed whether people have any preferences in the information they consume, we then turn to more subtle factors which affect adherence to misinformation in Section~\ref{sec:correcting}.
Here, we explore how memory, social norms, and a need to resolve ambiguity can assist the belief in misinformation.
Given this, we also discuss fact-checking, the situations under which it is most likely to work, and whether it can backfire in certain situations.
Finally, we focus on group dynamics in Section~\ref{sec:groups}.
We specifically explore how polarization can result in more misinformation-prone environments from the perspective of both one-sided information consumption (i.e. echo chambers), or cross-cutting, hostile interactions between different groups.
After this, we synthesize these two perspectives and discuss how they may combine to give rise to more polarity.
As a direct consequence, we also briefly discuss radicalization on online platforms as a form of extreme falsehoods about particular groups.
We conclude with recommendations for anti-misinformation interventions, as well as directions for future research, in Section~\ref{sec:discussion}.

\section{Confirming Prior Beliefs: Motivated cognition vs. lazy cognition account}\label{sec:confirming}

In cases where misinformation becomes adhesive, there must be cognitive processes that draw people to it. 
Two theoretical accounts have been put forward. 
One, called the motivated cognition account~\cite{taber_motivated_2006}, suggests that accuracy may be disregarded in favor of information that is more aligned with a person's beliefs and sense of identity (when identity/beliefs are at odds with accuracy). 
That is, information that is more ``preferable'' will win over information that is accurate. 
The other, called the lazy cognition account~\cite{pennycook_psychology_2021}, posits that accuracy is an important feature for users who parse information, however, cognitive overloads can divert attention away from it. 
According to this account, accuracy goals may only be downplayed because of information overload. 
Nonetheless, shifting attention back to accuracy can reduce the adherence to misinformation. 

In this section, we discuss these two accounts and conclude that users have implicit identity-maintenance goals when parsing information, but these can be offset by explicitly highlighting accuracy goals.

\subsection{Biased Assimilation}

Both of these accounts are fundamentally rooted in the idea of biased assimilation~\cite{lord_biased_1979}, which we briefly cover here. 
As mentioned, the assumption that increasing the diversity of information we are exposed to will necessarily translate to more balanced opinions may not hold. 
Most people display bias not only in the views they choose to get exposed to but also in the way they interpret congruent and incongruent views.
For example, experimental studies which have exposed participants to ostensibly real scientific research have found that they tend to judge the methodological rigor, and by extension the validity of the research, based on whether its conclusions agree with their pre-existing beliefs around topics such as capital punishment~\cite{lord_biased_1979} and sexuality~\cite{munro_scientific_2010}. 

Biased assimilation is the phenomenon of accepting congruent information while dismissing incongruent information as invalid, and most likely extends to news sources as well as views encountered online. 
As for research around diversifying exposure, Weeks et al.~\cite{weeks_incidental_2017} find that incidental exposure to counter-attitudinal information (i.e., what~\cite{aslay_maximizing_2018, gillani_me_2018} attempted to achieve) may drive \textit{selective} exposure to attitudinally congruent information, possibly due to a desire to justify the maintenance of existing beliefs.
Similarly, labeling news articles as conservative or liberal-leaning drives selective exposure even in the presence of adjacent credibility labels~\cite{gao_label_2018}.
That is, users who come across belief-disconfirming evidence may either actively seek out information to reaffirm their initial beliefs, or simply choose to directly discount it.
This may thus drive one-sided information consumption even when the information available to the user is fairly diverse and non-partisan (as has been found to be the case with, for example, Google search results~\cite{metaxa_search_2019}).

For instance, Bail et al.~\cite{bail_exposure_2018} asked 1,652 participants to identify their political ideologies (conservative or liberal), and had them follow a Twitter bot that posted content opposite to that ideology. 
In a follow-up questionnaire 1.5 months later, liberals reported being more liberal,  and conservatives reported being more conservative (although only the latter effect was statistically significant). 
Importantly, this effect was stronger for participants who did not unfollow the bot before the follow-up survey, demonstrating that, in this case, avoiding incidental exposure to counter-attitudinal information resulted in relatively less polarized views. 

Corroborating this, a meta-analysis of strictly controlled experiments on biased assimilation~\cite{ditto_at_2019} found that Americans of either a liberal or conservative persuasion are likely to interpret information in a way consistent with their beliefs, for example in punitive judgments of partisans' actions. 
The authors also found that Democrats believe Republicans, but not Democrats, are more biased than the average person and vice-versa. 
Upon a cross-partisan interaction, biased assimilation may thus be more pronounced as partisans may be more defensive against the purported bias of the other side (but see also~\cite{jost_political_2003, jost_ideological_2017, jost_politics_2017, ditto_partisan_2019} for an extended discussion on whether ideological asymmetries in bias exist). 
Because political ideologies can be viewed as aggregations of an individual's moral values~\cite{graham_liberals_2009, haidt_when_2007}, information going against such values may be resisted.
For example, calling for a lockdown of places of worship as a preventive measure for COVID-19 may be met with resistance from people who value religion highly, leading them to downplay the severity of the virus~\cite{sultana_dissemination_2021}.

Biased assimilation demonstrates that people are more accepting of belief-congruent, and more dismissive of belief-incongruent information. 
This means that misinformation that is in line with a general ideology or conceptual prior beliefs may be quite effective in convincing users with such an ideology or beliefs. 
Nonetheless, it is unclear whether there is an explicit preference for such misinformation (i.e., people willingly disregard accuracy in favor of informational compatibility with their beliefs) or whether this is a problem of an implicit (unconscious) inattention to accuracy.

\subsection{Motivated Cognition (Preference Account)}\label{subsec:motiv}

\subsubsection{Ideological motivation}

The motivated cognition account posits that people have a motivation to maintain their prior beliefs even when given information to the contrary.
For example, Taber and Lodge~\cite{taber_motivated_2006} found that their participants passively accepted arguments congruent with their political views, but generated explanations against incongruent ones. 
This in itself is not strictly an irrational response, since one can come to adopt a view precisely because they cannot counter-argue it. 

However, a motivating bias becomes apparent when considering that people are also selectively receptive to scientific evidence~\cite{lewandowsky_motivated_2016}. 
For example, they may deem scientists who oppose their views as less credible~\cite{kahan_misconceptions_2017}, or judge scientific rigor simply based on whether the science agrees with their views or not~\cite{lord_biased_1979, munro_scientific_2010}. 
Further, rational accounts cannot explain why people's existing views can become \textit{more} ingrained when encountering opposing information~\cite{bail_exposure_2018}, or why observable events seemingly evoke identity-protective judgments~\cite{ditto_at_2019}.

Kraft and colleagues~\cite{kraft_why_2015} suggest that motivated cognition occurs due to underlying emotional processes. 
This account builds on previous work which posits that opinion-laden topics are emotionally charged, and create a positive or negative emotional response upon encountering agreeable or disagreeable information (a phenomenon known as hot cognition)~\cite{lodge_automaticity_2005}. 
This emotional response then carries over into subsequent judgments, in what is called affective contagion. Positive emotions prime more acceptance, while negative emotions prime less acceptance of new information~\cite{erisen_affective_2014}.
The effective result is that belief-congruent information creates positive emotions, which subsequently increase the acceptance of this information; the opposite holds for belief-incongruent information. 
In a sense, this resembles a positive feedback loop.

\subsubsection{Social motivation}

Some authors have argued for additional social motivations in the cognition of information processing. Beyond individual identities, people also possess social identities stemming from the norms and values of the social groups to which they belong, such as their race, profession, or political ideology~\cite{jost_social_2004}. 
Because we are members of a virtually infinite number of social groups, which of these identities is most salient depends on social context, as well as the degree to which the individual identifies with that group. 
For example, racial identity will be more salient when someone comes under a racist attack, and one's nationality may be more important to them if they identify as a patriot. 
This forms ``ingroups,'' i.e., social groups we are part of, and ``outgroups.''

Due to the malleability of social identities based on social context, self-categorization into groups can even occur arbitrarily. 
For example, highlighting differences in traits as trivial as art preferences can lead people to act punitively towards outgroup members for no other reason beyond these differences~\cite{tajfel_social_1971}. 
This means that beliefs can be effectively converted into social identities, based on shared agreements with others (for example, groups of ``anti-vaxxers'' against groups of ``vaccine advocates''). 
Polarization can therefore be thought to arise because the very act of sharing, endorsing, or rejecting information and misinformation online can also be conceived of as intergroup conflict over which group holds the better ideology~\cite{slater_reinforcing_2007}. 

Van Bavel and Pereira~\cite{van_bavel_partisan_2018} develop a social identity-based model of political views, in which they argue that accuracy goals in information can contradict the value and belief goals of an individual's partisan identity. 
For example, Republican conservatism may bias them towards information that justifies existing systems~\cite{jost_system_2012} and is aligned with their moral values around loyalty to their nation~\cite{haidt_when_2007}. 
In that case, if accurate information goes against these goals, it is unlikely to be favored in the person's adoption of it~\cite{hart_feeling_2009}. 
Social identities introduce new motivations, in the form of group norms and values, which deviate from a focus on accuracy. 

\paragraph{Identity-protective tactics.} For social identities to be maintained when parsing incongruent information, tactics are employed to preserve ideological predispositions even when they are inaccurate.
Gillespie~\cite{gillespie_disruption_2020} identifies three such tactics: avoiding, delegitimizing, and limiting. 
Avoiding opposing information may include, among others, excluding outgroup members from a conversation or ignoring them when they speak. 
Delegitimizing resembles biased assimilation and may involve things such as stigmatizing the source so as to paint all outgroup members as a homogeneous mass with recycled arguments. 
Limiting methods are usually employed when the oppositional fact is difficult to counter-argue, thus rationalizations to diminish its impact are used; for example, claiming a transgression by an ingroup member occurred as an isolated incident or under difficult circumstances, where the same extenuations would not be granted to an outgroup member.

\paragraph{Conflict-driven information sharing.} If people indeed engage in identity-protective tactics, this points to the idea that social motivations promote polarization.
That is, information is often shared not because of its veracity, but to discredit the outgroup and establish the ingroup.
This makes information sharing combative, not deliberative.
For example, anti-Republican comments on news sources drive Democrats to rate their ingroup more favorably, but Republicans to rate Democrats less favorably~\cite{suhay_polarizing_2018}. 
Furthermore, research on climate change skeptics and believers finds that anger towards the outgroup and identification with the ingroup predicts the degree to which participants will engage in sociopolitical action (such as information sharing) in support of their respective group~\cite{bliuc_public_2015}. 
The two conflicting social identities are the primary drivers of polarity in information sharing behaviors.

\paragraph{Motivating impact of social identity threats.} The role of social motivations in information sharing and adherence acts mainly through threats to the ingroup image. 
In~\cite{rothschild_defensive_2020}, participants who witnessed an ingroup politician's scandal directed outrage towards the \textit{outgroup} more than did participants who witnessed a neutral or outgroup politician's scandal. 
This effect was highest for those who identified with their respective political parties most strongly. 
While this seems counter-intuitive at first, it makes sense in the context of the social motivation account; ingroup scandals threaten the ingroup's moral image, whereas outgroup scandals do not. 
Hence, the former lead to the adoption of identity-protective mechanisms.
Similarly, threatening gamers' identities by showing them studies that suggest a link between video games and aggression can cause them to denounce scientific findings~\cite{nauroth_social_2015}, while there is some tenuous evidence to suggest that partisans are more likely to denounce misinformation when it threatens their ingroup identity and increases their perceptions of media hostility towards them~\cite{cohen_correct_2020}.

\subsection{Lazy Cognition (Inattention Account)}

A recent stream of research has made the case that biased information processing is not an artifact of motivated reasoning, but rather, no reasoning at all~\cite{pennycook_psychology_2021}. 
That is, people do want to be accurate in the views they espouse.
For example, interviews with US university students reveal that, before sharing news on social media, they consider factors such as the credibility of the source and whether they have read the entire article as opposed to simply the title~\cite{wang_more_2020}.
However, people's attention is often shifted away from accuracy by unconscious processes that bring other goals to the surface (these could include processes discussed above, such as threats to an ingroup's moral image). 
Individuals with better analytical reasoning skills are more accurate in discerning fake from real news~\cite{pennycook_lazy_2019}, although they still demonstrate partial bias in rating belief-consistent news as overall more accurate than belief-inconsistent news.

It seems that this improved discernment of analytical individuals stems from a better ability to catch out misinformation, rather than knowing which information is true or not. 
In~\cite{pennycook_fighting_2020}, analytical reasoning primarily drove disbelief in fake COVID-19 headlines, rather than increasing belief in true ones. 
Moreover, when the importance of accuracy in information parsing was highlighted to participants the perceived accuracy of misinformation was reduced by 32.4\%, indicating that accuracy considerations may not be active in users by default, but can be activated externally. 
A similar pattern was observed in~\cite{bago_fake_2020}, where allowing participants to deliberate led to lower belief in fake headlines. 
However, whether participants deliberated or not did not have an impact on attitude polarization, as the difference between ratings of congruent and incongruent headlines was the same in either case.

This suggests that adherence to information, and especially misinformation, is not always driven by motivated or identity-preserving cognition. 
However, the prominence of online misinformation~\cite{vosoughi_spread_2018} also suggests that social media environments are not deliberative settings by default. 
This is exemplified by recent work which found that social media users consider information accuracy to be important, but this does not carry over to their news-sharing intentions (and behaviors) on social media. 
Instead, opinion-congruent headlines tend to be implicitly preferred over accurate ones. 
However, when attention is shifted to accuracy, for example by getting misinformation-prone people to think about the accuracy of unrelated news, the veracity of news shared on social media increases~\cite{pennycook_shifting_2021}.

\subsection{Take-Aways: Synthesizing the Two Accounts}

The main contention in which the two accounts differ is whether there is an explicit motivation to maintain existing beliefs or social identities, or whether the frequent forgoing of accuracy is simply a matter of inattention.
While it is beyond the scope of our review to cover all of their contesting points, there are very useful practical insights to be had from both.

People who parse information do value accuracy~\cite{pennycook_shifting_2021}, yet information processing can very often be an emotional process that acts outside the individual's conscious realization~\cite{kraft_why_2015}. 
These emotional processes can increase adherence to pre-existing beliefs, or lead people to use information in ways that protect the image of groups they are part of~\cite{van_bavel_partisan_2018}. 
In effect, polarizing information, especially when it threatens an ingroup identity~\cite{rothschild_defensive_2020, nauroth_social_2015}, can shift the focus away from accuracy. 
Therefore, anti-misinformation interventions must avoid assuming that accuracy is always the default goal of users, make pre-emptive efforts to establish this goal, and deliver new information in non-identity-threatening ways.

Furthermore, because the social context is crucial to which goals users may implicitly or explicitly adopt, measures should be taken to avoid defensiveness when addressing misinformation. 
For example, Pennycook et al.~\cite{pennycook_shifting_2021} prompted real Twitter users to share more accurate news by asking them to consider the veracity of news \textit{outside} contentious topics such as politics. 
Such approaches are more likely to work because they do not directly challenge the user. This preserves the user's perceived agency and protects against potential reactance~\cite{brehm_control_1993}.

Taking the highly polarized and misinformation-laden online environment (as will be discussed in Section~\ref{sec:groups}), these two accounts together also suggest that social media conversations over contentious issues resemble conflict, not deliberation. 
The role of analytical skills in this context is unclear; for example, while analytical individuals are better at detecting misinformation~\cite{pennycook_lazy_2019, pennycook_fighting_2020}, they are also better at arguing for their ingroup's perspective~\cite{kahan_ideology_2012}. 
The situations under which they would exercise one ability over another are a good future research direction, as is the degree to which accuracy could be embedded as an ingroup value. 
In that case, arguing for an ingroup perspective would run parallel to using accurate, or at least non-misleading, information to do so. 

In the context of the cognitive dissonance framework we described earlier, the work covered in this section suggests that users may find adopting belief- or identity-congruent misinformation less psychologically costly when they are experiencing identity threats, or when they believe their values may be under attack. 
In contrast, adopting such information may be made more psychologically costly when the environment is made deliberative and attention is shifted to accuracy, especially when users experience agency and freedom in expressing their views.

More specifically for research in computational social science, the insights discussed here can point towards a need to characterize different online communities in terms of how deliberative or combative they are.
This is because this categorization can have large implications for social dynamics, particularly how discussions are likely to evolve when members of different groups interact between themselves.
For example, more recently, Rajadesingan et al. \cite{rajadesingan_political_2021} demonstrated that political discussions take place to a very large extent in non-political Reddit forums, however, cross-partisan interactions tend to be less toxic there.
Advancing work in line with this to `map out' deliberative communities is important, as it could also help with predicting where misinformation is most likely to proliferate.
A shift of focus to examining communities from this lens could also assist with detecting linguistic features within deliberative communities to be implemented in more combative ones, as well as allowing CSCW researchers to study general platform architectures that may give rise to combative or deliberative information sharing.

\paragraph{Computational modeling of opinion updating.} Before moving to the next section, it is worth pointing out that a vast majority of research in the field of opinion dynamics, which is concerned with the modeling of individual opinion-updating, has so far mainly adopted the motivated cognition account. 
For example, some models have used bias parameters to capture the reduced likelihood that a user will update their opinion if the information they receive comes from a non-agreeable neighbor~\cite{dandekar_biased_2013}.
Others have applied repulsion and assimilation thresholds, where the probability that a user will update their opinion in a consistent or opposite direction with the information they encounter depends on their attitudinal distance from the neighbor providing the information~\cite{coates_unified_2018}.

These models do provide useful insights.
For example, some research suggests that reducing controversy on an issue is most effective when links are established between the nodes in different clusters which have the highest intra-cluster connectivity, but taking into account the small likelihood that these highly centralized nodes would follow other ``leader'' nodes from other clusters~\cite{garimella_reducing_2017}.

However, most of these models collapse group membership and stance on an issue into the same parameter. 
This should always result in polarization when two disagreeing agents interact, as disagreement (treated the same as network distance) is the determinant of opinion-updating probability. 
On the other hand, some authors have suggested that group identity cannot explain any variance in belief-updating beyond what prior beliefs explain~\cite{pennycook_psychology_2021}, and while we cannot fully assert that this is the case, we do see the value in creating models which separate stance from group identity (see, for example,~\citet{macy_polarization_2021}). 
This approach would enable studying whether de-polarization can be achieved through users who may have a different stance on a specific issue from the rest of their group (or are at least less extreme on their stance), since they may potentially have a greater opinion-updating influence on other users that are within their group (cluster). 
We identify this as another future direction.

\section{Correcting misinformation: What works}\label{sec:correcting}

Misinformation may be more flexible in adopting characteristics that people are drawn to relative to factual information.
Hills~\cite{hills_dark_2019} argues that information ``survives'' through cognitive selection, much like how evolution follows natural selection.
According to him, information is selected if it has the following characteristics: negative, belief-consistent, carrying predictive value, and social. 
Belief-consistency refers to the motivations described above, while negative information refers to a natural tendency for loss aversion which reflects an evolutionary vigilance mechanism~\cite{kahneman_prospect_1979}. 
Predictive information is that which provides insights about future events, but this is often derived in very biased ways, such that illusory correlations~\cite{tversky_availability_1973} or false causalities~\cite{matute_illusions_2015} are sometimes perceived. 
This characteristic is perhaps most relevant in the context of conspiratorial misinformation. 
The social characteristic means that people adopt information which they witness others adopting, which may signal that this information is acceptable in the social environment~\cite{ajzen_theory_1991}.
For example, news pieces which garner a significant number of likes on Facebook are perceived to be more credible, even if they may not be necessarily so~\cite{luo_credibility_2022}.

In many situations, the only advantage of true information over false information is its accuracy. 
However, attention to accuracy is rarely the primary response of users who encounter information online. 
Hills's account provides a good grounding of ways in which misinformation can attract users, which can subsequently increase its proliferation. 
In order to combat misinformation, therefore, we must either understand and disrupt the cognitive mechanisms which enable adherence to it or deliver factual information in ways that make it equally attractive. 
In this section, we explore some cognitive enablers of misinformation, as well as methods that have been successfully -- or unsuccessfully -- employed to combat it. 

\subsection{How misinformation operates on cognition}

The cognitive mechanisms through which misinformation impacts attitudes and opinions are too many to adequately cover and warrant a very extensive discussion on their own. 
However, we focus on three key mechanisms which may be of especial impact to anti-misinformation intervention efforts. 
These have to do with memory processes, social perceptions of misinformation's popularity, and how people may use misinformation as a means of resolving ambiguities.

\subsubsection{Memory and persistence of misinformation} 

\paragraph{Continued influence effect.} Once disseminated, misinformation can be very persistent in memory even when it is explicitly understood to be false. 
The continued influence effect means that people continue to use misinformation in their judgments, even when they are told that such information is completely false~\cite{walter_meta-analytic_2020}. 
Ecker and colleagues~\cite{ecker_explicit_2010} find that information provided to participants, for example, that the victims of a bus accident were elderly people, which is then retracted and stated to be false continues to shape how they interpret particular events. 
This is somewhat diminished when people are warned about the continued influence effect prior to reading any information, and is further diminished when it is combined with alternative explanations; for example, when the retraction goes beyond just stating that the victims were not elderly people, and adds that they were young hockey players (this pattern was also reported in~\cite{johnson_sources_1994}). 
Nonetheless, the continued influence is not fully suppressed in either case.

Misinformation can even affect judgments when people are explicitly warned of which information is true or false before listening to a story. 
For example, in one study, participants listened to a story about a robbery narrated by two voices; a male and a female. 
They were told at the beginning that the male voice always tells the truth, and the female voice always lies. 
Normally, participants should have completely disregarded the female voice. However, the researchers found that when the female voice attenuated the circumstances of the robbery, participants were less punitive to the robber; when it aggravated the circumstances, they were more punitive~\cite{pantazi_power_2018}. 
In further work,~\cite{brydges_working_2018} find that the continued influence effect is mostly due to restrictions in the processing capacity (working memory) of individuals which makes distractions difficult to ignore, rather than the mere storage of information in memory.

For this reason, researchers suggest it is always better to preempt misinformation rather than correct it, whenever possible~\cite{lewandowsky_debunking_2020} (in line with the ``prevention is better than cure'' axiom). 
Recently, advances have been made towards ``gamifying'' inoculation against misinformation. Inoculation theory proposes that individuals can build up cognitive immunity against certain types of arguments or information, by being exposed to weakened forms of them~\cite{mcguire_effectiveness_1961, mcguire_resistance_1961}. 
While initially proposed as an explanation of how people can become highly resistant to attitude change, it has been successfully used in getting people to discern fake headlines by exposing them to mild misinformation in an anti-misinformation game\footnote{\href{https://www.getbadnews.com}{https://www.getbadnews.com}} \cite{basol_good_2020, maertens_long-term_2020}.
Similarly, the game ``Fakey'' has shown promise in getting users to more accurately discern true news pieces in a dynamic, time-relevant news environment after repeated plays~\cite{micallef_fakey_2021}.
Achieving widespread public engagement with such games remains an open challenge. 

\paragraph{Misinformation effect.} Nonetheless, misinformation can act in more subtle ways. 
Especially when one is not attentive, misinformation is capable of altering their memory even about events they witnessed~\cite{loftus_planting_2005}.
For example, asking participants to estimate the speed of a car crashing into another in a video will elicit higher estimates if the collision is described as a ``smash'' rather than a ``contact''~\cite{loftus_reconstruction_1974}. 
The medium of communication also seems to affect this, for example with spoken form having a greater leading effect than written form in some cases~\cite{goldschmied_appraising_2017}. 
This has implications for information that may not be explicitly false but framed in a misleading manner, as well as what format (e.g. video vs. article) misinformation is most effective in.

The misinformation effect seemingly serves the maintenance of existing beliefs. 
Jones and colleagues~\cite{jones_believing_2017} showed their participants footage of a police arrest, and had them read a report by the arresting officer which did not align with the footage. 
When later asked about what they saw in the footage, pro-police participants believed the officer against what they saw and falsely recalled the suspect carrying a knife, whereas non-pro-police people did not make this recollection error. 
Therefore, even for witnessed events, a lack of attention to specific cues can cause memory faults that are consistent with a person's position on an issue.

\subsubsection{False consensus}

The perception of how others act online can be a very important driver for a user. 
Because of this, the proliferation of misinformation by often unwitting third parties is a particularly pertinent problem. 
People are susceptible to a false consensus bias~\cite{ross_false_1977}, which is a tendency to mistakenly believe that one's behaviors, attitudes, and beliefs are shared by the majority of the population. 
If someone is inclined to believe misinformation, and then witnesses others spreading it, then this bias may be amplified.

For example, a study conducted in Australia found that, while only 7.2\% of the approximately 5000 surveyed respondents believed climate change is fake, those respondents thought their belief was held by 50\% of the Australian population (on the contrary, those who believed in climate change slightly underestimated the true proportion of people who shared their beliefs)~\cite{leviston_your_2013}. 
Those most susceptible to false consensus were also the least likely to change their opinions. 
False consensus may be particularly pronounced when an individual is typical of their social environment (for example, being white in majority-white America)~\cite{yeager_moderation_2019}, which may suggest that restricting social environments by entering echo chambers can amplify false consensus perceptions.

Sharing misinformation reinforces false consensus. 
Laypeople are poor at distinguishing between the number of primary versus secondary sources which support a view, focusing instead  on the number of instances that argue for or against a position. 
In other words, whether five articles cite five different sources supporting a particular point, or if all five articles cite a single source, does not make much difference in the perceived credibility of this information. 
This effect persists even when it is hinted that the primary source is not an expert, when people are instructed to pay attention to which sources are cited, and even when they are asked to explicitly reason whether the number of primary sources affects credibility~\cite{yousif_illusion_2019}.
Therefore, a single piece of misinformation amplified by various channels can create the illusion that this information is credible.
Harvey et al.~\cite{harvey_internet_2018} demonstrated this problem by showing that 80\% of the 90 climate-change-denying blogs in their sample all cited a single person (who did not have sufficient expertise on the subject matter) to make the case that polar bears are not endangered.
Similarly, another study showed that scientific figures who oppose the general scientific consensus on COVID-19 issues are disproportionately amplified to a large extent on the Twitter platform~\cite{efstratiou_misrepresenting_2021}.

Importantly, Pennycook et al.~\cite{pennycook_fighting_2020} find that analytical individuals who are more likely to detect misinformation are also less likely to share news online, meaning vocal minorities who believe in misinformation may dominate the online environment and exacerbate false consensus. By extension, the perceived social acceptance to believe such false information can amplify adherence to it. Disproving this sense of consensus should partly adjust users' beliefs since people tend to behave in line with the norms they \textit{believe} govern their community~\cite{ajzen_theory_1991, goldstein_room_2008}. However, this must be done in subtle ways that do not motivate users to actively seek attitudinally congruent information at the expense of accurate information~\cite{weeks_incidental_2017}.

\subsubsection{Ambiguity resolution}\label{sec:ambiguity}

Very often, events may induce uncertainty or ambiguity in the public, which will urge them to seek out explanations.
Misinformation can fill these gaps in people's understanding~\cite{lewandowsky_beyond_2017, lewandowsky_debunking_2020}.
As in the case of motivated cognition, the need to resolve ambiguity can supersede informational veracity. When individuals are high in their need to find closure (i.e., are highly averse to ambiguity), they show a decreased ability to consider mixed information.
For example, they may disregard one side of the information in favor of the side which agrees with their prior beliefs in contexts such as vaccine efficacy~\cite{nan_biased_2015}.
Further, Pica et al.~\cite{pica_role_2014} find that those high in need for closure are more susceptible to false memories (as described in Section~\ref{sec:correcting}) because they suppress conflicting memories to resolve ambiguity. 

Similarly, if no ambiguity is perceived, people may feel that a piece of misinformation adequately explains an event.
Extremity on certain issues seems interlinked with a false perception that one understands something they do not. Fernbach et al.~\cite{fernbach_political_2013} found that inducing uncertainty by asking participants to provide mechanistic explanations of policies reduced their self-reported understanding of these policies, and led to lower position extremity on them as a result.
Misinformation may thus also exert its influence through a mistaken belief that the receiver understands a topic to a much greater extent than they actually do~\cite{rozenblit_misunderstood_2002}.
This is also related to overconfidence biases attributable to an underestimation of an issue's true complexity~\cite{kruger_unskilled_1999}.
Indeed, belief in conspiracy theories is closely related to overconfidence in one's ability to provide explanations for certain policies or events~\cite{vitriol_illusion_2018}, and many such conspiracies attempt to fill a gap in understanding with easy to parse, yet false, information.

Ambiguity resolution may explain recent findings around how willing people are to consult other sources with respect to the accuracy of news they are consuming. When news concern a novel topic, as was the case with COVID-19 in late 2019 to early 2020, people are more receptive to suggestions by an AI regarding news veracity. However, with established topics where belief systems have become more established and the initial ambiguity has been resolved, they become less receptive to such suggestions~\cite{horne_tailoring_2020}. Similarly, while major events such as shootings in the US spark widely spanning conversations, these conversations become quickly polarized into ideological groups over time~\cite{barbera_tweeting_2015}. Consulting mixed sources retains much of this ambiguity which is psychologically costly, whereas ``picking sides'' can resolve this cost.

\subsection{Fact-checking and corrections}

One of the most commonly utilized methods of combating misinformation is fact-checking and debunking.
This can be done by social media platforms that can directly label certain posts as false, through independent fact-checking organizations, or, occasionally, from other users themselves.
Besides the vast amount of time and staff resources that fact-checking requires, its effectiveness varies and it is seemingly most robust when the biases described above are taken into account.

Two recent meta-analyses have provided very useful insights about when fact-checks are most likely to work.
Chan et al.'s meta-analysis~\cite{chan_debunking_2017} found, across 20 different experiments 
(N = 6878), that debunking is least effective and misinformation is most persistent when people initially generate more explanations in line with the misinformation.
On the contrary, misinformation becomes less persistent and debunking more effective when people counter-argue the misinformation \textit{before} they are able to consolidate it.
Perhaps most importantly, debunking is most effective when it is detailed (i.e. provides alternative explanations) rather than vague (i.e. simply stating the misinformation is false), consistent with ambiguity resolution~\cite{lewandowsky_beyond_2017, lewandowsky_debunking_2020, johnson_sources_1994, ecker_explicit_2010}.
Walter et al.~\cite{walter_fact-checking_2020}, analyzing 30 experiments (N = 20,963), find that fact-checking is less effective when claims are only partially debunked, it can meet resistance by strongly identified partisans, and perhaps not surprisingly, is not as successful against false election campaign claims since election periods are characterized by elevated polarization and partisan bias.

Political misinformation appears to be especially intricate.
When politicians spread misinformation, their supporters are naturally more likely to believe it than their opponents~\cite{swire_processing_2017, neumann_labelling_2020}.
Interestingly, correcting misinformation shared by Donald Trump does reduce his supporters' belief in the misinformation, however, it does not change their voting intentions~\cite{swire_processing_2017} or attitudes towards him~\cite{nyhan_taking_2020}.
Consistent with the continued influence effect, belief in the corrected misinformation is restored after a week~\cite{swire_processing_2017}.
This is important because malicious actors may be able to falsely attribute misinformation to political figures and remain fairly confident in its longevity in the informational environment.
What follows may be the creation of a chaotic political discourse, with the aim to sow discord, undermine reasonable arguments, and disengage large numbers of the public from politics~\cite{pomerantsev_how_2014}, leaving the landscape occupied by the most polarized and misinformation-prone people.

Fact-checkers can reduce polarization on issues such as immigration and climate change, however false but attitudinally congruent information is still more adhesive than true, but incongruent information~\cite{hameleers_misinformation_2020}.
Media distrust and higher education levels seem to drive active searching for fact-checks to some extent~\cite{hameleers_misinformation_2020}.
However, the effects of debunking are multi-faceted and interact with various other environmental features. In realistic environments, it is likely that exposure to fact-checking needs to be consistent and relatively undisrupted by countering information.
In laboratory simulations of election environments, for example, participants advocate for their initially preferred candidates more strongly in the face of small amounts of negative information about them.
They do not update their beliefs until this negative information reaches a certain threshold, raising questions around the number of corrections that need to constitute the informational environment~\cite{redlawsk_affective_2010}.

Many studies have examined and established the effectiveness of fact-checks in controlled settings, however, their role in the context of distracting information has not received enough attention.
On social networks, users are unlikely to be consistently exposed to counter-attitudinal information~\cite{weeks_incidental_2017}.
Thus, the fact-checking impact may be attenuated if conflicting information distracts from or counteracts the corrections, and could even have adverse effects if the distracting information promotes distrust towards the fact-checker.
We identify the examination of fact-check effectiveness in continued exposure vs. disrupted exposure settings, as well as quantifying the proportion of fact-checks that must constitute the informational environment as potential research avenues, and explore the extent to which fact-checks can have an adverse effect in the next section.

\subsection{Backfire effects: Fact or fiction?}

Many scholars have expressed concerns regarding fact-check efficacy, due to potential ``backfire effects''.
These were first reported in a study by Nyhan and Reifler~\cite{nyhan_when_2010}, who found that showing claims made by politicians to partisans who endorsed them, alongside corrections to these claims, could \textit{increase}, rather than decrease the degree to which partisans believed misinformation.
This is based on the theory of motivated reasoning described above~\cite{taber_motivated_2006}. 

\paragraph{Failures in replicating the backfire effect.} The documentation of this effect spurred many replication attempts, however, there is a recent trend towards a consensus that this effect is overstated and very elusive~\cite{lewandowsky_debunking_2020}.
For example, Ecker et al.~\cite{ecker_people_2014} tested the impact of retracting news that falsely identified robbery suspects as Aboriginals.
If the backfire effect applied in this case, then participants who were prejudiced against Aboriginals should have rejected the retraction and amplified their belief in the misinformation that the robbers were Aboriginal. However, no backfire effects were found.
Both prejudiced and non-prejudiced participants updated their beliefs to the same extent, and both seemed to accept the retraction.
Although an extensive debate of the replicability of backfire effects is beyond the scope of our review, it does seem that their extent is very limited and contextual~(see \cite{swire-thompson_searching_2020} for a good discussion).
Nonetheless, there are useful insights from research that has probed this effect. 

Wood and Porter~\cite{wood_elusive_2019} attempted to test the backfire effect across five experiments where participants were either shown only a false statement by a political leader or the statement along with its correction.
Participants were more likely to believe misinformation that was in alignment with their own political ideologies, however, corrections were effective in reducing belief in it (with an effect size as large as that of ideology).
While the authors were not able to replicate the backfire effect, their findings did suggest that misinformation corrections may operate at surface levels.
For example, while both conservatives and liberals seemed to accept a correction which clarified that no weapons of mass destruction were found in Iraq prior to the 2003 invasion, conservatives (but not liberals) mistakenly reasoned that this was because Saddam Hussein managed to conceal them prior to abandoning the storage sites. 

This exemplifies an important point; corrections may indeed work in rejecting misinformation, however, this may happen at surface level.
As explained in Section~\ref{sec:ambiguity}, if gaps in understanding are not adequately addressed by a correction, they are vulnerable to becoming filled with further misinformation.
Therefore, measuring the \textit{holistic} impact of fact-checks on beliefs, and not just changes in beliefs for incredibly narrow issues, would allow for capturing the impact of fact-checking on cognitive patterns and adherence to misinformation in general (such that the root, and not the symptom, is addressed). 

\paragraph{Labels and implied truth.} In further work, Pennycook et al. provided evidence suggesting that corrections are effective in reducing the perceived accuracy of misinformation~\cite{pennycook_implied_2020}.
The backfire effect was not observed in its traditional sense, however, the authors made the case for an implied truth effect; in studies where misinformation was labeled as false, anything which was not labeled was considered as more accurate than baseline (even if false).
This effect was diminished when true information which had been verified was also tagged as such, as this clarified that unlabeled information simply had not been examined, and was thus not necessarily true.

Overall, backfire effects are most probably a very minimal risk, but they are still observed in some cases (e.g., when the target group is very strongly opinionated on the subject of the fact-check)~\cite{swire-thompson_searching_2020}.
Fact-checking may need to also prevent the generation of alternative explanations for a piece of misinformation that has been corrected~\cite{wood_elusive_2019}, meaning that corrections must be comprehensive and multi-faceted.
Interventions that follow a corrective approach should also consider that corrections can inadvertently increase the perceived accuracy of information that is false but not yet corrected~\cite{pennycook_implied_2020} (although this also needs to be verified in realistic social media contexts).
More anecdotally, backfire effects could occur (or at the very least, fact-checking could be much less effective) with people who distrust official government or scientific institutions.
Distrust in official institutions has been found to be a major driver of adherence to COVID-19 misinformation~\cite{enders_different_2020, pickles_covid-19_2021, roozenbeek_susceptibility_2020}, and therefore studying backfire effects in the context of this specific demographic may be worthwhile.

\paragraph{The role of messengers in the backfire effect.} Before we dismiss the backfire effect, we should note a very important consideration.
A vast majority of fact-checking research focuses on fact-checkers as official organizations or indirectly reassures that the correction provided in the research is true.
Much less attention has been paid to fact-checking by users, whether that is in the form of generating the fact-checking themselves or disseminating fact-checks by official sources.
In real-world settings, it is likely that users who will disseminate fact-checks against oppositional viewpoints account for a large proportion of the prevalence of misinformation corrections online.

People are more likely to trust misinformation when it comes from a source they support~\cite{swire_processing_2017, neumann_labelling_2020}, and this may extend to fact-checking as well.
A consistent factor that impacts upon the effectiveness of any intervention is the messenger that delivers it~\cite{dolan_influencing_2012}.
While backfire effects are elusive in the context of official fact-checkers, they may be much more prominent in cross-partisan interactions where one person attempts to correct another, due to the purported bias and mistrust in the other side~\cite{ditto_at_2019}. 

There is at least some evidence to suggest that users may categorize online sources into outgroups and ingroups~\cite{wilkins_one_2020}, however, the impact of this on the receptivity of corrections to misinformation remains something that has not explicitly been studied. Margolin et al.~\cite{margolin_political_2018} do find that Twitter users are approximately 2.5 times more likely to accept a fact-check if there is a reciprocal follow relationship between them and the account sharing the fact-check.
Furthermore, Parekh et al.~\cite{parekh_comparing_2020} find that, while fact-checks aimed at correcting misinformation are generally well-received in neutral political communities on Reddit, they garner negative reactions from users in more partisan communities, and they are often used combatively to establish a dominance of opinion.
It remains unanswered whether group memberships of the messenger also affect acceptance of fact-checks.

This is very possible because perceiving more similarities with the messenger is likely to reduce reactance and increase the chances that a person will comply with the message~\cite{silvia_deflecting_2005}.
For example, Munger~\cite{munger_tweetment_2017} finds that white male Twitter users targeting black users with racist harassment reduce their use of hateful language when they are confronted by a bot (purporting to be a real user), but only if the bot assumes the identity of a white man and also has high authority (high number of followers).
On the contrary, a subset of harassers who do not attempt to conceal their identity \textit{increase} their use of hateful language if they are confronted by a low-authority bot which assumes the identity of a black man.
Recently, Allen et al.~\cite{allen_birds_2022} have demonstrated that users of Twitter's crowd-sourced, fact-checking platform Birdwatch mostly tend to fact-check pieces which are ideologically opposed to them, and they are more likely to label fact-checks from other users as ``helpful'' if that user is part of the same political group as them (although crowd-sourced fact checking does indeed show promise in reducing the spread of misinformation).
Similarly, other work finds that users are much more likely to share fact-checks which are aligned with their own political views (although it is suggested that Democrat users are less biased than Republican users in that respect)~\cite{shin_partisan_2017}.

Therefore, we advise against fully dismissing the backfire effect in the context of user-to-user interactions, especially when fact-checks are weaponized to maintain ingroup image.
Future research could examine the impact of corrections shared in this context, particularly whether they polarize user beliefs, and whether they can drive distrust in fact-checking organizations.
This could come in the form of, for example, examining whether sharing of fact-checks is more prominent where topic controversiality is high, or monitoring the quality and bias of news consumed by users following interactions where they are exposed to fact-checks.

Relatedly, Mosleh et al.~\cite{mosleh_perverse_2021} tested the impact of fact-checking real users on Twitter using bot accounts which appeared like human users.
They found that, generally, fact-checking reduced the quality of news featured in these users' subsequent retweets (but not original tweets), increased their toxicity, and pushed them further towards their political ideology.
While the authors also attempted to test for the effect of the fact-checking (bot) accounts' political affiliation, their number of observations was too underpowered to detect a true effect.
Backfire effects may therefore be more likely ``in the wild'' rather than in laboratory experiments, however, larger-scale studies are required to examine the true impact of fact-checking on online platforms.

\subsection{Take-Aways: Considerations from Cognition Research}

In this section, we have covered some socio-cognitive enablers of misinformation.
Specifically, we discussed why misinformation is so adhesive in people's memory, how false perceptions of others' beliefs can amplify its dissemination, and how it is often used to resolve ambiguities in sub-optimal ways.
We also discussed the conditions under which corrections are likely to be most effective, as well as considerations on when they may backfire.

Returning to our cognitive dissonance account, we can conclude that the psychological cost of adhering to misinformation is reduced when people are unaware of the persistent effects of misinformation in memory~\cite{ecker_explicit_2010, johnson_sources_1994}.
Further, misinformation adherence is made less psychologically costly when corrections fail to resolve people's gaps in understanding different events~\cite{ecker_explicit_2010, chan_debunking_2017, lewandowsky_debunking_2020}, or when they are not exhaustive enough to prevent relying on alternative explanations~\cite{nyhan_when_2010, wood_elusive_2019}.
Oftentimes, a single explanation may be more impactful than multiple disjointed ones, especially if it satisfies the ``cognitive selection'' criteria (social, predictive, belief-consistent, negative;~\cite{hills_dark_2019}).
For example, some of the biggest amplifiers of COVID-19 anti-vaccine misinformation are involved in alternative medicine industries\footnote{\href{https://www.counterhate.com/disinformationdozen}{Center for Countering Digital Hate: The disinformation dozen}}.
Presenting this conflict of interest as an explanation of why they push anti-vaccine misinformation may be highly impactful, as it demonstrates clear cause and intention (predictive), is consistent with conspiratorial mindsets, is negative against the misinformers, and may gain enough traction to become social. 

People's psychological cost in adhering to misinformation may be increased when corrections comprehensively resolve their gaps in understanding, when they realize that the prominence of their misinformed opinions is fringe~\cite{harvey_internet_2018, leviston_your_2013, ross_false_1977, yousif_illusion_2019}, rather than widespread, and when they are indirectly made to understand that certain issues that they are misinformed on are much more complex than they initially thought~\cite{fernbach_political_2013, vitriol_illusion_2018}.
Generally, we can say with a fair degree of certainty that fact-checks and corrections are successful in increasing the psychological cost of adopting misinformation~\cite{swire-thompson_searching_2020}.
However, we advise caution in attempting to generalize this to corrections that come up in non-deliberative settings, especially when they come from an ideological opponent in a context where the focus is not on accuracy, but on identity maintenance.
We identify studying the role of combative fact-checking as a future research direction. In the next section, we discuss the potential of combative information sharing in a wider intergroup context.

\section{Internet communities and group perspectives}\label{sec:groups}

Throughout the review, we have alluded to the importance of group considerations and polarization in adherence to misinformation.
Indeed, scholars have established a correlation between users' susceptibility to sharing low-quality sources and their degree of partisanship or network clusteredness~\cite{nikolov_right_2021}.
Furthermore, classifier models for detecting misinformation-prone topics make use of topic controversiality as an important predictor~\cite{vicario_polarization_2019}.
This section focuses on analyzing these group-based processes in more depth through the lens of social psychological theory.
We integrate perspectives on online echo chambers and hostile interactions between oppositional users, concluding that one should be thought about in the context of the other.
We also discuss extreme forms of polarization, namely extremism and radicalization, as enablers of outgroup-directed misinformation which make the deterioration of an outgroup's image a much more important goal than accuracy.
As we have argued before, all of these phenomena can shift the information exchange setting from a deliberative to a combative one.

\subsection{Echo chambers and impact on online discourse}

Echo chambers are a common term in misinformation research and refer to closed-off online groups which do not interact with users that do not share their views.
This means the same views are consistently amplified (or ``echoed'') within the group, resulting in higher attitude polarization.
Because some unique groups may consume disproportionate amounts of misinformation (e.g., conspiratorial communities), echo chambers and their disruption constitute an important aspect of misinformation research.

\subsubsection{Documentation of echo chambers in online platforms}

The echo chamber effect has been observed on numerous occasions and in different ways.
A recent, large-scale analysis of 5.1B comments across 14 years of Reddit activity reports that users predominantly interact with political spaces that align with their ideology~\cite{waller_quantifying_2021}.
Similarly, network polarity on Twitter is higher for political than non-political topics~\cite{garimella_political_2018}, with people who show bipartisanship receiving lower engagement seemingly due to lower preference for non-polarized discourse.
For example, work examining discourse around the Black Lives Matter movement on Twitter finds that this discourse is segregated into a left-leaning and a right-leaning ``supercluster'', each of which amplifies members within those clusters~\cite{stewart_drawing_2017}.
Further, a field experiment causally establishes that people are three times more likely to follow another user on Twitter if they share the same political leaning~\cite{mosleh_shared_2021}.
Even for conversations that may initially cut across ideologies, such as discussions regarding mass shootings on Twitter, these diverge into polarized discourse over time~\cite{barbera_tweeting_2015}, while other events, such as referenda, can also lead users to form spontaneous, distinct communities based on the Twitter and Facebook pages they follow~\cite{del_vicario_public_2017}.

Echo chambers are also observed in the context of conspiratorial and anti-scientific communities.
These may be particularly challenging, since belief in one conspiracy increases the likelihood of believing in others due to conspiratorial cognitive patterns~\cite{douglas_psychology_2017,leman_beliefs_2013,stojanov_conspiracy_2019}.
For instance, believing in one piece of health misinformation increases the likelihood of believing in another~\cite{scherer_who_2021}.
Facebook users tend to like either anti-vaccination or pro-vaccination pages almost exclusively, such that users who leave likes on both types of page are extremely rare~\cite{zollo_dealing_2019,zollo_debunking_2017}.
Similar patterns are observed in terms of engagement where individual users' activity is predominantly concentrated either in conspiratorial or scientific pages~\cite{brugnoli_recursive_2019}, with this activity also being affected by the users' polarized neighborhoods.
Conspiracy echo chambers may also hinder fact-checking efforts, since as little as 1.2\% out of 9M conspiratorial users interacted with fact-checks which were aimed towards them in~\cite{zollo_debunking_2017}.

\subsubsection{Breaking the chamber: Cross-cutting interactions}

Although breaking echo chambers has been the subject of previous research~\cite{aslay_maximizing_2018,gillani_me_2018}, some scholars have suggested that echo chambers are not a prominent problem.
Rather, it is supported that people occasionally interact with oppositional users themselves, but it is precisely these interactions that drive polarization due to their hostile nature. 
For example, some authors support that there is reasonable diversity in the news sites that users consult~\cite{guess_less_2019,guess_why_2018}, and people generally report that they often encounter viewpoints they disagree with~\cite{dubois_echo_2018}.

On the Reddit platform, recent research has found that cross-cutting interactions between Trump and Clinton supporters on a somewhat neutral political forum (r/politics) are very common (and hostile)~\cite{de_francisci_morales_no_2021}, thus opposing the echo chamber narrative.
Further, Marchal~\cite{marchal_be_2021} finds that cross-cutting interactions between liberal and conservative users on r/politics tend to be more hostile, and such hostile interactions are more likely to be cut short. 
In similar work, Cinelli et al.~\cite{cinelli_echo_2021} report that echo chambers may be more prominent on network-based platforms such as Facebook or Twitter than other platforms such as Reddit or Gab (although note that, while Gab is also a microblogging platform like Twitter, it can be considered an alt-right echo chamber in itself~\cite{zannettou_what_2018}).
Such work would suggest that echo chambers are mostly the result of algorithmic filter bubbles rather than an organic preference for one-sided information consumption.
Another study finds that interactions between oppositional users on a controversial YouTube video are frequent, hostile, and cause further polarization because users have a desire to argue, rather than engage in coversation with, outgroup members~\cite{bliuc_you_2020}.

Hostile cross-cutting interactions are not always incidental. 
Datta and Adar~\cite{datta_extracting_2019} find that political communities on Reddit often target each other, predominantly over election periods.
Furthermore, Reddit communities may occasionally mount ``brigading attacks'' on others, as was the case with supporters of Donald Trump on the subreddit r/The\_Donald attacking another Reddit forum, r/politics~\cite{mills_pop-up_2018}.
Cross-cutting interactions may therefore take place explicitly in bad faith, or they may simply get derailed by the temper of the exchanging users.
In any case, these discussions are likely to sideline accuracy as a primary motivation, making them especially fertile grounds for adhering to misinformation.

\subsubsection{Towards integrating cross-cutting interactions and echo chamber narratives}

Hostile cross-cutting interactions are sometimes treated as a better, rather than a complementary, explanation of polarization and the spread of misinformation~\cite{de_francisci_morales_no_2021}.
Nonetheless, echo chambers are most likely indeed a very prominent issue.
Studies which examine cross-cutting interactions tend to focus on isolated incidents or communities~\cite{de_francisci_morales_no_2021,bliuc_you_2020}, whereas larger-scale studies find support for the echo chamber effect~\cite{waller_quantifying_2021}.
Furthermore, a systematic review of studies probing this effect finds that research utilizing digital trace data tends to find at least some evidence of echo chambers, while self-reported user data tends to downplay their existence~\cite{terren_echo_2021}.
This could be due to social desirability biases of surveyed participants~\cite{fisher_social_1993}, or due to users themselves not quantifying the degree to which they encounter cross-cutting information in an objective manner (e.g., because counter-attitudinal information may stand out more than attitude-consistent information).
Importantly, some research suggests that echo chambers are formed primarily due to users seeking out consistent viewpoints, and not so much due to them actively avoiding opposing ones~\cite{garrett_echo_2009}.

Given what we have covered, we argue that both echo chambers and hostile interactions are well-documented phenomena, but they are by no means mutually exclusive.
Indeed, we believe that treating the existence of echo chambers as a binary problem (i.e., exist or do not exist) is of little practical importance.
Rather, we believe that involvement in echo chambers should be conceived on a continuous spectrum and for each \emph{unique} user, allowing for the quantification of the relationship between information consumption diversity and hostile engagement with cross-cutting content.
For example, it could be the case that initial pushbacks against oppositional content due to high echo chamber involvement are reversed when a user further diversifies their information consumption (i.e., reaches a ``tipping point''~\cite{redlawsk_affective_2010}).
In other words, it is the proportion of interaction that occurs with like-minded vs. different-opinionated users that must be measured, not simply whether cross-cutting interactions occur to a sufficient extent.
Crucially, the study of echo chambers should situate the user within their wider consumption patterns, and not only within narrow communities of interest~\cite{morini_toward_2021}.

Furthermore, cross-cutting interactions need not be negative.
Much of animus-enabled misinformation may spread because of prejudice and stereotypes against another group, which reduces them to a homogeneous mass with a lack of individuality~\cite{frederic_heterogeneity_2018}.
However, positive contact can occur under certain preconditions such as the two interacting groups having shared goals and a respect for mutual values~\cite{allport_nature_1954}.
Under deliberative settings where these preconditions are satisfied, people seem to forgo their biases; those who base their opinions on evidence are resistant to opinion change by others, however those who hold their opinions based on insufficient evidence are amenable to change, even when their opinions are extreme~\cite{zhang_encountering_2019}.
To that end, there is a need to understand the situations where information-sharing is likely to be combative and not deliberative.
In the next subsection, we explore this question in more depth through the lens of intergroup prejudice, radicalization, and outgroup-directed misinformation.

\subsection{Outgroup-directed misinformation: Prejudice and radicalization}

As we have argued above, polarization can enable adherence to misinformation by drastically demoting accuracy goals.
This can come about through prejudice and dehumanization of another group, and it can lead to more extreme forms of polarization and group isolation.
Understanding some motivations behind radicalization and extremism can therefore help to better situate these phenomena in the context of combative information-sharing.

Groups often compete for resources~\cite{sherif_groups_1953}, and can experience what is known as relative deprivation when they have access to fewer resources than their competitors~\cite{carrillo_relative_2011,merton_social_1968}.
This can also occur symbolically.
For example, when a group feels that its values are threatened by another it may direct more prejudice towards that group~\cite{stephan_intergroup_2017}.
Relative deprivation is closely linked with extremism.
It can explain why anti-immigration attitudes can lead to voting for far-right candidates~\cite{urbanska_swaying_2018}, and it is related to espousing martyrdom~\cite{ozdemir_deprivation_2020} and support for violent action~\cite{obaidi_group-based_2019}; all of these scenarios are quite unlikely in a healthy information environment.
Both minority groups (due to social justice issues) and majority groups (due to threats to the status quo) can experience relative deprivation, and thus be led to extremism~\cite{kunst_understanding_2020}.

\subsubsection{Trends towards extreme behavior}

In the context of extremism, any information shared primarily aims towards painting a target outgroup as the villain~\cite{reicher_making_2008}, and not towards being accurate.
This can have substantial consequences.
For example, engaging with calls for collective action against an outgroup increases extremist vocabulary over time, and may lead to support of terrorist groups~\cite{smith_detecting_2020}.

A recent stream of research has suggested that extremism is on the rise, particularly towards the far-right.
This could be due to the ideological differences through which right-wing advocates favor simpler, easy-to-understand narratives (which extremist rhetoric may provide in the form of outgroup-directed misinformation)~\cite{jost_political_2003, jost_ideological_2017, jost_politics_2017}.
A study on YouTube~\cite{hosseinmardi_evaluating_2020} found that many users transitioned from consuming moderate-right content to consuming alt-right content, with a three-fold increase in alt-right content engagement between 2017 and 2019.
Ribeiro et al.~\cite{ribeiro_auditing_2020} also found that users could be led to consume extreme right-wing videos after consuming non-extreme or semi-extreme right-wing content, and similarly recorded a surge in engagement with these types of videos since 2015.

Content moderation has been put forward as one potential solution against radicalization, however, a study on banned Reddit communities which migrated to websites of their own raises some important considerations~\cite{horta_ribeiro_platform_2021}.
Although the size of these communities shrank following the bans, the migrations were marked by heavier use of toxic and outgroup-directed language (e.g., ``they'' pronouns).
The authors found that this was most likely due to the most extreme users being the ones who migrated to begin with.
Individually, users also became more tribalistic, toxic, hostile, and negative following the migrations. 
Therefore, while content moderation may reduce the size and reach of fringe communities, it may also ``box in'' the most extreme members who are then likely to radicalize the community even further.
This is consistent with a phenomenon known as risky shift, where the collective action of groups is more extreme than the average extremity of each individual group member~\cite{kogan_risky-shift_1967}.

\subsubsection{Rationalizing prejudice and dehumanization}

Much of extremist rhetoric centers around protecting the ingroup from the hatred of the outgroup, therefore radicalized people may believe they are reacting to something rather than being perpetrators~\cite{reicher_making_2008}.
Partisans in the US have been found to overvalue the degree of prejudice and dehumanization which is directed to them by outgroup partisans by a factor of approximately two~\cite{moore-berg_exaggerated_2020}, using this rationalization to justify greater prejudice and dehumanization towards the outgroup in turn.
Dehumanization could come in the form of believing that the other group is less civilized or evolved, thinking that the outgroup does not deserve as much as the average person, or viewing the outgroup as something less than human~\cite{moore-berg_prime_2020}.

Similarly, partisans perceive higher amounts of dislike, opposition, and reprimand of their own actions from outgroup partisans than these outgroup partisans actually report~\cite{lees_inaccurate_2020}; in other words, they have exaggerated meta-perceptions for outgroup perceptions of the ingroup.
This is primarily driven by a belief that outgroup actions are meant as obstructionist mechanisms against ingroup goals.
However, correcting these meta-perceptions can reduce belief in outgroups as obstructionists and improves intergroup attitudes~\cite{lees_inaccurate_2020}.
Evidently, mistaken meta-perceptions can drive a reduced willingness to engage with outgroup members, and increased prejudice between groups~\cite{moore-berg_exaggerated_2020}.
This, therefore, can maintain the attractiveness of outgroup-directed misinformation.

When people are led to believe that an outgroup dehumanizes them (e.g., in the context of Americans and Iranians, Israelis and Palestinians, Hungarians and Romani, etc.), they reciprocate this dehumanization and endorse more aggression towards this outgroup.
On the other hand, witnessing an outgroup humanizing (e.g., admiring) the ingroup also results in reciprocated humanization~\cite{kteily_they_2016}.
Importantly, people do display a self-centering bias in that they do not anticipate reciprocated dehumanization when they are the first to dehumanize an outgroup, suggesting that animosities targeted from an outgroup towards the ingroup are likely to be seen as provocations rather than responses to ingroup-initiated transgressions~\cite{kteily_they_2016}.
Perceived outgroup-imposed threats and endorsement of right-wing authoritarian ideas further make outgroup dehumanization more likely~\cite{kteily_darker_2017}.

Overall, animosities between groups can be exaggerated in the minds of their members (although see~\cite{abeles_perception_2019} for an example where disagreements are underestimated), leading to evocative cascades of further escalation.
In the information environment, this can create a highly polarized, highly combative setting where corrections and information more generally may be viewed as a mobilization against the ingroup, and drive radical collective action.

Correcting mistaken meta-perceptions may be a good avenue towards reducing extremism, polarization, and ultimately, combative misinformation spreading. Showing people statistics of the true outgroup's perception towards them has been effective in previous studies~\cite{lees_inaccurate_2020}, however, enabling positive contact between groups so that radicalized or polarized individuals can adjust their meta-perceptions first-hand may also be a good approach~\cite{allport_nature_1954}. More generally, we urge researchers to examine the prevalence and role of meta-perception distortions in extreme content.
This could come in the form of comparing how one group actually talks about another group, to the rhetoric of the latter group about what the former group purportedly says about them.

\subsection{Take-Aways: Group dynamics in misinformation}

As a conclusion to this section, misinformation is best situated within the perspective of online groups.
Following from the discussion in Section~\ref{sec:confirming} regarding the focus on accuracy and the focus on identity maintenance, intergroup dynamics are in a position to significantly affect this balance.
Therefore, in cases of intergroup conflict, we would expect very little focus on accuracy which would allow misinformation to dominate. 

In line with cognitive dissonance, we conclude that users may find it less psychologically costly to adhere to misinformation when they predominantly only engage with online echo chambers as this information will go unchallenged ~\cite{zollo_debunking_2017, shin_partisan_2017}.
However, even in cases of intergroup interaction, they will find it less costly to adhere to misinformation when these interactions are combative~\cite{vicario_polarization_2019}.
Misinformation adherence will also be made less psychologically costly when users have mistaken perceptions of how an outgroup views them, and can therefore rationalize their own animosity towards the outgroup as being reactive~\cite{urbanska_swaying_2018, ozdemir_deprivation_2020, obaidi_group-based_2019, kteily_darker_2017, lees_inaccurate_2020}. 

On the contrary, users may find it more psychologically costly to adopt misinformation when they are engaging in deliberative conversations~\cite{zhang_encountering_2019}, even when these are cross-cutting.
Additionally, if mistaken meta-perceptions are corrected, then they may find it harder to rationalize directing prejudice towards an outgroup~\cite{kteily_they_2016, lees_inaccurate_2020}, and thus find it more costly to adopt any misinformation which negatively targets that group. 

\section{Discussion}\label{sec:discussion}

In this work, we have drawn together studies from several related fields, namely social psychology, cognitive psychology, human-computer interaction, web data mining, and computational social science, and synthesized them into what we have interpreted as recommendations for fighting misinformation and open research questions.
We used the concept of cognitive dissonance in drawing together these studies, making the case that certain actions or environments make adherence to misinformation less or more psychologically costly.

In Section~\ref{sec:confirming}, we discussed how information is often chosen implicitly in a motivated fashion, however, this motivated cognition can be made more psychologically costly by shifting users' attention to accuracy.
In Section~\ref{sec:correcting}, we presented various subtler ways in which misinformation can operate on cognition, before making a case for the efficacy of debunking and the (limited) cases in which fact-checking can backfire.
We suggested that researchers should study whether fact-checks delivered by undesirable messengers may display pushback effects.
In Section~\ref{sec:groups}, we discussed echo chambers and hostile intergroup interactions as two main factors of adherence to misinformation.
We argued that the two are not mutually exclusive, and called for quantifying the relationship between informational diversity and hostility displayed in intergroup interactions.
We also discussed online radicalization in the context of social psychological theory as an enabler of outgroup-directed misinformation.

\subsection{Recommendations for anti-misinformation interventions}

Throughout the review, we derived a set of recommendations for researchers working in combating misinformation.
These are ways through which adherence to misinformation can be made more psychologically costly.
In this subsection, we provide a synopsis of the main recommendations we draw out in the ``take-aways'' of each main section.
The most major of these are summarized in Table~\ref{tab:recommend}. 

\begin{table*}[t]
    \small
    \centering
    \begin{tabular}{ @{}r@{}p{14cm}l }
    \toprule
\multicolumn{2}{l}{\textbf{Recommendation}} & \textbf{Publication} \\
    \midrule
    1.~ & Minimize the potential for identity attack & \cite{bliuc_public_2015, rothschild_defensive_2020, nauroth_social_2015} \\ 
    2.~ &Avoid cognitive overload to enable deliberation, e.g., reduce overall amount of information before balancing & \cite{bago_fake_2020} \\
    3.~ &Ensure that accuracy is primed before introducing new information & \cite{pennycook_fighting_2020, bago_fake_2020, pennycook_shifting_2021} \\
     4.~ &Wherever possible, inform people about the ``stickiness'' of misinformation in memory & \cite{lewandowsky_debunking_2020, ecker_explicit_2010} \\
     5.~ & Disrupt false consensus by providing feedback on how many people \textit{truly} espouse fringe views & \cite{ross_false_1977, leviston_your_2013, yousif_illusion_2019, harvey_internet_2018} \\
    6.~ &When using corrective approaches, be comprehensive about correcting multiple facets and avoid leaving gaps to prevent alternative explanations & \cite{nyhan_when_2010, wood_elusive_2019, ecker_explicit_2010, johnson_sources_1994} \\
    7.~ &Provide both correct and false information labels to prevent implied truth effects & \cite{pennycook_implied_2020} \\
    8.~ &Indirectly probe users' understanding on specific issues, so their confidence in any misinformation they espouse is reduced & \cite{fernbach_political_2013, vitriol_illusion_2018} \\
    9.~ &If promoting intergroup interactions, ensure the information exchange environment (e.g., by framing the topic) is deliberative, not combative & \cite{zhang_encountering_2019, zollo_debunking_2017, de_francisci_morales_no_2021, bliuc_you_2020} \\
    10.~ &Address mistaken assumptions that online groups may have about how other groups view them & \cite{lees_inaccurate_2020} \\ 
    \bottomrule
\end{tabular}
    \caption{Recommendations for anti-misinformation interventions.}
    \label{tab:recommend}
\end{table*}

From our discussion on motivated and lazy cognition in Section~\ref{sec:confirming}, the main recommendations revolve around promoting accuracy goals and demoting any other motivations when parsing information.
Providing subtle accuracy nudges is a promising approach in improving the quality of shared news~\cite{pennycook_shifting_2021} since accuracy is not always the default goal of users, especially in the fast-paced, information-heavy environment of social media~\cite{bago_fake_2020}.
All the while, this needs to be balanced with preventing users from ``hooking'' onto other motivations, such as protecting their social identities or their moral values from perceived attacks that accurate information may induce.
In that respect, the true utility of accurate information can be diminished or even reversed if these motivations are not considered.
Therefore, the choice of how accurate information is delivered (or whether it even should be delivered if no attempts have been made to make the informational environment deliberative) should always be preceded by a consideration of this tradeoff.

In Section~\ref{sec:correcting}, we unpack insights around how misinformation operates on human cognition.
We discuss the effects that misinformation can have on memory, particularly around how it can alter recollections of certain events.
We also cover the role of ambiguity, and how it can have both positive (e.g., by reducing people's confidence in their own misconstrued assessments) and negative effects (e.g., by reducing their ability and willingness to consider mixed evidence).
In line with this, the research we cover would suggest that people are made aware of these misinformation effects on memory~\cite{ecker_explicit_2010,lewandowsky_debunking_2020} and that corrections to misinformation are comprehensive enough to avoid further ambiguities (which may leave room to fill these ambiguities with further misinformation)~\cite{wood_elusive_2019}.
Further, in cases where users hold strong opinions on issues they do not fully understand, probing this (lack of) understanding may be a good indirect way of reducing the confidence in their opinion~\cite{fernbach_political_2013}.
This should make them more receptive to corrections.
In any case, the perceived social acceptability of information can make it more attractive~\cite{hills_dark_2019}.
Thus, people may often think that a piece of misinformation they believe is widely endorsed by others~\cite{leviston_your_2013,harvey_internet_2018,ross_false_1977}, not least because misinformation may be reproduced by multiple people while having originated from a single source~\cite{yousif_illusion_2019}.
Disrupting such false perceptions of consensus should therefore remove this social vector of attractiveness.

Finally, Section~\ref{sec:groups} focuses on group psychology and related phenomena; namely, echo chambers and hostile intergroup interactions.
We conclude that breaking echo chambers is certainly a worthwhile pursuit, however, it is equally imperative to avoid an ``out of the frying pan and into the fire'' situation.
Specifically, elevated hostilities where cross-cutting interactions occur are a frequently documented phenomenon~\cite{de_francisci_morales_no_2021,bliuc_you_2020,marchal_be_2021}.
These interactions are not productive, and information sharing in such contexts will almost never result in effective persuasion of the other party.
Due to this elevated polarization, identity-based and moral motivations are at the forefront, severely downgrading accuracy goals.
Therefore, ensuring that the information exchange environment is deliberative, and not combative, is our primary recommendation.
This may be achieved using anti-toxicity nudges, posting ``cooldown'' periods in highly active threads, etc.
Importantly, previous research also highlights the value of correcting intergroup meta-perceptions.
That is, groups often believe that outgroups paint them in a much worse light than actuality.
Thus, correcting these perceptions may reduce reciprocal prejudice between groups and prevent the devaluing of accuracy.

We stress that all of these recommendations should be realized indirectly wherever possible, so as to prevent reactance~\cite{brehm_control_1993, silvia_deflecting_2005}.
For example, Pennycook et al.~\cite{pennycook_shifting_2021} were able to get people to reduce the amount of political misinformation they shared online by messaging them on Twitter and asking them to rate the accuracy of information that had nothing to do with politics.
In this way, they primed accuracy in the users in an indirect manner.
Moreover, some scholars have suggested that combining anti-misinformation interventions such as content moderation, accuracy nudges, and banning of problematic accounts can be substantially more effective than employing either of these interventions by itself~\cite{bak-coleman_combining_2021}.
Therefore, it is likely that combining our recommendations may have a similar effect.
However, we also highlight the contextuality of misinformation adherence and, by extension, anti-misinformation interventions.
To that end, our recommendations may have differential impacts depending on the specific circumstances. 
Understanding the informational context within which these interventions operate is equally important.

More broadly, the recommendations we make are certainly not exhaustive and the methods through which they are delivered can substantially affect their efficacy.
For example, interventions which can be delivered as default settings on online platforms are more likely to have a significant impact than those requiring explicit user uptake.
Methods which aim towards changing the mentalities and norms governing platforms, such as inoculation-driven strategies, may also be more promising in the long run than event-specific responses.
We would like to stress that our recommendations mainly reflect intended actions and consequences, but not necessarily their methods of delivery.
Analyzing these methods of delivery to determine the most promising ways of administering each of our recommendations, either as an intervention or design affordance, is an endeavor as important as drawing out recommendations themselves.
However, it is not something that we have analyzed in depth here.
We hope that our current work can act as a first step towards creating a framework where the intended outcomes of anti-misinformation recommendations are paired with their most efficacious methods of delivery.

\subsection{Future research directions}

Throughout the paper, we have also identified various future research directions.
We believe that these are very important open questions that remain to be addressed before we can understand how to best implement our anti-misinformation efforts online, and summarize the main ones below.

\begin{enumerate}
    \item \textit{What are the contexts under which people would choose to exercise accuracy over identity-maintaining or motivated cognition?}
    \vskip 0.05cm
    From our discussion in Section~\ref{sec:confirming}, we have seen that people often do value accuracy in the information they consume and share.
    However, they may forgo these accuracy goals in the face of social or emotional distractions.
    Comparing the spread and prevalence of misinformation between highly combative spaces and deliberative spaces which promote evidence-based dialogue may provide a better understanding of the features and types of such distractions.
    \vskip 0.05cm
    \item \textit{Can we embed accuracy into the ingroup identity so that we can improve deliberation, and increase the harmony between valuing accuracy and maintaining an ingroup image?}
    \vskip 0.05cm
    In Sections~\ref{subsec:motiv} and~\ref{sec:groups} we discussed how social identities and social group memberships can create a motivation to prefer information which is approving of these identities and groups, even if not necessarily true.
    These groups are characterized by unique values and norms, and in some cases, accuracy constitutes such a value.
    An example would be scientific communities, where rigor and evidence are important for someone identifying as a scientist.
    For such a community, using misinformation to defend the ingroup identity would constitute a breach of the group's values.
    Future research can examine whether embedding similar values in other groups, for example political ones, can give accuracy an additional boost even when the ingroup image is threatened.
    This could come in the form of cases where community role models urge more caution in the information that other members consume.
    \vskip 0.05cm
    \item \textit{How can online communities be categorized into deliberative or combative ones? What impact does this categorization have on the nature of social interactions in these communities?}
    \vskip 0.05cm
    An important advancement to answer both of the above questions would be the characterization of not only hostile, but also constructive and deliberative communities online.
    As we have seen in Section~\ref{sec:groups}, the vast majority of research has focused on hostile communities and interactions.
    A framework for characterizing communities in this way would also enable scholars in the field to better isolate the causes behind the circulation of misinformation.
    For example, if constructive intergroup interactions in deliberative communities coincide with lower misinformation prevalence, then a more direct link can be established between the motivations behind sharing misinformation in non-deliberative spaces.
    Importantly, the study of such communities could concern existing online spaces (e.g., Reddit forums dedicated to civil discussion which can be described as ``anti-echo chambers''~\cite{guest_anti-echo_2018}) or emergent civil networks following particular events.
    \vskip 0.05cm
    \item \textit{Can polarization be reduced by increasing the number of groups (polarities) that are involved in an issue?}
    \vskip 0.05cm
    Much of the research we covered focused on bi-polar topics; for example politics (left vs. right)~\cite{garimella_political_2018} or views on science (conspiratorial vs. pro-science)~\cite{brugnoli_recursive_2019}.
    It would be worthwhile for future studies to examine topics that consist of multiple groups with distinct views.
    This would be useful not only for quantifying the degree of polarization there, but also for observing whether some groups can act as bridges between others.
    \vskip 0.05cm
    \item \textit{What is the impact of treating standing on an issue as separate to group membership/social identity, such that attempts are made to reduce attitude polarization through users within, rather than between different communities?}
    \vskip 0.05cm
    As we discussed in the final part of Section~\ref{sec:confirming}, many models in opinion dynamics treat stance on an issue and group membership as the same thing.
    That is, the probability of whether some information will be accepted or not depends on how ``ideologically close'' the messenger of this information is to the receiver.
    It would be useful for future research to empirically examine unique situations where two users may be part of the same group, but hold different opinions on an issue; for example, two Conservative users who disagree on gun laws.
    If users are more receptive to oppositional information which, paradoxically, comes from others similar to them, then this could have significant implications for leveraging the ingroup identity to create more deliberation online.
    \vskip 0.05cm
    \item \textit{How effective are fact-checks in the context of disrupted exposure, against the context of consistent exposure? What is the proportion of fact-checks required in an informational environment to achieve true attitude change in the network?}
    \vskip 0.05cm
    To better contextualize the effectiveness of fact-checking discussed in Section~\ref{sec:correcting}, it would be beneficial for future research to examine fact-checking as a practice on social media platforms, and not just the impact of individual, disjointed fact-check pieces.
    This could be done by incorporating decay factors into the study of fact-checking and see if fact-checks are more effective for users who interact with them in continued succession, or by examining average changes in stance or sentiment across entire networks as a function of the number of fact-checks in these networks.
    \vskip 0.05cm
    \item \textit{Are backfire effects against misinformation corrections more prominent when they come from an outgroup member? If so, is this because such corrections may be viewed as combative information sharing?}
    \vskip 0.05cm
    This question directly relates to the possibility we raise in Section~\ref{sec:correcting}.
    The backfire potential of fact-checks has been found to be very low, but this has mostly been demonstrated in situations where the fact-check messenger is somewhat neutral (e.g. a fact-checking organization).
    When the fact-check is shared by a partisan messenger, however, opposing partisans may view this as an attempt to invalidate their groups and become more entrenched in the misinformation.
    More evidence is needed with respect to this in real social networks.
    \vskip 0.05cm
    \item \textit{Can corrections used in user-to-user, non-deliberative, identity-maintaining ways drive distrust towards fact-checking organizations?}
    \vskip 0.05cm
    As an extension of question 7, researchers may also wish to study whether trust in fact-checking organizations decreases over time for users who consistently have their opinions challenged.
    This would provide further nuances regarding the risks and benefits of fact-checking misinformation.
    \vskip 0.05cm
    \item \textit{What is the relationship between the degree of engagement with outgroup users and the hostility displayed in cross-cutting conversations when echo chamber involvement is conceived of as a continuous, rather than binary, variable?}
    \vskip 0.05cm
    In Section~\ref{sec:groups}, we discussed how polarization may be enabled both by echo chambers and hostile intergroup interactions.
    While both are phenomena that have been recorded on social media, there has been a notable lack of efforts to integrate the two.
    We urge researchers to probe this in more detail, by examining echo chambers at the user level; that is, the extent to which different users have one-sided information consumption patterns, and how this subsequently affects their interactions with others online.
    \vskip 0.05cm
    \item \textit{To what extent does extremist rhetoric distort the accuracy of meta-perceptions towards an outgroup? Can radicalization be attenuated by addressing mistaken meta-perceptions?}
    \vskip 0.05cm
    While exaggerated meta-perceptions have been found in laboratory settings~\cite{moore-berg_exaggerated_2020,lees_inaccurate_2020}, studying this phenomenon using social media data is also worthwhile. 
    Future research can examine situations in which hate directed towards other groups is defended under the guise of the other group purportedly initiating these hostilities.
    An example would be comparing extremist political movements' narratives around their treatment from more moderate movements to the actual vernacular used in moderate movements to describe these extreme movements.
\end{enumerate}

\section{Conclusion}

In this paper, we approached the problem of misinformation adherence from multiple disciplines.
We discussed insights from cognitive psychology with respect to how misinformation can affect or be affected by information parsing mechanisms, as well as social psychology with respect to how groups and social identities can introduce more motivations and complexity in assimilating information.
We unpacked insights from this body of work in the context of research conducted within human-computer interaction, web data mining, and computational social science.

We believe that our inter-disciplinary synthesis will help CSCW researchers when designing anti-misinformation interventions and misinformation-resilient affordances, in addition to providing several important future directions.

\descr{Acknowledgments.} This research has been partially funded by the UK EPSRC grant EP/S022503/1, which supports the UCL Centre for Doctoral Training in Cybersecurity, and the UK's National Research Centre on Privacy, Harm Reduction, and Adversarial Influence Online (REPHRAIN, UKRI grant: EP/V011189/1). Any opinions, conclusions, or recommendations expressed in this work are those of the authors and do not necessarily reflect the views of the UK EPSRC.

\small
\bibliographystyle{ACM-Reference-Format}
\bibliography{references}


\begin{thebibliography}{177}


\ifx \showCODEN    \undefined \def \showCODEN     #1{\unskip}     \fi
\ifx \showDOI      \undefined \def \showDOI       #1{#1}\fi
\ifx \showISBNx    \undefined \def \showISBNx     #1{\unskip}     \fi
\ifx \showISBNxiii \undefined \def \showISBNxiii  #1{\unskip}     \fi
\ifx \showISSN     \undefined \def \showISSN      #1{\unskip}     \fi
\ifx \showLCCN     \undefined \def \showLCCN      #1{\unskip}     \fi
\ifx \shownote     \undefined \def \shownote      #1{#1}          \fi
\ifx \showarticletitle \undefined \def \showarticletitle #1{#1}   \fi
\ifx \showURL      \undefined \def \showURL       {\relax}        \fi
\providecommand\bibfield[2]{#2}
\providecommand\bibinfo[2]{#2}
\providecommand\natexlab[1]{#1}
\providecommand\showeprint[2][]{arXiv:#2}

\bibitem[\protect\citeauthoryear{Abeles, Howe, Krosnick, and MacInnis}{Abeles
  et~al\mbox{.}}{2019}]%
        {abeles_perception_2019}
\bibfield{author}{\bibinfo{person}{Adina~T. Abeles}, \bibinfo{person}{Lauren~C.
  Howe}, \bibinfo{person}{Jon~A. Krosnick}, {and} \bibinfo{person}{Bo
  MacInnis}.} \bibinfo{year}{2019}\natexlab{}.
\newblock \showarticletitle{Perception of public opinion on global warming and
  the role of opinion deviance}.
\newblock \bibinfo{journal}{\emph{Journal of Environmental Psychology}}
  \bibinfo{volume}{63} (\bibinfo{date}{June} \bibinfo{year}{2019}),
  \bibinfo{pages}{118--129}.
\newblock
\showISSN{02724944}
\urldef\tempurl%
\url{https://doi.org/10.1016/j.jenvp.2019.04.001}
\showDOI{\tempurl}


\bibitem[\protect\citeauthoryear{Ajzen}{Ajzen}{1991}]%
        {ajzen_theory_1991}
\bibfield{author}{\bibinfo{person}{Icek Ajzen}.}
  \bibinfo{year}{1991}\natexlab{}.
\newblock \showarticletitle{The theory of planned behavior}.
\newblock \bibinfo{journal}{\emph{Organizational Behavior and Human Decision
  Processes}} \bibinfo{volume}{50}, \bibinfo{number}{2} (\bibinfo{date}{Dec.}
  \bibinfo{year}{1991}), \bibinfo{pages}{179--211}.
\newblock
\showISSN{0749-5978}
\urldef\tempurl%
\url{https://doi.org/10.1016/0749-5978(91)90020-T}
\showDOI{\tempurl}


\bibitem[\protect\citeauthoryear{Allen, Howland, Mobius, Rothschild, and
  Watts}{Allen et~al\mbox{.}}{2020}]%
        {allen_evaluating_2020}
\bibfield{author}{\bibinfo{person}{Jennifer Allen}, \bibinfo{person}{Baird
  Howland}, \bibinfo{person}{Markus Mobius}, \bibinfo{person}{David
  Rothschild}, {and} \bibinfo{person}{Duncan~J. Watts}.}
  \bibinfo{year}{2020}\natexlab{}.
\newblock \showarticletitle{Evaluating the fake news problem at the scale of
  the information ecosystem}.
\newblock \bibinfo{journal}{\emph{Science Advances}} \bibinfo{volume}{6},
  \bibinfo{number}{14} (\bibinfo{year}{2020}), \bibinfo{pages}{eaay3539}.
\newblock
\urldef\tempurl%
\url{https://doi.org/10.1126/sciadv.aay3539}
\showDOI{\tempurl}


\bibitem[\protect\citeauthoryear{Allen, Martel, and Rand}{Allen
  et~al\mbox{.}}{2022}]%
        {allen_birds_2022}
\bibfield{author}{\bibinfo{person}{Jennifer Nancy~Lee Allen},
  \bibinfo{person}{Cameron Martel}, {and} \bibinfo{person}{David Rand}.}
  \bibinfo{year}{2022}\natexlab{}.
\newblock \bibinfo{booktitle}{\emph{Birds of a feather don’t fact-check each
  other: {Partisanship} and the evaluation of news in {Twitter}’s {Birdwatch}
  crowdsourced fact-checking program}}.
\newblock \bibinfo{type}{{T}echnical {R}eport}.
  \bibinfo{institution}{PsyArXiv}.
\newblock
\urldef\tempurl%
\url{https://doi.org/10.31234/osf.io/57e3q}
\showDOI{\tempurl}


\bibitem[\protect\citeauthoryear{Allport}{Allport}{1954}]%
        {allport_nature_1954}
\bibfield{author}{\bibinfo{person}{Gordon~W. Allport}.}
  \bibinfo{year}{1954}\natexlab{}.
\newblock \bibinfo{booktitle}{\emph{The nature of prejudice}}.
\newblock \bibinfo{publisher}{Addison-Wesley}, \bibinfo{address}{Oxford,
  England}.
\newblock


\bibitem[\protect\citeauthoryear{Aslay, Matakos, Galbrun, and Gionis}{Aslay
  et~al\mbox{.}}{2018}]%
        {aslay_maximizing_2018}
\bibfield{author}{\bibinfo{person}{C. Aslay}, \bibinfo{person}{A. Matakos},
  \bibinfo{person}{E. Galbrun}, {and} \bibinfo{person}{A. Gionis}.}
  \bibinfo{year}{2018}\natexlab{}.
\newblock \showarticletitle{Maximizing the {Diversity} of {Exposure} in a
  {Social} {Network}}. In \bibinfo{booktitle}{\emph{2018 {IEEE} {International}
  {Conference} on {Data} {Mining} ({ICDM})}}. \bibinfo{pages}{863--868}.
\newblock
\urldef\tempurl%
\url{https://doi.org/10.1109/ICDM.2018.00102}
\showDOI{\tempurl}


\bibitem[\protect\citeauthoryear{Bago, Rand, and Pennycook}{Bago
  et~al\mbox{.}}{2020}]%
        {bago_fake_2020}
\bibfield{author}{\bibinfo{person}{Bence Bago}, \bibinfo{person}{David~G.
  Rand}, {and} \bibinfo{person}{Gordon Pennycook}.}
  \bibinfo{year}{2020}\natexlab{}.
\newblock \showarticletitle{Fake news, fast and slow: {Deliberation} reduces
  belief in false (but not true) news headlines.}
\newblock \bibinfo{journal}{\emph{Journal of Experimental Psychology: General}}
  \bibinfo{volume}{149}, \bibinfo{number}{8} (\bibinfo{year}{2020}),
  \bibinfo{pages}{1608}.
\newblock
\showISSN{1939-2222}
\urldef\tempurl%
\url{https://doi.org/10.1037/xge0000729}
\showDOI{\tempurl}


\bibitem[\protect\citeauthoryear{Bail, Argyle, Brown, Bumpus, Chen, Hunzaker,
  Lee, Mann, Merhout, and Volfovsky}{Bail et~al\mbox{.}}{2018}]%
        {bail_exposure_2018}
\bibfield{author}{\bibinfo{person}{Christopher~A. Bail},
  \bibinfo{person}{Lisa~P. Argyle}, \bibinfo{person}{Taylor~W. Brown},
  \bibinfo{person}{John~P. Bumpus}, \bibinfo{person}{Haohan Chen},
  \bibinfo{person}{M.~B.~Fallin Hunzaker}, \bibinfo{person}{Jaemin Lee},
  \bibinfo{person}{Marcus Mann}, \bibinfo{person}{Friedolin Merhout}, {and}
  \bibinfo{person}{Alexander Volfovsky}.} \bibinfo{year}{2018}\natexlab{}.
\newblock \showarticletitle{Exposure to opposing views on social media can
  increase political polarization}.
\newblock \bibinfo{journal}{\emph{Proceedings of the National Academy of
  Sciences}} \bibinfo{volume}{115}, \bibinfo{number}{37} (\bibinfo{date}{Sept.}
  \bibinfo{year}{2018}), \bibinfo{pages}{9216--9221}.
\newblock
\showISSN{0027-8424, 1091-6490}
\urldef\tempurl%
\url{https://doi.org/10.1073/pnas.1804840115}
\showDOI{\tempurl}


\bibitem[\protect\citeauthoryear{Bak-Coleman, Kennedy, Wack, Beers, Schafer,
  Spiro, Starbird, and West}{Bak-Coleman et~al\mbox{.}}{2021}]%
        {bak-coleman_combining_2021}
\bibfield{author}{\bibinfo{person}{Joseph Bak-Coleman}, \bibinfo{person}{Ian
  Kennedy}, \bibinfo{person}{Morgan Wack}, \bibinfo{person}{Andrew Beers},
  \bibinfo{person}{Joseph~Scott Schafer}, \bibinfo{person}{Emma Spiro},
  \bibinfo{person}{Kate Starbird}, {and} \bibinfo{person}{Jevin West}.}
  \bibinfo{year}{2021}\natexlab{}.
\newblock \bibinfo{booktitle}{\emph{Combining interventions to reduce the
  spread of viral misinformation}}.
\newblock \bibinfo{type}{{T}echnical {R}eport}.
  \bibinfo{institution}{SocArXiv}.
\newblock
\urldef\tempurl%
\url{https://doi.org/10.31235/osf.io/4jtvm}
\showDOI{\tempurl}


\bibitem[\protect\citeauthoryear{Barberá, Jost, Nagler, Tucker, and
  Bonneau}{Barberá et~al\mbox{.}}{2015}]%
        {barbera_tweeting_2015}
\bibfield{author}{\bibinfo{person}{Pablo Barberá}, \bibinfo{person}{John~T.
  Jost}, \bibinfo{person}{Jonathan Nagler}, \bibinfo{person}{Joshua~A. Tucker},
  {and} \bibinfo{person}{Richard Bonneau}.} \bibinfo{year}{2015}\natexlab{}.
\newblock \showarticletitle{Tweeting {From} {Left} to {Right}: {Is} {Online}
  {Political} {Communication} {More} {Than} an {Echo} {Chamber}?}
\newblock \bibinfo{journal}{\emph{Psychological Science}} \bibinfo{volume}{26},
  \bibinfo{number}{10} (\bibinfo{date}{Oct.} \bibinfo{year}{2015}),
  \bibinfo{pages}{1531--1542}.
\newblock
\showISSN{0956-7976, 1467-9280}
\urldef\tempurl%
\url{https://doi.org/10.1177/0956797615594620}
\showDOI{\tempurl}


\bibitem[\protect\citeauthoryear{Basol, Roozenbeek, and van~der Linden}{Basol
  et~al\mbox{.}}{2020}]%
        {basol_good_2020}
\bibfield{author}{\bibinfo{person}{Melisa Basol}, \bibinfo{person}{Jon
  Roozenbeek}, {and} \bibinfo{person}{Sander van~der Linden}.}
  \bibinfo{year}{2020}\natexlab{}.
\newblock \showarticletitle{Good {News} about {Bad} {News}: {Gamified}
  {Inoculation} {Boosts} {Confidence} and {Cognitive} {Immunity} {Against}
  {Fake} {News}}.
\newblock \bibinfo{journal}{\emph{Journal of Cognition}} \bibinfo{volume}{3},
  \bibinfo{number}{1} (\bibinfo{year}{2020}).
\newblock
\showISSN{2514-4820}
\urldef\tempurl%
\url{https://doi.org/10.5334/joc.91}
\showDOI{\tempurl}


\bibitem[\protect\citeauthoryear{Bliuc, McGarty, Thomas, Lala, Berndsen, and
  Misajon}{Bliuc et~al\mbox{.}}{2015}]%
        {bliuc_public_2015}
\bibfield{author}{\bibinfo{person}{Ana-Maria Bliuc}, \bibinfo{person}{Craig
  McGarty}, \bibinfo{person}{Emma~F. Thomas}, \bibinfo{person}{Girish Lala},
  \bibinfo{person}{Mariette Berndsen}, {and} \bibinfo{person}{RoseAnne
  Misajon}.} \bibinfo{year}{2015}\natexlab{}.
\newblock \showarticletitle{Public division about climate change rooted in
  conflicting socio-political identities}.
\newblock \bibinfo{journal}{\emph{Nature Climate Change}} \bibinfo{volume}{5},
  \bibinfo{number}{3} (\bibinfo{date}{March} \bibinfo{year}{2015}),
  \bibinfo{pages}{226--229}.
\newblock
\showISSN{1758-678X, 1758-6798}
\urldef\tempurl%
\url{https://doi.org/10.1038/nclimate2507}
\showDOI{\tempurl}


\bibitem[\protect\citeauthoryear{Bliuc, Smith, and Moynihan}{Bliuc
  et~al\mbox{.}}{2020}]%
        {bliuc_you_2020}
\bibfield{author}{\bibinfo{person}{Ana-Maria Bliuc}, \bibinfo{person}{Laura G~E
  Smith}, {and} \bibinfo{person}{Tina Moynihan}.}
  \bibinfo{year}{2020}\natexlab{}.
\newblock \showarticletitle{“{You} wouldn’t celebrate {September} 11”:
  {Testing} online polarisation between opposing ideological camps on
  {YouTube}}.
\newblock \bibinfo{journal}{\emph{Group Processes \& Intergroup Relations}}
  \bibinfo{volume}{23}, \bibinfo{number}{6} (\bibinfo{year}{2020}),
  \bibinfo{pages}{827--844}.
\newblock


\bibitem[\protect\citeauthoryear{Brehm}{Brehm}{1993}]%
        {brehm_control_1993}
\bibfield{author}{\bibinfo{person}{Jack~W. Brehm}.}
  \bibinfo{year}{1993}\natexlab{}.
\newblock \showarticletitle{Control, {Its} {Loss}, and {Psychological}
  {Reactance}}.
\newblock In \bibinfo{booktitle}{\emph{Control {Motivation} and {Social}
  {Cognition}}}, \bibfield{editor}{\bibinfo{person}{Gifford Weary},
  \bibinfo{person}{Faith Gleicher}, {and} \bibinfo{person}{Kerry~L. Marsh}}
  (Eds.). \bibinfo{publisher}{Springer}, \bibinfo{address}{New York, NY},
  \bibinfo{pages}{3--30}.
\newblock
\showISBNx{978-1-4613-8309-3}
\urldef\tempurl%
\url{https://doi.org/10.1007/978-1-4613-8309-3_1}
\showDOI{\tempurl}


\bibitem[\protect\citeauthoryear{Brugnoli, Cinelli, Quattrociocchi, and
  Scala}{Brugnoli et~al\mbox{.}}{2019}]%
        {brugnoli_recursive_2019}
\bibfield{author}{\bibinfo{person}{Emanuele Brugnoli}, \bibinfo{person}{Matteo
  Cinelli}, \bibinfo{person}{Walter Quattrociocchi}, {and}
  \bibinfo{person}{Antonio Scala}.} \bibinfo{year}{2019}\natexlab{}.
\newblock \showarticletitle{Recursive patterns in online echo chambers}.
\newblock \bibinfo{journal}{\emph{Scientific Reports}} \bibinfo{volume}{9},
  \bibinfo{number}{1} (\bibinfo{date}{Dec.} \bibinfo{year}{2019}),
  \bibinfo{pages}{20118}.
\newblock
\showISSN{2045-2322}
\urldef\tempurl%
\url{https://doi.org/10.1038/s41598-019-56191-7}
\showDOI{\tempurl}


\bibitem[\protect\citeauthoryear{Brydges, Gignac, and Ecker}{Brydges
  et~al\mbox{.}}{2018}]%
        {brydges_working_2018}
\bibfield{author}{\bibinfo{person}{Christopher~R. Brydges},
  \bibinfo{person}{Gilles~E. Gignac}, {and} \bibinfo{person}{Ullrich K.~H.
  Ecker}.} \bibinfo{year}{2018}\natexlab{}.
\newblock \showarticletitle{Working memory capacity, short-term memory
  capacity, and the continued influence effect: {A} latent-variable analysis}.
\newblock \bibinfo{journal}{\emph{Intelligence}}  \bibinfo{volume}{69}
  (\bibinfo{date}{July} \bibinfo{year}{2018}), \bibinfo{pages}{117--122}.
\newblock
\showISSN{0160-2896}
\urldef\tempurl%
\url{https://doi.org/10.1016/j.intell.2018.03.009}
\showDOI{\tempurl}


\bibitem[\protect\citeauthoryear{Carrillo, Corning, Dennehy, and
  Crosby}{Carrillo et~al\mbox{.}}{2011}]%
        {carrillo_relative_2011}
\bibfield{author}{\bibinfo{person}{Jenny Carrillo},
  \bibinfo{person}{Alexandra~F. Corning}, \bibinfo{person}{Tara~C. Dennehy},
  {and} \bibinfo{person}{Faye~J. Crosby}.} \bibinfo{year}{2011}\natexlab{}.
\newblock \showarticletitle{Relative deprivation: {Understanding} the dynamics
  of discontent}.
\newblock In \bibinfo{booktitle}{\emph{Theories in social psychology}}.
  \bibinfo{publisher}{Wiley Blackwell}, \bibinfo{pages}{140--160}.
\newblock
\showISBNx{978-1-4443-3123-3 978-1-4443-3122-6 978-1-4443-4209-3}


\bibitem[\protect\citeauthoryear{Caulfield, Spring, and Sasse}{Caulfield
  et~al\mbox{.}}{2019}]%
        {caulfield_why_2019}
\bibfield{author}{\bibinfo{person}{Tristan Caulfield},
  \bibinfo{person}{Jonathan~M. Spring}, {and} \bibinfo{person}{M.~Angela
  Sasse}.} \bibinfo{year}{2019}\natexlab{}.
\newblock \showarticletitle{Why {Jenny} can't figure out which of these
  messages is a covert information operation}. In
  \bibinfo{booktitle}{\emph{Proceedings of the {New} {Security} {Paradigms}
  {Workshop}}}. \bibinfo{publisher}{ACM}, \bibinfo{address}{San Carlos Costa
  Rica}, \bibinfo{pages}{118--128}.
\newblock
\showISBNx{978-1-4503-7647-1}
\urldef\tempurl%
\url{https://doi.org/10.1145/3368860.3368870}
\showDOI{\tempurl}


\bibitem[\protect\citeauthoryear{Chan, Jones, Hall~Jamieson, and
  Albarracín}{Chan et~al\mbox{.}}{2017}]%
        {chan_debunking_2017}
\bibfield{author}{\bibinfo{person}{Man-pui~Sally Chan},
  \bibinfo{person}{Christopher~R. Jones}, \bibinfo{person}{Kathleen
  Hall~Jamieson}, {and} \bibinfo{person}{Dolores Albarracín}.}
  \bibinfo{year}{2017}\natexlab{}.
\newblock \showarticletitle{Debunking: {A} {Meta}-{Analysis} of the
  {Psychological} {Efficacy} of {Messages} {Countering} {Misinformation}}.
\newblock \bibinfo{journal}{\emph{Psychological Science}} \bibinfo{volume}{28},
  \bibinfo{number}{11} (\bibinfo{date}{Nov.} \bibinfo{year}{2017}),
  \bibinfo{pages}{1531--1546}.
\newblock
\showISSN{0956-7976}
\urldef\tempurl%
\url{https://doi.org/10.1177/0956797617714579}
\showDOI{\tempurl}


\bibitem[\protect\citeauthoryear{Cinelli, Morales, Galeazzi, Quattrociocchi,
  and Starnini}{Cinelli et~al\mbox{.}}{2021}]%
        {cinelli_echo_2021}
\bibfield{author}{\bibinfo{person}{Matteo Cinelli}, \bibinfo{person}{Gianmarco
  De~Francisci Morales}, \bibinfo{person}{Alessandro Galeazzi},
  \bibinfo{person}{Walter Quattrociocchi}, {and} \bibinfo{person}{Michele
  Starnini}.} \bibinfo{year}{2021}\natexlab{}.
\newblock \showarticletitle{The echo chamber effect on social media}.
\newblock \bibinfo{journal}{\emph{Proceedings of the National Academy of
  Sciences}} \bibinfo{volume}{118}, \bibinfo{number}{9} (\bibinfo{date}{March}
  \bibinfo{year}{2021}).
\newblock
\showISSN{0027-8424, 1091-6490}
\urldef\tempurl%
\url{https://doi.org/10.1073/pnas.2023301118}
\showDOI{\tempurl}


\bibitem[\protect\citeauthoryear{Coates, Han, and Kleerekoper}{Coates
  et~al\mbox{.}}{2018}]%
        {coates_unified_2018}
\bibfield{author}{\bibinfo{person}{Adam Coates}, \bibinfo{person}{Liangxiu
  Han}, {and} \bibinfo{person}{Anthony Kleerekoper}.}
  \bibinfo{year}{2018}\natexlab{}.
\newblock \showarticletitle{A {Unified} {Framework} for {Opinion} {Dynamics}}.
\newblock  (\bibinfo{year}{2018}), \bibinfo{pages}{9}.
\newblock


\bibitem[\protect\citeauthoryear{Cohen, Seate, Kromka, Sutherland, Thomas,
  Skerda, and Nicholson}{Cohen et~al\mbox{.}}{2020}]%
        {cohen_correct_2020}
\bibfield{author}{\bibinfo{person}{Elizabeth~L. Cohen},
  \bibinfo{person}{Anita~Atwell Seate}, \bibinfo{person}{Stephen~M. Kromka},
  \bibinfo{person}{Andrew Sutherland}, \bibinfo{person}{Matthew Thomas},
  \bibinfo{person}{Karissa Skerda}, {and} \bibinfo{person}{Andrew Nicholson}.}
  \bibinfo{year}{2020}\natexlab{}.
\newblock \showarticletitle{To correct or not to correct? {Social} identity
  threats increase willingness to denounce fake news through presumed media
  influence and hostile media perceptions}.
\newblock \bibinfo{journal}{\emph{Communication Research Reports}}
  \bibinfo{volume}{37}, \bibinfo{number}{5} (\bibinfo{date}{Oct.}
  \bibinfo{year}{2020}), \bibinfo{pages}{263--275}.
\newblock
\showISSN{0882-4096}
\urldef\tempurl%
\url{https://doi.org/10.1080/08824096.2020.1841622}
\showDOI{\tempurl}


\bibitem[\protect\citeauthoryear{Dandekar, Goel, and Lee}{Dandekar
  et~al\mbox{.}}{2013}]%
        {dandekar_biased_2013}
\bibfield{author}{\bibinfo{person}{P. Dandekar}, \bibinfo{person}{A. Goel},
  {and} \bibinfo{person}{D.~T. Lee}.} \bibinfo{year}{2013}\natexlab{}.
\newblock \showarticletitle{Biased assimilation, homophily, and the dynamics of
  polarization}.
\newblock \bibinfo{journal}{\emph{Proceedings of the National Academy of
  Sciences}} \bibinfo{volume}{110}, \bibinfo{number}{15} (\bibinfo{date}{April}
  \bibinfo{year}{2013}), \bibinfo{pages}{5791--5796}.
\newblock
\showISSN{0027-8424, 1091-6490}
\urldef\tempurl%
\url{https://doi.org/10.1073/pnas.1217220110}
\showDOI{\tempurl}


\bibitem[\protect\citeauthoryear{Datta and Adar}{Datta and Adar}{2019}]%
        {datta_extracting_2019}
\bibfield{author}{\bibinfo{person}{Srayan Datta} {and} \bibinfo{person}{Eytan
  Adar}.} \bibinfo{year}{2019}\natexlab{}.
\newblock \showarticletitle{Extracting {Inter}-{Community} {Conflicts} in
  {Reddit}}.
\newblock \bibinfo{journal}{\emph{Proceedings of the International AAAI
  Conference on Web and Social Media}}  \bibinfo{volume}{13}
  (\bibinfo{date}{July} \bibinfo{year}{2019}), \bibinfo{pages}{146--157}.
\newblock
\showISSN{2334-0770}
\urldef\tempurl%
\url{https://ojs.aaai.org/index.php/ICWSM/article/view/3217}
\showURL{%
\tempurl}


\bibitem[\protect\citeauthoryear{Davenport and Beck}{Davenport and
  Beck}{2002}]%
        {davenport_attention_2002}
\bibfield{author}{\bibinfo{person}{Thomas~H. Davenport} {and}
  \bibinfo{person}{John~C. Beck}.} \bibinfo{year}{2002}\natexlab{}.
\newblock \bibinfo{booktitle}{\emph{Attention {Economy}: {Understanding} the
  {New} {Currency} of {Business}} (\bibinfo{edition}{new ed edition} ed.)}.
\newblock \bibinfo{publisher}{Harvard Business Review Press},
  \bibinfo{address}{Boston, Mass}.
\newblock
\showISBNx{978-1-57851-871-5}


\bibitem[\protect\citeauthoryear{De~Francisci~Morales, Monti, and
  Starnini}{De~Francisci~Morales et~al\mbox{.}}{2021}]%
        {de_francisci_morales_no_2021}
\bibfield{author}{\bibinfo{person}{Gianmarco De~Francisci~Morales},
  \bibinfo{person}{Corrado Monti}, {and} \bibinfo{person}{Michele Starnini}.}
  \bibinfo{year}{2021}\natexlab{}.
\newblock \showarticletitle{No echo in the chambers of political interactions
  on {Reddit}}.
\newblock \bibinfo{journal}{\emph{Scientific Reports}} \bibinfo{volume}{11},
  \bibinfo{number}{1} (\bibinfo{date}{Dec.} \bibinfo{year}{2021}),
  \bibinfo{pages}{2818}.
\newblock
\showISSN{2045-2322}
\urldef\tempurl%
\url{https://doi.org/10.1038/s41598-021-81531-x}
\showDOI{\tempurl}


\bibitem[\protect\citeauthoryear{Del~Vicario, Gaito, Quattrociocchi, Zignani,
  and Zollo}{Del~Vicario et~al\mbox{.}}{2017}]%
        {del_vicario_public_2017}
\bibfield{author}{\bibinfo{person}{Michela Del~Vicario},
  \bibinfo{person}{Sabrina Gaito}, \bibinfo{person}{Walter Quattrociocchi},
  \bibinfo{person}{Matteo Zignani}, {and} \bibinfo{person}{Fabiana Zollo}.}
  \bibinfo{year}{2017}\natexlab{}.
\newblock \showarticletitle{Public discourse and news consumption on online
  social media: {A} quantitative, cross-platform analysis of the {Italian}
  {Referendum}}.
\newblock \bibinfo{journal}{\emph{arXiv:1702.06016 [physics]}}
  (\bibinfo{date}{June} \bibinfo{year}{2017}).
\newblock
\urldef\tempurl%
\url{http://arxiv.org/abs/1702.06016}
\showURL{%
\tempurl}


\bibitem[\protect\citeauthoryear{Ditto, Clark, Liu, Wojcik, Chen, Grady,
  Celniker, and Zinger}{Ditto et~al\mbox{.}}{2019a}]%
        {ditto_partisan_2019}
\bibfield{author}{\bibinfo{person}{Peter~H. Ditto}, \bibinfo{person}{Cory~J.
  Clark}, \bibinfo{person}{Brittany~S. Liu}, \bibinfo{person}{Sean~P. Wojcik},
  \bibinfo{person}{Eric~E. Chen}, \bibinfo{person}{Rebecca~H. Grady},
  \bibinfo{person}{Jared~B. Celniker}, {and} \bibinfo{person}{Joanne~F.
  Zinger}.} \bibinfo{year}{2019}\natexlab{a}.
\newblock \showarticletitle{Partisan {Bias} and {Its} {Discontents}}.
\newblock \bibinfo{journal}{\emph{Perspectives on Psychological Science}}
  \bibinfo{volume}{14}, \bibinfo{number}{2} (\bibinfo{date}{March}
  \bibinfo{year}{2019}), \bibinfo{pages}{304--316}.
\newblock
\showISSN{1745-6916, 1745-6924}
\urldef\tempurl%
\url{https://doi.org/10.1177/1745691618817753}
\showDOI{\tempurl}


\bibitem[\protect\citeauthoryear{Ditto, Liu, Clark, Wojcik, Chen, Grady,
  Celniker, and Zinger}{Ditto et~al\mbox{.}}{2019b}]%
        {ditto_at_2019}
\bibfield{author}{\bibinfo{person}{Peter~H. Ditto},
  \bibinfo{person}{Brittany~S. Liu}, \bibinfo{person}{Cory~J. Clark},
  \bibinfo{person}{Sean~P. Wojcik}, \bibinfo{person}{Eric~E. Chen},
  \bibinfo{person}{Rebecca~H. Grady}, \bibinfo{person}{Jared~B. Celniker},
  {and} \bibinfo{person}{Joanne~F. Zinger}.} \bibinfo{year}{2019}\natexlab{b}.
\newblock \showarticletitle{At {Least} {Bias} {Is} {Bipartisan}: {A}
  {Meta}-{Analytic} {Comparison} of {Partisan} {Bias} in {Liberals} and
  {Conservatives}}.
\newblock \bibinfo{journal}{\emph{Perspectives on Psychological Science}}
  \bibinfo{volume}{14}, \bibinfo{number}{2} (\bibinfo{date}{March}
  \bibinfo{year}{2019}), \bibinfo{pages}{273--291}.
\newblock
\showISSN{1745-6916, 1745-6924}
\urldef\tempurl%
\url{https://doi.org/10.1177/1745691617746796}
\showDOI{\tempurl}


\bibitem[\protect\citeauthoryear{Dolan, Hallsworth, Halpern, King, Metcalfe,
  and Vlaev}{Dolan et~al\mbox{.}}{2012}]%
        {dolan_influencing_2012}
\bibfield{author}{\bibinfo{person}{P. Dolan}, \bibinfo{person}{M. Hallsworth},
  \bibinfo{person}{D. Halpern}, \bibinfo{person}{D. King}, \bibinfo{person}{R.
  Metcalfe}, {and} \bibinfo{person}{I. Vlaev}.}
  \bibinfo{year}{2012}\natexlab{}.
\newblock \showarticletitle{Influencing behaviour: {The} mindspace way}.
\newblock \bibinfo{journal}{\emph{Journal of Economic Psychology}}
  \bibinfo{volume}{33}, \bibinfo{number}{1} (\bibinfo{date}{Feb.}
  \bibinfo{year}{2012}), \bibinfo{pages}{264--277}.
\newblock
\showISSN{0167-4870}
\urldef\tempurl%
\url{https://doi.org/10.1016/j.joep.2011.10.009}
\showDOI{\tempurl}


\bibitem[\protect\citeauthoryear{Douglas, Sutton, and Cichocka}{Douglas
  et~al\mbox{.}}{2017}]%
        {douglas_psychology_2017}
\bibfield{author}{\bibinfo{person}{Karen~M. Douglas},
  \bibinfo{person}{Robbie~M. Sutton}, {and} \bibinfo{person}{Aleksandra
  Cichocka}.} \bibinfo{year}{2017}\natexlab{}.
\newblock \showarticletitle{The {Psychology} of {Conspiracy} {Theories}}.
\newblock \bibinfo{journal}{\emph{Current Directions in Psychological Science}}
  \bibinfo{volume}{26}, \bibinfo{number}{6} (\bibinfo{date}{Dec.}
  \bibinfo{year}{2017}), \bibinfo{pages}{538--542}.
\newblock
\showISSN{0963-7214}
\urldef\tempurl%
\url{https://doi.org/10.1177/0963721417718261}
\showDOI{\tempurl}


\bibitem[\protect\citeauthoryear{Dubois and Blank}{Dubois and Blank}{2018}]%
        {dubois_echo_2018}
\bibfield{author}{\bibinfo{person}{Elizabeth Dubois} {and}
  \bibinfo{person}{Grant Blank}.} \bibinfo{year}{2018}\natexlab{}.
\newblock \showarticletitle{The echo chamber is overstated: the moderating
  effect of political interest and diverse media}.
\newblock \bibinfo{journal}{\emph{Information, Communication \& Society}}
  \bibinfo{volume}{21}, \bibinfo{number}{5} (\bibinfo{date}{May}
  \bibinfo{year}{2018}), \bibinfo{pages}{729--745}.
\newblock
\showISSN{1369-118X}
\urldef\tempurl%
\url{https://doi.org/10.1080/1369118X.2018.1428656}
\showDOI{\tempurl}


\bibitem[\protect\citeauthoryear{Ecker, Lewandowsky, Fenton, and Martin}{Ecker
  et~al\mbox{.}}{2014}]%
        {ecker_people_2014}
\bibfield{author}{\bibinfo{person}{Ullrich K.~H. Ecker},
  \bibinfo{person}{Stephan Lewandowsky}, \bibinfo{person}{Olivia Fenton}, {and}
  \bibinfo{person}{Kelsey Martin}.} \bibinfo{year}{2014}\natexlab{}.
\newblock \showarticletitle{Do people keep believing because they want to?
  {Preexisting} attitudes and the continued influence of misinformation}.
\newblock \bibinfo{journal}{\emph{Memory \& Cognition}} \bibinfo{volume}{42},
  \bibinfo{number}{2} (\bibinfo{date}{Feb.} \bibinfo{year}{2014}),
  \bibinfo{pages}{292--304}.
\newblock
\showISSN{0090-502X, 1532-5946}
\urldef\tempurl%
\url{https://doi.org/10.3758/s13421-013-0358-x}
\showDOI{\tempurl}


\bibitem[\protect\citeauthoryear{Ecker, Lewandowsky, and Tang}{Ecker
  et~al\mbox{.}}{2010}]%
        {ecker_explicit_2010}
\bibfield{author}{\bibinfo{person}{Ullrich K.~H. Ecker},
  \bibinfo{person}{Stephan Lewandowsky}, {and} \bibinfo{person}{David T.~W.
  Tang}.} \bibinfo{year}{2010}\natexlab{}.
\newblock \showarticletitle{Explicit warnings reduce but do not eliminate the
  continued influence of misinformation}.
\newblock \bibinfo{journal}{\emph{Memory \& Cognition}} \bibinfo{volume}{38},
  \bibinfo{number}{8} (\bibinfo{date}{Dec.} \bibinfo{year}{2010}),
  \bibinfo{pages}{1087--1100}.
\newblock
\showISSN{0090-502X, 1532-5946}
\urldef\tempurl%
\url{https://doi.org/10.3758/MC.38.8.1087}
\showDOI{\tempurl}


\bibitem[\protect\citeauthoryear{Efstratiou and Caulfield}{Efstratiou and
  Caulfield}{2021}]%
        {efstratiou_misrepresenting_2021}
\bibfield{author}{\bibinfo{person}{Alexandros Efstratiou} {and}
  \bibinfo{person}{Tristan Caulfield}.} \bibinfo{year}{2021}\natexlab{}.
\newblock \showarticletitle{Misrepresenting {Scientific} {Consensus} on
  {COVID}-19: {The} {Amplification} of {Dissenting} {Scientists} on {Twitter}}.
\newblock \bibinfo{journal}{\emph{arXiv:2111.10594 [physics]}}
  (\bibinfo{date}{Nov.} \bibinfo{year}{2021}).
\newblock
\urldef\tempurl%
\url{http://arxiv.org/abs/2111.10594}
\showURL{%
\tempurl}


\bibitem[\protect\citeauthoryear{Enders, Uscinski, Klofstad, and Stoler}{Enders
  et~al\mbox{.}}{2020}]%
        {enders_different_2020}
\bibfield{author}{\bibinfo{person}{Adam~M. Enders}, \bibinfo{person}{Joseph~E.
  Uscinski}, \bibinfo{person}{Casey Klofstad}, {and} \bibinfo{person}{Justin
  Stoler}.} \bibinfo{year}{2020}\natexlab{}.
\newblock \showarticletitle{The different forms of {COVID}-19 misinformation
  and their consequences}.
\newblock \bibinfo{journal}{\emph{The Harvard Kennedy School Misinformation
  Review}} (\bibinfo{date}{Nov.} \bibinfo{year}{2020}).
\newblock
\urldef\tempurl%
\url{https://dash.harvard.edu/handle/1/37366466}
\showURL{%
\tempurl}


\bibitem[\protect\citeauthoryear{Erisen, Lodge, and Taber}{Erisen
  et~al\mbox{.}}{2014}]%
        {erisen_affective_2014}
\bibfield{author}{\bibinfo{person}{Cengiz Erisen}, \bibinfo{person}{Milton
  Lodge}, {and} \bibinfo{person}{Charles~S. Taber}.}
  \bibinfo{year}{2014}\natexlab{}.
\newblock \showarticletitle{Affective {Contagion} in {Effortful} {Political}
  {Thinking}}.
\newblock \bibinfo{journal}{\emph{Political Psychology}} \bibinfo{volume}{35},
  \bibinfo{number}{2} (\bibinfo{year}{2014}), \bibinfo{pages}{187--206}.
\newblock
\showISSN{1467-9221}
\urldef\tempurl%
\url{https://doi.org/10.1111/j.1467-9221.2012.00937.x}
\showDOI{\tempurl}


\bibitem[\protect\citeauthoryear{Evans}{Evans}{2003}]%
        {evans_two_2003}
\bibfield{author}{\bibinfo{person}{Jonathan St. B.~T. Evans}.}
  \bibinfo{year}{2003}\natexlab{}.
\newblock \showarticletitle{In two minds: dual-process accounts of reasoning}.
\newblock \bibinfo{journal}{\emph{Trends in Cognitive Sciences}}
  \bibinfo{volume}{7}, \bibinfo{number}{10} (\bibinfo{date}{Oct.}
  \bibinfo{year}{2003}), \bibinfo{pages}{454--459}.
\newblock
\showISSN{1364-6613}
\urldef\tempurl%
\url{https://doi.org/10.1016/j.tics.2003.08.012}
\showDOI{\tempurl}


\bibitem[\protect\citeauthoryear{Fernbach, Rogers, Fox, and Sloman}{Fernbach
  et~al\mbox{.}}{2013}]%
        {fernbach_political_2013}
\bibfield{author}{\bibinfo{person}{Philip~M Fernbach}, \bibinfo{person}{Todd
  Rogers}, \bibinfo{person}{Craig~R Fox}, {and} \bibinfo{person}{Steven~A
  Sloman}.} \bibinfo{year}{2013}\natexlab{}.
\newblock \showarticletitle{Political {Extremism} {Is} {Supported} by an
  {Illusion} of {Understanding}}.
\newblock \bibinfo{journal}{\emph{Psychological Science}} \bibinfo{volume}{24},
  \bibinfo{number}{6} (\bibinfo{year}{2013}), \bibinfo{pages}{939--946}.
\newblock


\bibitem[\protect\citeauthoryear{Festinger}{Festinger}{1957}]%
        {festinger_theory_1957}
\bibfield{author}{\bibinfo{person}{Leon Festinger}.}
  \bibinfo{year}{1957}\natexlab{}.
\newblock \bibinfo{booktitle}{\emph{A {Theory} of {Cognitive} {Dissonance}}}.
\newblock \bibinfo{publisher}{Stanford University Press}.
\newblock
\showISBNx{978-0-8047-0911-8}


\bibitem[\protect\citeauthoryear{Festinger and Carlsmith}{Festinger and
  Carlsmith}{1959}]%
        {festinger_cognitive_1959}
\bibfield{author}{\bibinfo{person}{Leon Festinger} {and}
  \bibinfo{person}{James~M. Carlsmith}.} \bibinfo{year}{1959}\natexlab{}.
\newblock \showarticletitle{Cognitive consequences of forced compliance}.
\newblock \bibinfo{journal}{\emph{The Journal of Abnormal and Social
  Psychology}} \bibinfo{volume}{58}, \bibinfo{number}{2}
  (\bibinfo{year}{1959}), \bibinfo{pages}{203--210}.
\newblock
\showISSN{0096-851X(Print)}
\urldef\tempurl%
\url{https://doi.org/10.1037/h0041593}
\showDOI{\tempurl}


\bibitem[\protect\citeauthoryear{Fisher}{Fisher}{1993}]%
        {fisher_social_1993}
\bibfield{author}{\bibinfo{person}{Robert~J. Fisher}.}
  \bibinfo{year}{1993}\natexlab{}.
\newblock \showarticletitle{Social {Desirability} {Bias} and the {Validity} of
  {Indirect} {Questioning}}.
\newblock \bibinfo{journal}{\emph{Journal of Consumer Research}}
  \bibinfo{volume}{20}, \bibinfo{number}{2} (\bibinfo{date}{Sept.}
  \bibinfo{year}{1993}), \bibinfo{pages}{303--315}.
\newblock
\showISSN{0093-5301}
\urldef\tempurl%
\url{https://doi.org/10.1086/209351}
\showDOI{\tempurl}


\bibitem[\protect\citeauthoryear{Frederic and Falomir-Pichastor}{Frederic and
  Falomir-Pichastor}{2018}]%
        {frederic_heterogeneity_2018}
\bibfield{author}{\bibinfo{person}{Natasha~S. Frederic} {and}
  \bibinfo{person}{Juan~M. Falomir-Pichastor}.}
  \bibinfo{year}{2018}\natexlab{}.
\newblock \showarticletitle{Heterogeneity of {Ingroup} {Identity} and
  {Anti}-{Immigrant} {Prejudice}: {The} {Moderating} {Role} of {RWA} and
  {Outgroup} {Homogeneity}}.
\newblock \bibinfo{journal}{\emph{International Review of Social Psychology}}
  \bibinfo{volume}{31}, \bibinfo{number}{1} (\bibinfo{date}{April}
  \bibinfo{year}{2018}), \bibinfo{pages}{13}.
\newblock
\showISSN{2397-8570}
\urldef\tempurl%
\url{https://doi.org/10.5334/irsp.152}
\showDOI{\tempurl}


\bibitem[\protect\citeauthoryear{Gao, Xiao, Karahalios, and Fu}{Gao
  et~al\mbox{.}}{2018}]%
        {gao_label_2018}
\bibfield{author}{\bibinfo{person}{Mingkun Gao}, \bibinfo{person}{Ziang Xiao},
  \bibinfo{person}{Karrie Karahalios}, {and} \bibinfo{person}{Wai-Tat Fu}.}
  \bibinfo{year}{2018}\natexlab{}.
\newblock \showarticletitle{To {Label} or {Not} to {Label}: {The} {Effect} of
  {Stance} and {Credibility} {Labels} on {Readers}' {Selection} and
  {Perception} of {News} {Articles}}.
\newblock \bibinfo{journal}{\emph{Proceedings of the ACM on Human-Computer
  Interaction}} \bibinfo{volume}{2}, \bibinfo{number}{CSCW}
  (\bibinfo{date}{Nov.} \bibinfo{year}{2018}), \bibinfo{pages}{55:1--55:16}.
\newblock
\urldef\tempurl%
\url{https://doi.org/10.1145/3274324}
\showDOI{\tempurl}


\bibitem[\protect\citeauthoryear{Garimella, De~Francisci~Morales, Gionis, and
  Mathioudakis}{Garimella et~al\mbox{.}}{2017}]%
        {garimella_reducing_2017}
\bibfield{author}{\bibinfo{person}{Kiran Garimella}, \bibinfo{person}{Gianmarco
  De~Francisci~Morales}, \bibinfo{person}{Aristides Gionis}, {and}
  \bibinfo{person}{Michael Mathioudakis}.} \bibinfo{year}{2017}\natexlab{}.
\newblock \showarticletitle{Reducing {Controversy} by {Connecting} {Opposing}
  {Views}}. In \bibinfo{booktitle}{\emph{Proceedings of the {Tenth} {ACM}
  {International} {Conference} on {Web} {Search} and {Data} {Mining}}}
  \emph{(\bibinfo{series}{{WSDM} '17})}. \bibinfo{publisher}{Association for
  Computing Machinery}, \bibinfo{address}{New York, NY, USA},
  \bibinfo{pages}{81--90}.
\newblock
\showISBNx{978-1-4503-4675-7}
\urldef\tempurl%
\url{https://doi.org/10.1145/3018661.3018703}
\showDOI{\tempurl}


\bibitem[\protect\citeauthoryear{Garimella, De~Francisci~Morales, Gionis, and
  Mathioudakis}{Garimella et~al\mbox{.}}{2018}]%
        {garimella_political_2018}
\bibfield{author}{\bibinfo{person}{Kiran Garimella}, \bibinfo{person}{Gianmarco
  De~Francisci~Morales}, \bibinfo{person}{Aristides Gionis}, {and}
  \bibinfo{person}{Michael Mathioudakis}.} \bibinfo{year}{2018}\natexlab{}.
\newblock \showarticletitle{Political {Discourse} on {Social} {Media}: {Echo}
  {Chambers}, {Gatekeepers}, and the {Price} of {Bipartisanship}}. In
  \bibinfo{booktitle}{\emph{Proceedings of the 2018 {World} {Wide} {Web}
  {Conference} on {World} {Wide} {Web} - {WWW} '18}}. \bibinfo{publisher}{ACM
  Press}, \bibinfo{address}{Lyon, France}, \bibinfo{pages}{913--922}.
\newblock
\showISBNx{978-1-4503-5639-8}
\urldef\tempurl%
\url{https://doi.org/10.1145/3178876.3186139}
\showDOI{\tempurl}


\bibitem[\protect\citeauthoryear{Garrett}{Garrett}{2009}]%
        {garrett_echo_2009}
\bibfield{author}{\bibinfo{person}{R.~Kelly Garrett}.}
  \bibinfo{year}{2009}\natexlab{}.
\newblock \showarticletitle{Echo chambers online?: {Politically} motivated
  selective exposure among {Internet} news users}.
\newblock \bibinfo{journal}{\emph{Journal of Computer-Mediated Communication}}
  \bibinfo{volume}{14}, \bibinfo{number}{2} (\bibinfo{date}{Jan.}
  \bibinfo{year}{2009}), \bibinfo{pages}{265--285}.
\newblock
\showISSN{1083-6101}
\urldef\tempurl%
\url{https://doi.org/10.1111/j.1083-6101.2009.01440.x}
\showDOI{\tempurl}


\bibitem[\protect\citeauthoryear{Gillani, Yuan, Saveski, Vosoughi, and
  Roy}{Gillani et~al\mbox{.}}{2018}]%
        {gillani_me_2018}
\bibfield{author}{\bibinfo{person}{Nabeel Gillani}, \bibinfo{person}{Ann Yuan},
  \bibinfo{person}{Martin Saveski}, \bibinfo{person}{Soroush Vosoughi}, {and}
  \bibinfo{person}{Deb Roy}.} \bibinfo{year}{2018}\natexlab{}.
\newblock \showarticletitle{Me, {My} {Echo} {Chamber}, and {I}: {Introspection}
  on {Social} {Media} {Polarization}}. In \bibinfo{booktitle}{\emph{Proceedings
  of the 2018 {World} {Wide} {Web} {Conference}}} \emph{(\bibinfo{series}{{WWW}
  '18})}. \bibinfo{publisher}{International World Wide Web Conferences Steering
  Committee}, \bibinfo{address}{Republic and Canton of Geneva, CHE},
  \bibinfo{pages}{823--831}.
\newblock
\showISBNx{978-1-4503-5639-8}
\urldef\tempurl%
\url{https://doi.org/10.1145/3178876.3186130}
\showDOI{\tempurl}


\bibitem[\protect\citeauthoryear{Gillespie}{Gillespie}{2020}]%
        {gillespie_disruption_2020}
\bibfield{author}{\bibinfo{person}{Alex Gillespie}.}
  \bibinfo{year}{2020}\natexlab{}.
\newblock \showarticletitle{Disruption, {Self}-{Presentation}, and {Defensive}
  {Tactics} at the {Threshold} of {Learning}}.
\newblock \bibinfo{journal}{\emph{Review of General Psychology}}
  (\bibinfo{date}{April} \bibinfo{year}{2020}),
  \bibinfo{pages}{108926802091425}.
\newblock
\showISSN{1089-2680, 1939-1552}
\urldef\tempurl%
\url{https://doi.org/10.1177/1089268020914258}
\showDOI{\tempurl}


\bibitem[\protect\citeauthoryear{Goldschmied, Sheptock, Kim, and
  Galily}{Goldschmied et~al\mbox{.}}{2017}]%
        {goldschmied_appraising_2017}
\bibfield{author}{\bibinfo{person}{Nadav Goldschmied}, \bibinfo{person}{Mark
  Sheptock}, \bibinfo{person}{Kacey Kim}, {and} \bibinfo{person}{Yair Galily}.}
  \bibinfo{year}{2017}\natexlab{}.
\newblock \showarticletitle{Appraising {Loftus} and {Palmer} (1974) post-event
  information versus concurrent commentary in the context of sport}.
\newblock \bibinfo{journal}{\emph{The Quarterly Journal of Experimental
  Psychology}} \bibinfo{volume}{70}, \bibinfo{number}{11} (\bibinfo{date}{Nov.}
  \bibinfo{year}{2017}), \bibinfo{pages}{2347--2356}.
\newblock
\showISSN{1747-0218}
\urldef\tempurl%
\url{https://doi.org/10.1080/17470218.2016.1237980}
\showDOI{\tempurl}


\bibitem[\protect\citeauthoryear{Goldstein, Cialdini, and
  Griskevicius}{Goldstein et~al\mbox{.}}{2008}]%
        {goldstein_room_2008}
\bibfield{author}{\bibinfo{person}{Noah~J. Goldstein},
  \bibinfo{person}{Robert~B. Cialdini}, {and} \bibinfo{person}{Vladas
  Griskevicius}.} \bibinfo{year}{2008}\natexlab{}.
\newblock \showarticletitle{A {Room} with a {Viewpoint}: {Using} {Social}
  {Norms} to {Motivate} {Environmental} {Conservation} in {Hotels}}.
\newblock \bibinfo{journal}{\emph{Journal of Consumer Research}}
  \bibinfo{volume}{35}, \bibinfo{number}{3} (\bibinfo{date}{Oct.}
  \bibinfo{year}{2008}), \bibinfo{pages}{472--482}.
\newblock
\showISSN{0093-5301}
\urldef\tempurl%
\url{https://doi.org/10.1086/586910}
\showDOI{\tempurl}


\bibitem[\protect\citeauthoryear{Graham, Haidt, and Nosek}{Graham
  et~al\mbox{.}}{2009}]%
        {graham_liberals_2009}
\bibfield{author}{\bibinfo{person}{Jesse Graham}, \bibinfo{person}{Jonathan
  Haidt}, {and} \bibinfo{person}{Brian~A. Nosek}.}
  \bibinfo{year}{2009}\natexlab{}.
\newblock \showarticletitle{Liberals and conservatives rely on different sets
  of moral foundations.}
\newblock \bibinfo{journal}{\emph{Journal of Personality and Social
  Psychology}} \bibinfo{volume}{96}, \bibinfo{number}{5} (\bibinfo{date}{May}
  \bibinfo{year}{2009}), \bibinfo{pages}{1029--1046}.
\newblock
\showISSN{1939-1315, 0022-3514}
\urldef\tempurl%
\url{https://doi.org/10.1037/a0015141}
\showDOI{\tempurl}


\bibitem[\protect\citeauthoryear{Guess, Nagler, and Tucker}{Guess
  et~al\mbox{.}}{2019}]%
        {guess_less_2019}
\bibfield{author}{\bibinfo{person}{Andrew Guess}, \bibinfo{person}{Jonathan
  Nagler}, {and} \bibinfo{person}{Joshua Tucker}.}
  \bibinfo{year}{2019}\natexlab{}.
\newblock \showarticletitle{Less than you think: {Prevalence} and predictors of
  fake news dissemination on {Facebook}}.
\newblock \bibinfo{journal}{\emph{Science Advances}} \bibinfo{volume}{5},
  \bibinfo{number}{1} (\bibinfo{year}{2019}), \bibinfo{pages}{eaau4586}.
\newblock
\urldef\tempurl%
\url{https://doi.org/10.1126/sciadv.aau4586}
\showDOI{\tempurl}


\bibitem[\protect\citeauthoryear{Guess, Nyhan, Lyons, and Reifler}{Guess
  et~al\mbox{.}}{2018}]%
        {guess_why_2018}
\bibfield{author}{\bibinfo{person}{Andrew Guess}, \bibinfo{person}{Brendan
  Nyhan}, \bibinfo{person}{Benjamin Lyons}, {and} \bibinfo{person}{Jason
  Reifler}.} \bibinfo{year}{2018}\natexlab{}.
\newblock \showarticletitle{Why selective exposure to like-minded political
  news is less prevalent than you think}.
\newblock \bibinfo{journal}{\emph{Knight Foundation}} (\bibinfo{year}{2018}),
  \bibinfo{pages}{25}.
\newblock


\bibitem[\protect\citeauthoryear{Guest}{Guest}{2018}]%
        {guest_anti-echo_2018}
\bibfield{author}{\bibinfo{person}{Ella Guest}.}
  \bibinfo{year}{2018}\natexlab{}.
\newblock \showarticletitle{({Anti}-){Echo} {Chamber} {Participation}:
  {Examining} {Contributor} {Activity} {Beyond} the {Chamber}}. In
  \bibinfo{booktitle}{\emph{Proceedings of the 9th {International} {Conference}
  on {Social} {Media} and {Society}}}. \bibinfo{publisher}{ACM},
  \bibinfo{address}{Copenhagen Denmark}, \bibinfo{pages}{301--304}.
\newblock
\showISBNx{978-1-4503-6334-1}
\urldef\tempurl%
\url{https://doi.org/10.1145/3217804.3217933}
\showDOI{\tempurl}


\bibitem[\protect\citeauthoryear{Haidt and Graham}{Haidt and Graham}{2007}]%
        {haidt_when_2007}
\bibfield{author}{\bibinfo{person}{Jonathan Haidt} {and} \bibinfo{person}{Jesse
  Graham}.} \bibinfo{year}{2007}\natexlab{}.
\newblock \showarticletitle{When {Morality} {Opposes} {Justice}:
  {Conservatives} {Have} {Moral} {Intuitions} that {Liberals} may not
  {Recognize}}.
\newblock \bibinfo{journal}{\emph{Social Justice Research}}
  \bibinfo{volume}{20}, \bibinfo{number}{1} (\bibinfo{date}{March}
  \bibinfo{year}{2007}), \bibinfo{pages}{98--116}.
\newblock
\showISSN{1573-6725}
\urldef\tempurl%
\url{https://doi.org/10.1007/s11211-007-0034-z}
\showDOI{\tempurl}


\bibitem[\protect\citeauthoryear{Hameleers and van~der Meer}{Hameleers and
  van~der Meer}{2020}]%
        {hameleers_misinformation_2020}
\bibfield{author}{\bibinfo{person}{Michael Hameleers} {and}
  \bibinfo{person}{Toni G. L.~A. van~der Meer}.}
  \bibinfo{year}{2020}\natexlab{}.
\newblock \showarticletitle{Misinformation and {Polarization} in a
  {High}-{Choice} {Media} {Environment}: {How} {Effective} {Are} {Political}
  {Fact}-{Checkers}?}
\newblock \bibinfo{journal}{\emph{Communication Research}}
  \bibinfo{volume}{47}, \bibinfo{number}{2} (\bibinfo{date}{March}
  \bibinfo{year}{2020}), \bibinfo{pages}{227--250}.
\newblock
\showISSN{0093-6502, 1552-3810}
\urldef\tempurl%
\url{https://doi.org/10.1177/0093650218819671}
\showDOI{\tempurl}


\bibitem[\protect\citeauthoryear{Hart, Albarracín, Eagly, Brechan, Lindberg,
  and Merrill}{Hart et~al\mbox{.}}{2009}]%
        {hart_feeling_2009}
\bibfield{author}{\bibinfo{person}{William Hart}, \bibinfo{person}{Dolores
  Albarracín}, \bibinfo{person}{Alice~H. Eagly}, \bibinfo{person}{Inge
  Brechan}, \bibinfo{person}{Matthew~J. Lindberg}, {and} \bibinfo{person}{Lisa
  Merrill}.} \bibinfo{year}{2009}\natexlab{}.
\newblock \showarticletitle{Feeling {Validated} {Versus} {Being} {Correct}:{A}
  {Meta}-{Analysis} of {Selective} {Exposure} to {Information}}.
\newblock \bibinfo{journal}{\emph{Psychological bulletin}}
  \bibinfo{volume}{135}, \bibinfo{number}{4} (\bibinfo{date}{July}
  \bibinfo{year}{2009}), \bibinfo{pages}{555--588}.
\newblock
\showISSN{0033-2909}
\urldef\tempurl%
\url{https://doi.org/10.1037/a0015701}
\showDOI{\tempurl}


\bibitem[\protect\citeauthoryear{Harvey, van~den Berg, Ellers, Kampen,
  Crowther, Roessingh, Verheggen, Nuijten, Post, Lewandowsky, Stirling,
  Balgopal, Amstrup, and Mann}{Harvey et~al\mbox{.}}{2018}]%
        {harvey_internet_2018}
\bibfield{author}{\bibinfo{person}{Jeffrey~A Harvey}, \bibinfo{person}{Daphne
  van~den Berg}, \bibinfo{person}{Jacintha Ellers}, \bibinfo{person}{Remko
  Kampen}, \bibinfo{person}{Thomas~W Crowther}, \bibinfo{person}{Peter
  Roessingh}, \bibinfo{person}{Bart Verheggen}, \bibinfo{person}{Rascha J~M
  Nuijten}, \bibinfo{person}{Eric Post}, \bibinfo{person}{Stephan Lewandowsky},
  \bibinfo{person}{Ian Stirling}, \bibinfo{person}{Meena Balgopal},
  \bibinfo{person}{Steven~C Amstrup}, {and} \bibinfo{person}{Michael~E Mann}.}
  \bibinfo{year}{2018}\natexlab{}.
\newblock \showarticletitle{Internet {Blogs}, {Polar} {Bears}, and
  {Climate}-{Change} {Denial} by {Proxy}}.
\newblock \bibinfo{journal}{\emph{BioScience}} \bibinfo{volume}{68},
  \bibinfo{number}{4} (\bibinfo{date}{April} \bibinfo{year}{2018}),
  \bibinfo{pages}{281--287}.
\newblock
\showISSN{0006-3568, 1525-3244}
\urldef\tempurl%
\url{https://doi.org/10.1093/biosci/bix133}
\showDOI{\tempurl}


\bibitem[\protect\citeauthoryear{Hills}{Hills}{2019}]%
        {hills_dark_2019}
\bibfield{author}{\bibinfo{person}{Thomas~T. Hills}.}
  \bibinfo{year}{2019}\natexlab{}.
\newblock \showarticletitle{The {Dark} {Side} of {Information}
  {Proliferation}}.
\newblock \bibinfo{journal}{\emph{Perspectives on Psychological Science}}
  \bibinfo{volume}{14}, \bibinfo{number}{3} (\bibinfo{date}{May}
  \bibinfo{year}{2019}), \bibinfo{pages}{323--330}.
\newblock
\showISSN{1745-6916, 1745-6924}
\urldef\tempurl%
\url{https://doi.org/10.1177/1745691618803647}
\showDOI{\tempurl}


\bibitem[\protect\citeauthoryear{Horne, Nevo, Adali, Manikonda, and
  Arrington}{Horne et~al\mbox{.}}{2020}]%
        {horne_tailoring_2020}
\bibfield{author}{\bibinfo{person}{Benjamin~D. Horne}, \bibinfo{person}{Dorit
  Nevo}, \bibinfo{person}{Sibel Adali}, \bibinfo{person}{Lydia Manikonda},
  {and} \bibinfo{person}{Clare Arrington}.} \bibinfo{year}{2020}\natexlab{}.
\newblock \showarticletitle{Tailoring heuristics and timing {AI} interventions
  for supporting news veracity assessments}.
\newblock \bibinfo{journal}{\emph{Computers in Human Behavior Reports}}
  \bibinfo{volume}{2} (\bibinfo{date}{Aug.} \bibinfo{year}{2020}),
  \bibinfo{pages}{100043}.
\newblock
\showISSN{2451-9588}
\urldef\tempurl%
\url{https://doi.org/10.1016/j.chbr.2020.100043}
\showDOI{\tempurl}


\bibitem[\protect\citeauthoryear{Horta~Ribeiro, Jhaver, Zannettou, Blackburn,
  Stringhini, De~Cristofaro, and West}{Horta~Ribeiro et~al\mbox{.}}{2021}]%
        {horta_ribeiro_platform_2021}
\bibfield{author}{\bibinfo{person}{Manoel Horta~Ribeiro},
  \bibinfo{person}{Shagun Jhaver}, \bibinfo{person}{Savvas Zannettou},
  \bibinfo{person}{Jeremy Blackburn}, \bibinfo{person}{Gianluca Stringhini},
  \bibinfo{person}{Emiliano De~Cristofaro}, {and} \bibinfo{person}{Robert
  West}.} \bibinfo{year}{2021}\natexlab{}.
\newblock \showarticletitle{Do {Platform} {Migrations} {Compromise} {Content}
  {Moderation}? {Evidence} from r/{The}\_Donald and r/{Incels}}.
\newblock \bibinfo{journal}{\emph{Proceedings of the ACM on Human-Computer
  Interaction}} \bibinfo{volume}{5}, \bibinfo{number}{CSCW2}
  (\bibinfo{date}{Oct.} \bibinfo{year}{2021}), \bibinfo{pages}{316:1--316:24}.
\newblock
\urldef\tempurl%
\url{https://doi.org/10.1145/3476057}
\showDOI{\tempurl}


\bibitem[\protect\citeauthoryear{Hosseinmardi, Ghasemian, Clauset, Rothschild,
  Mobius, and Watts}{Hosseinmardi et~al\mbox{.}}{2020}]%
        {hosseinmardi_evaluating_2020}
\bibfield{author}{\bibinfo{person}{Homa Hosseinmardi}, \bibinfo{person}{Amir
  Ghasemian}, \bibinfo{person}{Aaron Clauset}, \bibinfo{person}{David~M.
  Rothschild}, \bibinfo{person}{Markus Mobius}, {and}
  \bibinfo{person}{Duncan~J. Watts}.} \bibinfo{year}{2020}\natexlab{}.
\newblock \showarticletitle{Evaluating the scale, growth, and origins of
  right-wing echo chambers on {YouTube}}.
\newblock \bibinfo{journal}{\emph{arXiv:2011.12843 [cs]}} (\bibinfo{date}{Nov.}
  \bibinfo{year}{2020}).
\newblock
\urldef\tempurl%
\url{http://arxiv.org/abs/2011.12843}
\showURL{%
\tempurl}


\bibitem[\protect\citeauthoryear{Jiang and Wilson}{Jiang and Wilson}{2018}]%
        {jiang_linguistic_2018}
\bibfield{author}{\bibinfo{person}{Shan Jiang} {and} \bibinfo{person}{Christo
  Wilson}.} \bibinfo{year}{2018}\natexlab{}.
\newblock \showarticletitle{Linguistic {Signals} under {Misinformation} and
  {Fact}-{Checking}: {Evidence} from {User} {Comments} on {Social} {Media}}.
\newblock \bibinfo{journal}{\emph{Proceedings of the ACM on Human-Computer
  Interaction}} \bibinfo{volume}{2}, \bibinfo{number}{CSCW}
  (\bibinfo{date}{Nov.} \bibinfo{year}{2018}), \bibinfo{pages}{82:1--82:23}.
\newblock
\urldef\tempurl%
\url{https://doi.org/10.1145/3274351}
\showDOI{\tempurl}


\bibitem[\protect\citeauthoryear{Johnson and Seifert}{Johnson and
  Seifert}{1994}]%
        {johnson_sources_1994}
\bibfield{author}{\bibinfo{person}{Hollyn~M. Johnson} {and}
  \bibinfo{person}{Colleen~M. Seifert}.} \bibinfo{year}{1994}\natexlab{}.
\newblock \showarticletitle{Sources of the continued influence effect: {When}
  misinformation in memory affects later inferences.}
\newblock \bibinfo{journal}{\emph{Journal of Experimental Psychology: Learning,
  Memory, and Cognition}} \bibinfo{volume}{20}, \bibinfo{number}{6}
  (\bibinfo{date}{Nov.} \bibinfo{year}{1994}), \bibinfo{pages}{1420--1436}.
\newblock
\showISSN{1939-1285, 0278-7393}
\urldef\tempurl%
\url{https://doi.org/10.1037/0278-7393.20.6.1420}
\showDOI{\tempurl}


\bibitem[\protect\citeauthoryear{Jones, Crozier, and Strange}{Jones
  et~al\mbox{.}}{2017}]%
        {jones_believing_2017}
\bibfield{author}{\bibinfo{person}{Kristyn~A. Jones},
  \bibinfo{person}{William~E. Crozier}, {and} \bibinfo{person}{Deryn Strange}.}
  \bibinfo{year}{2017}\natexlab{}.
\newblock \showarticletitle{Believing is {Seeing}: {Biased} {Viewing} of
  {Body}-{Worn} {Camera} {Footage}}.
\newblock \bibinfo{journal}{\emph{Journal of Applied Research in Memory and
  Cognition}} \bibinfo{volume}{6}, \bibinfo{number}{4} (\bibinfo{date}{Dec.}
  \bibinfo{year}{2017}), \bibinfo{pages}{460--474}.
\newblock
\showISSN{22113681}
\urldef\tempurl%
\url{https://doi.org/10.1016/j.jarmac.2017.07.007}
\showDOI{\tempurl}


\bibitem[\protect\citeauthoryear{Jost}{Jost}{2017}]%
        {jost_ideological_2017}
\bibfield{author}{\bibinfo{person}{John~T. Jost}.}
  \bibinfo{year}{2017}\natexlab{}.
\newblock \showarticletitle{Ideological {Asymmetries} and the {Essence} of
  {Political} {Psychology}}.
\newblock \bibinfo{journal}{\emph{Political Psychology}} \bibinfo{volume}{38},
  \bibinfo{number}{2} (\bibinfo{year}{2017}), \bibinfo{pages}{167--208}.
\newblock
\showISSN{1467-9221}
\urldef\tempurl%
\url{https://doi.org/10.1111/pops.12407}
\showDOI{\tempurl}


\bibitem[\protect\citeauthoryear{Jost, Glaser, Kruglanski, and Sulloway}{Jost
  et~al\mbox{.}}{2003}]%
        {jost_political_2003}
\bibfield{author}{\bibinfo{person}{John~T. Jost}, \bibinfo{person}{Jack
  Glaser}, \bibinfo{person}{Arie~W. Kruglanski}, {and}
  \bibinfo{person}{Frank~J. Sulloway}.} \bibinfo{year}{2003}\natexlab{}.
\newblock \showarticletitle{Political conservatism as motivated social
  cognition.}
\newblock \bibinfo{journal}{\emph{Psychological Bulletin}}
  \bibinfo{volume}{129}, \bibinfo{number}{3} (\bibinfo{year}{2003}),
  \bibinfo{pages}{339--375}.
\newblock
\showISSN{1939-1455, 0033-2909}
\urldef\tempurl%
\url{https://doi.org/10.1037/0033-2909.129.3.339}
\showDOI{\tempurl}


\bibitem[\protect\citeauthoryear{Jost, Stern, Rule, and Sterling}{Jost
  et~al\mbox{.}}{2017}]%
        {jost_politics_2017}
\bibfield{author}{\bibinfo{person}{John~T. Jost}, \bibinfo{person}{Chadly
  Stern}, \bibinfo{person}{Nicholas~O. Rule}, {and} \bibinfo{person}{Joanna
  Sterling}.} \bibinfo{year}{2017}\natexlab{}.
\newblock \showarticletitle{The {Politics} of {Fear}: {Is} {There} an
  {Ideological} {Asymmetry} in {Existential} {Motivation}?}
\newblock \bibinfo{journal}{\emph{Social Cognition}} \bibinfo{volume}{35},
  \bibinfo{number}{4} (\bibinfo{date}{July} \bibinfo{year}{2017}),
  \bibinfo{pages}{324--353}.
\newblock
\showISSN{0278-016X}
\urldef\tempurl%
\url{https://doi.org/10.1521/soco.2017.35.4.324}
\showDOI{\tempurl}


\bibitem[\protect\citeauthoryear{Jost and van~der Toorn}{Jost and van~der
  Toorn}{2012}]%
        {jost_system_2012}
\bibfield{author}{\bibinfo{person}{John~T. Jost} {and}
  \bibinfo{person}{Jojanneke van~der Toorn}.} \bibinfo{year}{2012}\natexlab{}.
\newblock \showarticletitle{System justification theory}.
\newblock In \bibinfo{booktitle}{\emph{Handbook of theories of social
  psychology, {Vol}. 2}}. \bibinfo{publisher}{Sage Publications Ltd},
  \bibinfo{address}{Thousand Oaks, CA}, \bibinfo{pages}{313--343}.
\newblock
\showISBNx{978-0-85702-961-4}
\urldef\tempurl%
\url{https://doi.org/10.4135/9781446249222.n42}
\showDOI{\tempurl}


\bibitem[\protect\citeauthoryear{Juul and Ugander}{Juul and Ugander}{2021}]%
        {juul_comparing_2021}
\bibfield{author}{\bibinfo{person}{Jonas~L. Juul} {and} \bibinfo{person}{Johan
  Ugander}.} \bibinfo{year}{2021}\natexlab{}.
\newblock \showarticletitle{Comparing information diffusion mechanisms by
  matching on cascade size}.
\newblock \bibinfo{journal}{\emph{Proceedings of the National Academy of
  Sciences}} \bibinfo{volume}{118}, \bibinfo{number}{46} (\bibinfo{date}{Nov.}
  \bibinfo{year}{2021}), \bibinfo{pages}{e2100786118}.
\newblock
\urldef\tempurl%
\url{https://doi.org/10.1073/pnas.2100786118}
\showDOI{\tempurl}


\bibitem[\protect\citeauthoryear{Kahan}{Kahan}{2012}]%
        {kahan_ideology_2012}
\bibfield{author}{\bibinfo{person}{Dan~M. Kahan}.}
  \bibinfo{year}{2012}\natexlab{}.
\newblock \bibinfo{booktitle}{\emph{Ideology, {Motivated} {Reasoning}, and
  {Cognitive} {Reflection}: {An} {Experimental} {Study}}}.
\newblock \bibinfo{type}{{SSRN} {Scholarly} {Paper}} ID 2182588.
  \bibinfo{institution}{Social Science Research Network},
  \bibinfo{address}{Rochester, NY}.
\newblock
\urldef\tempurl%
\url{https://doi.org/10.2139/ssrn.2182588}
\showDOI{\tempurl}


\bibitem[\protect\citeauthoryear{Kahan}{Kahan}{2017}]%
        {kahan_misconceptions_2017}
\bibfield{author}{\bibinfo{person}{Dan~M. Kahan}.}
  \bibinfo{year}{2017}\natexlab{}.
\newblock \bibinfo{booktitle}{\emph{Misconceptions, {Misinformation}, and the
  {Logic} of {Identity}-{Protective} {Cognition}}}.
\newblock \bibinfo{type}{{SSRN} {Scholarly} {Paper}} ID 2973067.
  \bibinfo{institution}{Social Science Research Network},
  \bibinfo{address}{Rochester, NY}.
\newblock
\urldef\tempurl%
\url{https://doi.org/10.2139/ssrn.2973067}
\showDOI{\tempurl}


\bibitem[\protect\citeauthoryear{Kahneman and Tversky}{Kahneman and
  Tversky}{1979}]%
        {kahneman_prospect_1979}
\bibfield{author}{\bibinfo{person}{Daniel Kahneman} {and} \bibinfo{person}{Amos
  Tversky}.} \bibinfo{year}{1979}\natexlab{}.
\newblock \showarticletitle{Prospect {Theory}: {An} {Analysis} of {Decision}
  under {Risk}}.
\newblock \bibinfo{journal}{\emph{Econometrica}} \bibinfo{volume}{47},
  \bibinfo{number}{2} (\bibinfo{year}{1979}), \bibinfo{pages}{263--291}.
\newblock
\showISSN{0012-9682}
\urldef\tempurl%
\url{https://doi.org/10.2307/1914185}
\showDOI{\tempurl}


\bibitem[\protect\citeauthoryear{Kogan and Wallach}{Kogan and Wallach}{1967}]%
        {kogan_risky-shift_1967}
\bibfield{author}{\bibinfo{person}{Nathan Kogan} {and}
  \bibinfo{person}{Michael~A Wallach}.} \bibinfo{year}{1967}\natexlab{}.
\newblock \showarticletitle{Risky-shift phenomenon in small decision-making
  groups: {A} test of the information-exchange hypothesis}.
\newblock \bibinfo{journal}{\emph{Journal of Experimental Social Psychology}}
  \bibinfo{volume}{3}, \bibinfo{number}{1} (\bibinfo{date}{Jan.}
  \bibinfo{year}{1967}), \bibinfo{pages}{75--84}.
\newblock
\showISSN{0022-1031}
\urldef\tempurl%
\url{https://doi.org/10.1016/0022-1031(67)90038-8}
\showDOI{\tempurl}


\bibitem[\protect\citeauthoryear{Kraft, Lodge, and Taber}{Kraft
  et~al\mbox{.}}{2015}]%
        {kraft_why_2015}
\bibfield{author}{\bibinfo{person}{Patrick~W. Kraft}, \bibinfo{person}{Milton
  Lodge}, {and} \bibinfo{person}{Charles~S. Taber}.}
  \bibinfo{year}{2015}\natexlab{}.
\newblock \showarticletitle{Why {People} “{Don}’t {Trust} the
  {Evidence}”: {Motivated} {Reasoning} and {Scientific} {Beliefs}}.
\newblock \bibinfo{journal}{\emph{The ANNALS of the American Academy of
  Political and Social Science}} \bibinfo{volume}{658}, \bibinfo{number}{1}
  (\bibinfo{date}{March} \bibinfo{year}{2015}), \bibinfo{pages}{121--133}.
\newblock
\showISSN{0002-7162, 1552-3349}
\urldef\tempurl%
\url{https://doi.org/10.1177/0002716214554758}
\showDOI{\tempurl}


\bibitem[\protect\citeauthoryear{Kruger and Dunning}{Kruger and
  Dunning}{1999}]%
        {kruger_unskilled_1999}
\bibfield{author}{\bibinfo{person}{Justin Kruger} {and} \bibinfo{person}{David
  Dunning}.} \bibinfo{year}{1999}\natexlab{}.
\newblock \showarticletitle{Unskilled and {Unaware} of {It}: {How}
  {Difficulties} in {Recognizing} {One}'s {Own} {Incompetence} {Lead} to
  {Inflated} {Self}-{Assessments}}.
\newblock \bibinfo{journal}{\emph{Journal of Personality and Social
  Psychology}} \bibinfo{volume}{77}, \bibinfo{number}{6}
  (\bibinfo{year}{1999}), \bibinfo{pages}{1121--1134}.
\newblock


\bibitem[\protect\citeauthoryear{Kteily, Hodson, and Bruneau}{Kteily
  et~al\mbox{.}}{2016}]%
        {kteily_they_2016}
\bibfield{author}{\bibinfo{person}{Nour Kteily}, \bibinfo{person}{Gordon
  Hodson}, {and} \bibinfo{person}{Emile Bruneau}.}
  \bibinfo{year}{2016}\natexlab{}.
\newblock \showarticletitle{They see us as less than human:
  {Metadehumanization} predicts intergroup conflict via reciprocal
  dehumanization.}
\newblock \bibinfo{journal}{\emph{Journal of Personality and Social
  Psychology}} \bibinfo{volume}{110}, \bibinfo{number}{3}
  (\bibinfo{year}{2016}), \bibinfo{pages}{343--370}.
\newblock
\showISSN{1939-1315, 0022-3514}
\urldef\tempurl%
\url{https://doi.org/10.1037/pspa0000044}
\showDOI{\tempurl}


\bibitem[\protect\citeauthoryear{Kteily and Bruneau}{Kteily and
  Bruneau}{2017}]%
        {kteily_darker_2017}
\bibfield{author}{\bibinfo{person}{Nour~S. Kteily} {and} \bibinfo{person}{Emile
  Bruneau}.} \bibinfo{year}{2017}\natexlab{}.
\newblock \showarticletitle{Darker {Demons} of {Our} {Nature}: {The} {Need} to
  ({Re}){Focus} {Attention} on {Blatant} {Forms} of {Dehumanization}}.
\newblock \bibinfo{journal}{\emph{Current Directions in Psychological Science}}
  \bibinfo{volume}{26}, \bibinfo{number}{6} (\bibinfo{date}{Dec.}
  \bibinfo{year}{2017}), \bibinfo{pages}{487--494}.
\newblock
\showISSN{0963-7214}
\urldef\tempurl%
\url{https://doi.org/10.1177/0963721417708230}
\showDOI{\tempurl}


\bibitem[\protect\citeauthoryear{Kunst and Obaidi}{Kunst and Obaidi}{2020}]%
        {kunst_understanding_2020}
\bibfield{author}{\bibinfo{person}{Jonas~R Kunst} {and} \bibinfo{person}{Milan
  Obaidi}.} \bibinfo{year}{2020}\natexlab{}.
\newblock \showarticletitle{Understanding violent extremism in the 21st
  century: the (re)emerging role of relative deprivation}.
\newblock \bibinfo{journal}{\emph{Current Opinion in Psychology}}
  \bibinfo{volume}{35} (\bibinfo{date}{Oct.} \bibinfo{year}{2020}),
  \bibinfo{pages}{55--59}.
\newblock
\showISSN{2352-250X}
\urldef\tempurl%
\url{https://doi.org/10.1016/j.copsyc.2020.03.010}
\showDOI{\tempurl}


\bibitem[\protect\citeauthoryear{Lees and Cikara}{Lees and Cikara}{2020}]%
        {lees_inaccurate_2020}
\bibfield{author}{\bibinfo{person}{Jeffrey Lees} {and} \bibinfo{person}{Mina
  Cikara}.} \bibinfo{year}{2020}\natexlab{}.
\newblock \showarticletitle{Inaccurate group meta-perceptions drive negative
  out-group attributions in competitive contexts}.
\newblock \bibinfo{journal}{\emph{Nature Human Behaviour}} \bibinfo{volume}{4},
  \bibinfo{number}{3} (\bibinfo{date}{March} \bibinfo{year}{2020}),
  \bibinfo{pages}{279--286}.
\newblock
\showISSN{2397-3374}
\urldef\tempurl%
\url{https://doi.org/10.1038/s41562-019-0766-4}
\showDOI{\tempurl}


\bibitem[\protect\citeauthoryear{Leman and Cinnirella}{Leman and
  Cinnirella}{2013}]%
        {leman_beliefs_2013}
\bibfield{author}{\bibinfo{person}{Patrick~John Leman} {and}
  \bibinfo{person}{Marco Cinnirella}.} \bibinfo{year}{2013}\natexlab{}.
\newblock \showarticletitle{Beliefs in conspiracy theories and the need for
  cognitive closure}.
\newblock \bibinfo{journal}{\emph{Frontiers in Psychology}}
  \bibinfo{volume}{4} (\bibinfo{year}{2013}).
\newblock
\showISSN{1664-1078}
\urldef\tempurl%
\url{https://doi.org/10.3389/fpsyg.2013.00378}
\showDOI{\tempurl}


\bibitem[\protect\citeauthoryear{Leviston, Walker, and Morwinski}{Leviston
  et~al\mbox{.}}{2013}]%
        {leviston_your_2013}
\bibfield{author}{\bibinfo{person}{Z. Leviston}, \bibinfo{person}{I. Walker},
  {and} \bibinfo{person}{S. Morwinski}.} \bibinfo{year}{2013}\natexlab{}.
\newblock \showarticletitle{Your opinion on climate change might not be as
  common as you think}.
\newblock \bibinfo{journal}{\emph{Nature Climate Change}} \bibinfo{volume}{3},
  \bibinfo{number}{4} (\bibinfo{date}{April} \bibinfo{year}{2013}),
  \bibinfo{pages}{334--337}.
\newblock
\showISSN{1758-678X, 1758-6798}
\urldef\tempurl%
\url{https://doi.org/10.1038/nclimate1743}
\showDOI{\tempurl}


\bibitem[\protect\citeauthoryear{Lewandowsky, Cook, and Lombardi}{Lewandowsky
  et~al\mbox{.}}{2020}]%
        {lewandowsky_debunking_2020}
\bibfield{author}{\bibinfo{person}{Stephan Lewandowsky}, \bibinfo{person}{John
  Cook}, {and} \bibinfo{person}{Doug Lombardi}.}
  \bibinfo{year}{2020}\natexlab{}.
\newblock \bibinfo{title}{Debunking {Handbook} 2020}.
\newblock
\newblock
\urldef\tempurl%
\url{https://doi.org/10.17910/B7.1182}
\showDOI{\tempurl}


\bibitem[\protect\citeauthoryear{Lewandowsky, Ecker, and Cook}{Lewandowsky
  et~al\mbox{.}}{2017}]%
        {lewandowsky_beyond_2017}
\bibfield{author}{\bibinfo{person}{Stephan Lewandowsky},
  \bibinfo{person}{Ullrich~K.H. Ecker}, {and} \bibinfo{person}{John Cook}.}
  \bibinfo{year}{2017}\natexlab{}.
\newblock \showarticletitle{Beyond {Misinformation}: {Understanding} and
  {Coping} with the “{Post}-{Truth}” {Era}}.
\newblock \bibinfo{journal}{\emph{Journal of Applied Research in Memory and
  Cognition}} \bibinfo{volume}{6}, \bibinfo{number}{4} (\bibinfo{date}{Dec.}
  \bibinfo{year}{2017}), \bibinfo{pages}{353--369}.
\newblock
\showISSN{22113681}
\urldef\tempurl%
\url{https://doi.org/10.1016/j.jarmac.2017.07.008}
\showDOI{\tempurl}


\bibitem[\protect\citeauthoryear{Lewandowsky and Oberauer}{Lewandowsky and
  Oberauer}{2016}]%
        {lewandowsky_motivated_2016}
\bibfield{author}{\bibinfo{person}{Stephan Lewandowsky} {and}
  \bibinfo{person}{Klaus Oberauer}.} \bibinfo{year}{2016}\natexlab{}.
\newblock \showarticletitle{Motivated {Rejection} of {Science}}.
\newblock \bibinfo{journal}{\emph{Current Directions in Psychological Science}}
  \bibinfo{volume}{25}, \bibinfo{number}{4} (\bibinfo{date}{Aug.}
  \bibinfo{year}{2016}), \bibinfo{pages}{217--222}.
\newblock
\showISSN{0963-7214}
\urldef\tempurl%
\url{https://doi.org/10.1177/0963721416654436}
\showDOI{\tempurl}


\bibitem[\protect\citeauthoryear{Lodge and Taber}{Lodge and Taber}{2005}]%
        {lodge_automaticity_2005}
\bibfield{author}{\bibinfo{person}{Milton Lodge} {and}
  \bibinfo{person}{Charles~S. Taber}.} \bibinfo{year}{2005}\natexlab{}.
\newblock \showarticletitle{The {Automaticity} of {Affect} for {Political}
  {Leaders}, {Groups}, and {Issues}: {An} {Experimental} {Test} of the {Hot}
  {Cognition} {Hypothesis}}.
\newblock \bibinfo{journal}{\emph{Political Psychology}} \bibinfo{volume}{26},
  \bibinfo{number}{3} (\bibinfo{year}{2005}), \bibinfo{pages}{455--482}.
\newblock
\showISSN{1467-9221}
\urldef\tempurl%
\url{https://doi.org/10.1111/j.1467-9221.2005.00426.x}
\showDOI{\tempurl}


\bibitem[\protect\citeauthoryear{Loftus}{Loftus}{2005}]%
        {loftus_planting_2005}
\bibfield{author}{\bibinfo{person}{Elizabeth~F. Loftus}.}
  \bibinfo{year}{2005}\natexlab{}.
\newblock \showarticletitle{Planting misinformation in the human mind: {A}
  30-year investigation of the malleability of memory}.
\newblock \bibinfo{journal}{\emph{Learning \& Memory}} \bibinfo{volume}{12},
  \bibinfo{number}{4} (\bibinfo{date}{July} \bibinfo{year}{2005}),
  \bibinfo{pages}{361--366}.
\newblock
\showISSN{1072-0502, 1549-5485}
\urldef\tempurl%
\url{https://doi.org/10.1101/lm.94705}
\showDOI{\tempurl}


\bibitem[\protect\citeauthoryear{Loftus and Palmer}{Loftus and Palmer}{1974}]%
        {loftus_reconstruction_1974}
\bibfield{author}{\bibinfo{person}{Elizabeth~F. Loftus} {and}
  \bibinfo{person}{John~C. Palmer}.} \bibinfo{year}{1974}\natexlab{}.
\newblock \showarticletitle{Reconstruction of automobile destruction: {An}
  example of the interaction between language and memory}.
\newblock \bibinfo{journal}{\emph{Journal of Verbal Learning and Verbal
  Behavior}} \bibinfo{volume}{13}, \bibinfo{number}{5} (\bibinfo{date}{Oct.}
  \bibinfo{year}{1974}), \bibinfo{pages}{585--589}.
\newblock
\showISSN{0022-5371}
\urldef\tempurl%
\url{https://doi.org/10.1016/S0022-5371(74)80011-3}
\showDOI{\tempurl}


\bibitem[\protect\citeauthoryear{Lord, Ross, and Lepper}{Lord
  et~al\mbox{.}}{1979}]%
        {lord_biased_1979}
\bibfield{author}{\bibinfo{person}{Charles~G. Lord}, \bibinfo{person}{Lee
  Ross}, {and} \bibinfo{person}{Mark~R. Lepper}.}
  \bibinfo{year}{1979}\natexlab{}.
\newblock \showarticletitle{Biased assimilation and attitude polarization:
  {The} effects of prior theories on subsequently considered evidence}.
\newblock \bibinfo{journal}{\emph{Journal of Personality and Social
  Psychology}} \bibinfo{volume}{37}, \bibinfo{number}{11}
  (\bibinfo{year}{1979}), \bibinfo{pages}{2098--2109}.
\newblock
\showISSN{1939-1315(Electronic),0022-3514(Print)}
\urldef\tempurl%
\url{https://doi.org/10.1037/0022-3514.37.11.2098}
\showDOI{\tempurl}


\bibitem[\protect\citeauthoryear{Luo, Hancock, and Markowitz}{Luo
  et~al\mbox{.}}{2022}]%
        {luo_credibility_2022}
\bibfield{author}{\bibinfo{person}{Mufan Luo}, \bibinfo{person}{Jeffrey~T.
  Hancock}, {and} \bibinfo{person}{David~M. Markowitz}.}
  \bibinfo{year}{2022}\natexlab{}.
\newblock \showarticletitle{Credibility {Perceptions} and {Detection}
  {Accuracy} of {Fake} {News} {Headlines} on {Social} {Media}: {Effects} of
  {Truth}-{Bias} and {Endorsement} {Cues}}.
\newblock \bibinfo{journal}{\emph{Communication Research}}
  \bibinfo{volume}{49}, \bibinfo{number}{2} (\bibinfo{date}{March}
  \bibinfo{year}{2022}), \bibinfo{pages}{171--195}.
\newblock
\showISSN{0093-6502}
\urldef\tempurl%
\url{https://doi.org/10.1177/0093650220921321}
\showDOI{\tempurl}


\bibitem[\protect\citeauthoryear{Macy, Ma, Tabin, Gao, and Szymanski}{Macy
  et~al\mbox{.}}{2021}]%
        {macy_polarization_2021}
\bibfield{author}{\bibinfo{person}{Michael~W. Macy}, \bibinfo{person}{Manqing
  Ma}, \bibinfo{person}{Daniel~R. Tabin}, \bibinfo{person}{Jianxi Gao}, {and}
  \bibinfo{person}{Boleslaw~K. Szymanski}.} \bibinfo{year}{2021}\natexlab{}.
\newblock \showarticletitle{Polarization and tipping points}.
\newblock \bibinfo{journal}{\emph{Proceedings of the National Academy of
  Sciences}} \bibinfo{volume}{118}, \bibinfo{number}{50} (\bibinfo{date}{Dec.}
  \bibinfo{year}{2021}), \bibinfo{pages}{e2102144118}.
\newblock
\urldef\tempurl%
\url{https://doi.org/10.1073/pnas.2102144118}
\showDOI{\tempurl}


\bibitem[\protect\citeauthoryear{Maertens, Roozenbeek, Basol, and van~der
  Linden}{Maertens et~al\mbox{.}}{2020}]%
        {maertens_long-term_2020}
\bibfield{author}{\bibinfo{person}{Rakoen Maertens}, \bibinfo{person}{Jon
  Roozenbeek}, \bibinfo{person}{Melisa Basol}, {and} \bibinfo{person}{Sander
  van~der Linden}.} \bibinfo{year}{2020}\natexlab{}.
\newblock \showarticletitle{Long-term effectiveness of inoculation against
  misinformation: {Three} longitudinal experiments}.
\newblock \bibinfo{journal}{\emph{Journal of Experimental Psychology: Applied}}
  (\bibinfo{year}{2020}), \bibinfo{pages}{No Pagination Specified--No
  Pagination Specified}.
\newblock
\showISSN{1939-2192(Electronic),1076-898X(Print)}
\urldef\tempurl%
\url{https://doi.org/10.1037/xap0000315}
\showDOI{\tempurl}


\bibitem[\protect\citeauthoryear{Marchal}{Marchal}{2021}]%
        {marchal_be_2021}
\bibfield{author}{\bibinfo{person}{Nahema Marchal}.}
  \bibinfo{year}{2021}\natexlab{}.
\newblock \showarticletitle{“{Be} {Nice} or {Leave} {Me} {Alone}”: {An}
  {Intergroup} {Perspective} on {Affective} {Polarization} in {Online}
  {Political} {Discussions}}.
\newblock \bibinfo{journal}{\emph{Communication Research}}
  (\bibinfo{date}{Sept.} \bibinfo{year}{2021}),
  \bibinfo{pages}{009365022110425}.
\newblock
\showISSN{0093-6502, 1552-3810}
\urldef\tempurl%
\url{https://doi.org/10.1177/00936502211042516}
\showDOI{\tempurl}


\bibitem[\protect\citeauthoryear{Margolin, Hannak, and Weber}{Margolin
  et~al\mbox{.}}{2018}]%
        {margolin_political_2018}
\bibfield{author}{\bibinfo{person}{Drew~B. Margolin}, \bibinfo{person}{Aniko
  Hannak}, {and} \bibinfo{person}{Ingmar Weber}.}
  \bibinfo{year}{2018}\natexlab{}.
\newblock \showarticletitle{Political {Fact}-{Checking} on {Twitter}: {When}
  {Do} {Corrections} {Have} an {Effect}?}
\newblock \bibinfo{journal}{\emph{Political Communication}}
  \bibinfo{volume}{35}, \bibinfo{number}{2} (\bibinfo{date}{April}
  \bibinfo{year}{2018}), \bibinfo{pages}{196--219}.
\newblock
\showISSN{1058-4609}
\urldef\tempurl%
\url{https://doi.org/10.1080/10584609.2017.1334018}
\showDOI{\tempurl}


\bibitem[\protect\citeauthoryear{Matute, Blanco, Yarritu, Díaz-Lago, Vadillo,
  and Barberia}{Matute et~al\mbox{.}}{2015}]%
        {matute_illusions_2015}
\bibfield{author}{\bibinfo{person}{Helena Matute}, \bibinfo{person}{Fernando
  Blanco}, \bibinfo{person}{Ion Yarritu}, \bibinfo{person}{Marcos Díaz-Lago},
  \bibinfo{person}{Miguel~A. Vadillo}, {and} \bibinfo{person}{Itxaso
  Barberia}.} \bibinfo{year}{2015}\natexlab{}.
\newblock \showarticletitle{Illusions of causality: how they bias our everyday
  thinking and how they could be reduced}.
\newblock \bibinfo{journal}{\emph{Frontiers in Psychology}}
  \bibinfo{volume}{6} (\bibinfo{date}{July} \bibinfo{year}{2015}).
\newblock
\showISSN{1664-1078}
\urldef\tempurl%
\url{https://doi.org/10.3389/fpsyg.2015.00888}
\showDOI{\tempurl}


\bibitem[\protect\citeauthoryear{McGuire}{McGuire}{1961a}]%
        {mcguire_effectiveness_1961}
\bibfield{author}{\bibinfo{person}{William~J. McGuire}.}
  \bibinfo{year}{1961}\natexlab{a}.
\newblock \showarticletitle{The {Effectiveness} of {Supportive} and
  {Refutational} {Defenses} in {Immunizing} and {Restoring} {Beliefs} {Against}
  {Persuasion}}.
\newblock \bibinfo{journal}{\emph{Sociometry}} \bibinfo{volume}{24},
  \bibinfo{number}{2} (\bibinfo{date}{June} \bibinfo{year}{1961}),
  \bibinfo{pages}{184}.
\newblock
\showISSN{00380431}
\urldef\tempurl%
\url{https://doi.org/10.2307/2786067}
\showDOI{\tempurl}


\bibitem[\protect\citeauthoryear{McGuire}{McGuire}{1961b}]%
        {mcguire_resistance_1961}
\bibfield{author}{\bibinfo{person}{W.~J. McGuire}.}
  \bibinfo{year}{1961}\natexlab{b}.
\newblock \showarticletitle{Resistance to persuasion conferred by active and
  passive prior refutation of the same and alternative counterarguments.}
\newblock \bibinfo{journal}{\emph{The Journal of Abnormal and Social
  Psychology}} \bibinfo{volume}{63}, \bibinfo{number}{2} (\bibinfo{date}{Sept.}
  \bibinfo{year}{1961}), \bibinfo{pages}{326--332}.
\newblock
\showISSN{0096-851X}
\urldef\tempurl%
\url{https://doi.org/10.1037/h0048344}
\showDOI{\tempurl}


\bibitem[\protect\citeauthoryear{Merton}{Merton}{1968}]%
        {merton_social_1968}
\bibfield{author}{\bibinfo{person}{Robert~King Merton}.}
  \bibinfo{year}{1968}\natexlab{}.
\newblock \bibinfo{booktitle}{\emph{Social {Theory} and {Social} {Structure}}}.
\newblock \bibinfo{publisher}{Simon and Schuster}.
\newblock
\showISBNx{978-0-02-921130-4}


\bibitem[\protect\citeauthoryear{Metaxa, Park, Landay, and Hancock}{Metaxa
  et~al\mbox{.}}{2019}]%
        {metaxa_search_2019}
\bibfield{author}{\bibinfo{person}{Danaë Metaxa}, \bibinfo{person}{Joon~Sung
  Park}, \bibinfo{person}{James~A. Landay}, {and} \bibinfo{person}{Jeff
  Hancock}.} \bibinfo{year}{2019}\natexlab{}.
\newblock \showarticletitle{Search {Media} and {Elections}: {A} {Longitudinal}
  {Investigation} of {Political} {Search} {Results}}.
\newblock \bibinfo{journal}{\emph{Proceedings of the ACM on Human-Computer
  Interaction}} \bibinfo{volume}{3}, \bibinfo{number}{CSCW}
  (\bibinfo{date}{Nov.} \bibinfo{year}{2019}), \bibinfo{pages}{129:1--129:17}.
\newblock
\urldef\tempurl%
\url{https://doi.org/10.1145/3359231}
\showDOI{\tempurl}


\bibitem[\protect\citeauthoryear{Micallef, Avram, Menczer, and Patil}{Micallef
  et~al\mbox{.}}{2021}]%
        {micallef_fakey_2021}
\bibfield{author}{\bibinfo{person}{Nicholas Micallef}, \bibinfo{person}{Mihai
  Avram}, \bibinfo{person}{Filippo Menczer}, {and} \bibinfo{person}{Sameer
  Patil}.} \bibinfo{year}{2021}\natexlab{}.
\newblock \showarticletitle{Fakey: {A} {Game} {Intervention} to {Improve}
  {News} {Literacy} on {Social} {Media}}.
\newblock \bibinfo{journal}{\emph{Proceedings of the ACM on Human-Computer
  Interaction}} \bibinfo{volume}{5}, \bibinfo{number}{CSCW1}
  (\bibinfo{date}{April} \bibinfo{year}{2021}), \bibinfo{pages}{6:1--6:27}.
\newblock
\urldef\tempurl%
\url{https://doi.org/10.1145/3449080}
\showDOI{\tempurl}


\bibitem[\protect\citeauthoryear{Mills}{Mills}{2018}]%
        {mills_pop-up_2018}
\bibfield{author}{\bibinfo{person}{Richard~A. Mills}.}
  \bibinfo{year}{2018}\natexlab{}.
\newblock \showarticletitle{Pop-up political advocacy communities on
  reddit.com: {SandersForPresident} and {The} {Donald}}.
\newblock \bibinfo{journal}{\emph{AI \& SOCIETY}} \bibinfo{volume}{33},
  \bibinfo{number}{1} (\bibinfo{date}{Feb.} \bibinfo{year}{2018}),
  \bibinfo{pages}{39--54}.
\newblock
\showISSN{1435-5655}
\urldef\tempurl%
\url{https://doi.org/10.1007/s00146-017-0712-9}
\showDOI{\tempurl}


\bibitem[\protect\citeauthoryear{Moore-Berg, Ankori-Karlinsky, Hameiri, and
  Bruneau}{Moore-Berg et~al\mbox{.}}{2020a}]%
        {moore-berg_exaggerated_2020}
\bibfield{author}{\bibinfo{person}{Samantha~L. Moore-Berg},
  \bibinfo{person}{Lee-Or Ankori-Karlinsky}, \bibinfo{person}{Boaz Hameiri},
  {and} \bibinfo{person}{Emile Bruneau}.} \bibinfo{year}{2020}\natexlab{a}.
\newblock \showarticletitle{Exaggerated meta-perceptions predict intergroup
  hostility between {American} political partisans}.
\newblock \bibinfo{journal}{\emph{Proceedings of the National Academy of
  Sciences}} \bibinfo{volume}{117}, \bibinfo{number}{26} (\bibinfo{date}{June}
  \bibinfo{year}{2020}), \bibinfo{pages}{14864--14872}.
\newblock
\showISSN{0027-8424, 1091-6490}
\urldef\tempurl%
\url{https://doi.org/10.1073/pnas.2001263117}
\showDOI{\tempurl}


\bibitem[\protect\citeauthoryear{Moore-Berg, Hameiri, and Bruneau}{Moore-Berg
  et~al\mbox{.}}{2020b}]%
        {moore-berg_prime_2020}
\bibfield{author}{\bibinfo{person}{Samantha~L Moore-Berg},
  \bibinfo{person}{Boaz Hameiri}, {and} \bibinfo{person}{Emile Bruneau}.}
  \bibinfo{year}{2020}\natexlab{b}.
\newblock \showarticletitle{The prime psychological suspects of toxic political
  polarization}.
\newblock \bibinfo{journal}{\emph{Current Opinion in Behavioral Sciences}}
  \bibinfo{volume}{34} (\bibinfo{date}{Aug.} \bibinfo{year}{2020}),
  \bibinfo{pages}{199--204}.
\newblock
\showISSN{23521546}
\urldef\tempurl%
\url{https://doi.org/10.1016/j.cobeha.2020.05.001}
\showDOI{\tempurl}


\bibitem[\protect\citeauthoryear{Morini, Pollacci, and Rossetti}{Morini
  et~al\mbox{.}}{2021}]%
        {morini_toward_2021}
\bibfield{author}{\bibinfo{person}{Virginia Morini}, \bibinfo{person}{Laura
  Pollacci}, {and} \bibinfo{person}{Giulio Rossetti}.}
  \bibinfo{year}{2021}\natexlab{}.
\newblock \showarticletitle{Toward a {Standard} {Approach} for {Echo} {Chamber}
  {Detection}: {Reddit} {Case} {Study}}.
\newblock \bibinfo{journal}{\emph{Applied Sciences}} \bibinfo{volume}{11},
  \bibinfo{number}{12} (\bibinfo{date}{Jan.} \bibinfo{year}{2021}),
  \bibinfo{pages}{5390}.
\newblock
\showISSN{2076-3417}
\urldef\tempurl%
\url{https://doi.org/10.3390/app11125390}
\showDOI{\tempurl}


\bibitem[\protect\citeauthoryear{Mosleh, Martel, Eckles, and Rand}{Mosleh
  et~al\mbox{.}}{2021a}]%
        {mosleh_perverse_2021}
\bibfield{author}{\bibinfo{person}{Mohsen Mosleh}, \bibinfo{person}{Cameron
  Martel}, \bibinfo{person}{Dean Eckles}, {and} \bibinfo{person}{David Rand}.}
  \bibinfo{year}{2021}\natexlab{a}.
\newblock \showarticletitle{Perverse {Downstream} {Consequences} of
  {Debunking}: {Being} {Corrected} by {Another} {User} for {Posting} {False}
  {Political} {News} {Increases} {Subsequent} {Sharing} of {Low} {Quality},
  {Partisan}, and {Toxic} {Content} in a {Twitter} {Field} {Experiment}}.
\newblock In \bibinfo{booktitle}{\emph{Proceedings of the 2021 {CHI}
  {Conference} on {Human} {Factors} in {Computing} {Systems}}}. Number 182.
  \bibinfo{publisher}{Association for Computing Machinery},
  \bibinfo{address}{New York, NY, USA}, \bibinfo{pages}{1--13}.
\newblock
\showISBNx{978-1-4503-8096-6}
\urldef\tempurl%
\url{https://doi.org/10.1145/3411764.3445642}
\showURL{%
\tempurl}


\bibitem[\protect\citeauthoryear{Mosleh, Martel, Eckles, and Rand}{Mosleh
  et~al\mbox{.}}{2021b}]%
        {mosleh_shared_2021}
\bibfield{author}{\bibinfo{person}{Mohsen Mosleh}, \bibinfo{person}{Cameron
  Martel}, \bibinfo{person}{Dean Eckles}, {and} \bibinfo{person}{David~G.
  Rand}.} \bibinfo{year}{2021}\natexlab{b}.
\newblock \showarticletitle{Shared partisanship dramatically increases social
  tie formation in a {Twitter} field experiment}.
\newblock \bibinfo{journal}{\emph{Proceedings of the National Academy of
  Sciences}} \bibinfo{volume}{118}, \bibinfo{number}{7} (\bibinfo{date}{Feb.}
  \bibinfo{year}{2021}).
\newblock
\showISSN{0027-8424, 1091-6490}
\urldef\tempurl%
\url{https://doi.org/10.1073/pnas.2022761118}
\showDOI{\tempurl}


\bibitem[\protect\citeauthoryear{Munger}{Munger}{2017}]%
        {munger_tweetment_2017}
\bibfield{author}{\bibinfo{person}{Kevin Munger}.}
  \bibinfo{year}{2017}\natexlab{}.
\newblock \showarticletitle{Tweetment {Effects} on the {Tweeted}:
  {Experimentally} {Reducing} {Racist} {Harassment}}.
\newblock \bibinfo{journal}{\emph{Political Behavior}} \bibinfo{volume}{39},
  \bibinfo{number}{3} (\bibinfo{date}{Sept.} \bibinfo{year}{2017}),
  \bibinfo{pages}{629--649}.
\newblock
\showISSN{1573-6687}
\urldef\tempurl%
\url{https://doi.org/10.1007/s11109-016-9373-5}
\showDOI{\tempurl}


\bibitem[\protect\citeauthoryear{Munro}{Munro}{2010}]%
        {munro_scientific_2010}
\bibfield{author}{\bibinfo{person}{Geoffrey~D. Munro}.}
  \bibinfo{year}{2010}\natexlab{}.
\newblock \showarticletitle{The {Scientific} {Impotence} {Excuse}:
  {Discounting} {Belief}-{Threatening} {Scientific} {Abstracts}}.
\newblock \bibinfo{journal}{\emph{Journal of Applied Social Psychology}}
  \bibinfo{volume}{40}, \bibinfo{number}{3} (\bibinfo{year}{2010}),
  \bibinfo{pages}{579--600}.
\newblock
\showISSN{1559-1816}
\urldef\tempurl%
\url{https://doi.org/10.1111/j.1559-1816.2010.00588.x}
\showDOI{\tempurl}


\bibitem[\protect\citeauthoryear{Nan and Daily}{Nan and Daily}{2015}]%
        {nan_biased_2015}
\bibfield{author}{\bibinfo{person}{Xiaoli Nan} {and} \bibinfo{person}{Kelly
  Daily}.} \bibinfo{year}{2015}\natexlab{}.
\newblock \showarticletitle{Biased {Assimilation} and {Need} for {Closure}:
  {Examining} the {Effects} of {Mixed} {Blogs} on {Vaccine}-{Related}
  {Beliefs}}.
\newblock \bibinfo{journal}{\emph{Journal of Health Communication}}
  \bibinfo{volume}{20}, \bibinfo{number}{4} (\bibinfo{date}{April}
  \bibinfo{year}{2015}), \bibinfo{pages}{462--471}.
\newblock
\showISSN{1081-0730}
\urldef\tempurl%
\url{https://doi.org/10.1080/10810730.2014.989343}
\showDOI{\tempurl}


\bibitem[\protect\citeauthoryear{Nauroth, Gollwitzer, Bender, and
  Rothmund}{Nauroth et~al\mbox{.}}{2015}]%
        {nauroth_social_2015}
\bibfield{author}{\bibinfo{person}{Peter Nauroth}, \bibinfo{person}{Mario
  Gollwitzer}, \bibinfo{person}{Jens Bender}, {and} \bibinfo{person}{Tobias
  Rothmund}.} \bibinfo{year}{2015}\natexlab{}.
\newblock \showarticletitle{Social {Identity} {Threat} {Motivates}
  {Science}-{Discrediting} {Online} {Comments}}.
\newblock \bibinfo{journal}{\emph{PLOS ONE}} \bibinfo{volume}{10},
  \bibinfo{number}{2} (\bibinfo{date}{Feb.} \bibinfo{year}{2015}),
  \bibinfo{pages}{e0117476}.
\newblock
\showISSN{1932-6203}
\urldef\tempurl%
\url{https://doi.org/10.1371/journal.pone.0117476}
\showDOI{\tempurl}


\bibitem[\protect\citeauthoryear{Neumann, Thielmann, and Pfattheicher}{Neumann
  et~al\mbox{.}}{2020}]%
        {neumann_labelling_2020}
\bibfield{author}{\bibinfo{person}{Henrike Neumann}, \bibinfo{person}{Isabel
  Thielmann}, {and} \bibinfo{person}{Stefan Pfattheicher}.}
  \bibinfo{year}{2020}\natexlab{}.
\newblock \showarticletitle{Labelling affects agreement with political
  statements of right-wing populist parties}.
\newblock \bibinfo{journal}{\emph{PLOS ONE}} \bibinfo{volume}{15},
  \bibinfo{number}{11} (\bibinfo{date}{Nov.} \bibinfo{year}{2020}),
  \bibinfo{pages}{e0239772}.
\newblock
\showISSN{1932-6203}
\urldef\tempurl%
\url{https://doi.org/10.1371/journal.pone.0239772}
\showDOI{\tempurl}


\bibitem[\protect\citeauthoryear{Nikolov, Flammini, and Menczer}{Nikolov
  et~al\mbox{.}}{2021}]%
        {nikolov_right_2021}
\bibfield{author}{\bibinfo{person}{Dimitar Nikolov},
  \bibinfo{person}{Alessandro Flammini}, {and} \bibinfo{person}{Filippo
  Menczer}.} \bibinfo{year}{2021}\natexlab{}.
\newblock \showarticletitle{Right and left, partisanship predicts (asymmetric)
  vulnerability to misinformation}.
\newblock \bibinfo{journal}{\emph{Harvard Kennedy School Misinformation
  Review}} (\bibinfo{date}{Feb.} \bibinfo{year}{2021}).
\newblock
\urldef\tempurl%
\url{https://doi.org/10.37016/mr-2020-55}
\showDOI{\tempurl}


\bibitem[\protect\citeauthoryear{Nyhan, Porter, Reifler, and Wood}{Nyhan
  et~al\mbox{.}}{2020}]%
        {nyhan_taking_2020}
\bibfield{author}{\bibinfo{person}{Brendan Nyhan}, \bibinfo{person}{Ethan
  Porter}, \bibinfo{person}{Jason Reifler}, {and} \bibinfo{person}{Thomas~J.
  Wood}.} \bibinfo{year}{2020}\natexlab{}.
\newblock \showarticletitle{Taking {Fact}-{Checks} {Literally} {But} {Not}
  {Seriously}? {The} {Effects} of {Journalistic} {Fact}-{Checking} on {Factual}
  {Beliefs} and {Candidate} {Favorability}}.
\newblock \bibinfo{journal}{\emph{Political Behavior}} \bibinfo{volume}{42},
  \bibinfo{number}{3} (\bibinfo{date}{Sept.} \bibinfo{year}{2020}),
  \bibinfo{pages}{939--960}.
\newblock
\showISSN{0190-9320, 1573-6687}
\urldef\tempurl%
\url{https://doi.org/10.1007/s11109-019-09528-x}
\showDOI{\tempurl}


\bibitem[\protect\citeauthoryear{Nyhan and Reifler}{Nyhan and Reifler}{2010}]%
        {nyhan_when_2010}
\bibfield{author}{\bibinfo{person}{Brendan Nyhan} {and} \bibinfo{person}{Jason
  Reifler}.} \bibinfo{year}{2010}\natexlab{}.
\newblock \showarticletitle{When {Corrections} {Fail}: {The} {Persistence} of
  {Political} {Misperceptions}}.
\newblock \bibinfo{journal}{\emph{Political Behavior}} \bibinfo{volume}{32},
  \bibinfo{number}{2} (\bibinfo{date}{June} \bibinfo{year}{2010}),
  \bibinfo{pages}{303--330}.
\newblock
\showISSN{1573-6687}
\urldef\tempurl%
\url{https://doi.org/10.1007/s11109-010-9112-2}
\showDOI{\tempurl}


\bibitem[\protect\citeauthoryear{Obaidi, Bergh, Akrami, and Anjum}{Obaidi
  et~al\mbox{.}}{2019}]%
        {obaidi_group-based_2019}
\bibfield{author}{\bibinfo{person}{Milan Obaidi}, \bibinfo{person}{Robin
  Bergh}, \bibinfo{person}{Nazar Akrami}, {and} \bibinfo{person}{Gulnaz
  Anjum}.} \bibinfo{year}{2019}\natexlab{}.
\newblock \showarticletitle{Group-{Based} {Relative} {Deprivation} {Explains}
  {Endorsement} of {Extremism} {Among} {Western}-{Born} {Muslims}}.
\newblock \bibinfo{journal}{\emph{Psychological Science}} \bibinfo{volume}{30},
  \bibinfo{number}{4} (\bibinfo{date}{April} \bibinfo{year}{2019}),
  \bibinfo{pages}{596--605}.
\newblock
\showISSN{0956-7976}
\urldef\tempurl%
\url{https://doi.org/10.1177/0956797619834879}
\showDOI{\tempurl}


\bibitem[\protect\citeauthoryear{Pantazi, Kissine, and Klein}{Pantazi
  et~al\mbox{.}}{2018}]%
        {pantazi_power_2018}
\bibfield{author}{\bibinfo{person}{Myrto Pantazi}, \bibinfo{person}{Mikhail
  Kissine}, {and} \bibinfo{person}{Olivier Klein}.}
  \bibinfo{year}{2018}\natexlab{}.
\newblock \showarticletitle{The {Power} of the {Truth} {Bias}: {False}
  {Information} {Affects} {Memory} and {Judgment} {Even} in the {Absence} of
  {Distraction}}.
\newblock \bibinfo{journal}{\emph{Social Cognition}} \bibinfo{volume}{36},
  \bibinfo{number}{2} (\bibinfo{date}{March} \bibinfo{year}{2018}),
  \bibinfo{pages}{167--198}.
\newblock
\showISSN{0278-016X}
\urldef\tempurl%
\url{https://doi.org/10.1521/soco.2018.36.2.167}
\showDOI{\tempurl}


\bibitem[\protect\citeauthoryear{Parekh, Margolin, and Ruths}{Parekh
  et~al\mbox{.}}{2020}]%
        {parekh_comparing_2020}
\bibfield{author}{\bibinfo{person}{Deven Parekh}, \bibinfo{person}{Drew
  Margolin}, {and} \bibinfo{person}{Derek Ruths}.}
  \bibinfo{year}{2020}\natexlab{}.
\newblock \showarticletitle{Comparing {Audience} {Appreciation} to
  {Fact}-{Checking} {Across} {Political} {Communities} on {Reddit}}. In
  \bibinfo{booktitle}{\emph{12th {ACM} {Conference} on {Web} {Science}}}.
  \bibinfo{publisher}{ACM}, \bibinfo{address}{Southampton United Kingdom},
  \bibinfo{pages}{144--154}.
\newblock
\showISBNx{978-1-4503-7989-2}
\urldef\tempurl%
\url{https://doi.org/10.1145/3394231.3397904}
\showDOI{\tempurl}


\bibitem[\protect\citeauthoryear{Pearson and Knobloch-Westerwick}{Pearson and
  Knobloch-Westerwick}{2019}]%
        {pearson_is_2019}
\bibfield{author}{\bibinfo{person}{George David~Hooke Pearson} {and}
  \bibinfo{person}{Silvia Knobloch-Westerwick}.}
  \bibinfo{year}{2019}\natexlab{}.
\newblock \showarticletitle{Is the {Confirmation} {Bias} {Bubble} {Larger}
  {Online}? {Pre}-{Election} {Confirmation} {Bias} in {Selective} {Exposure} to
  {Online} versus {Print} {Political} {Information}}.
\newblock \bibinfo{journal}{\emph{Mass Communication and Society}}
  \bibinfo{volume}{22}, \bibinfo{number}{4} (\bibinfo{date}{July}
  \bibinfo{year}{2019}), \bibinfo{pages}{466--486}.
\newblock
\showISSN{1520-5436}
\urldef\tempurl%
\url{https://doi.org/10.1080/15205436.2019.1599956}
\showDOI{\tempurl}


\bibitem[\protect\citeauthoryear{Pennycook, Bear, Collins, and Rand}{Pennycook
  et~al\mbox{.}}{2020a}]%
        {pennycook_implied_2020}
\bibfield{author}{\bibinfo{person}{Gordon Pennycook}, \bibinfo{person}{Adam
  Bear}, \bibinfo{person}{Evan~T. Collins}, {and} \bibinfo{person}{David~G.
  Rand}.} \bibinfo{year}{2020}\natexlab{a}.
\newblock \showarticletitle{The {Implied} {Truth} {Effect}: {Attaching}
  {Warnings} to a {Subset} of {Fake} {News} {Headlines} {Increases} {Perceived}
  {Accuracy} of {Headlines} {Without} {Warnings}}.
\newblock \bibinfo{journal}{\emph{Management Science}} \bibinfo{volume}{66},
  \bibinfo{number}{11} (\bibinfo{date}{Nov.} \bibinfo{year}{2020}),
  \bibinfo{pages}{4944--4957}.
\newblock
\showISSN{0025-1909, 1526-5501}
\urldef\tempurl%
\url{https://doi.org/10.1287/mnsc.2019.3478}
\showDOI{\tempurl}


\bibitem[\protect\citeauthoryear{Pennycook, Epstein, Mosleh, Arechar, Eckles,
  and Rand}{Pennycook et~al\mbox{.}}{2021}]%
        {pennycook_shifting_2021}
\bibfield{author}{\bibinfo{person}{Gordon Pennycook}, \bibinfo{person}{Ziv
  Epstein}, \bibinfo{person}{Mohsen Mosleh}, \bibinfo{person}{Antonio~A.
  Arechar}, \bibinfo{person}{Dean Eckles}, {and} \bibinfo{person}{David~G.
  Rand}.} \bibinfo{year}{2021}\natexlab{}.
\newblock \showarticletitle{Shifting attention to accuracy can reduce
  misinformation online}.
\newblock \bibinfo{journal}{\emph{Nature}} (\bibinfo{date}{March}
  \bibinfo{year}{2021}), \bibinfo{pages}{1--6}.
\newblock
\showISSN{1476-4687}
\urldef\tempurl%
\url{https://doi.org/10.1038/s41586-021-03344-2}
\showDOI{\tempurl}


\bibitem[\protect\citeauthoryear{Pennycook, McPhetres, Zhang, Lu, and
  Rand}{Pennycook et~al\mbox{.}}{2020b}]%
        {pennycook_fighting_2020}
\bibfield{author}{\bibinfo{person}{Gordon Pennycook}, \bibinfo{person}{Jonathon
  McPhetres}, \bibinfo{person}{Yunhao Zhang}, \bibinfo{person}{Jackson~G. Lu},
  {and} \bibinfo{person}{David~G. Rand}.} \bibinfo{year}{2020}\natexlab{b}.
\newblock \showarticletitle{Fighting {COVID}-19 {Misinformation} on {Social}
  {Media}: {Experimental} {Evidence} for a {Scalable} {Accuracy}-{Nudge}
  {Intervention}}.
\newblock \bibinfo{journal}{\emph{Psychological Science}} \bibinfo{volume}{31},
  \bibinfo{number}{7} (\bibinfo{date}{July} \bibinfo{year}{2020}),
  \bibinfo{pages}{770--780}.
\newblock
\showISSN{0956-7976}
\urldef\tempurl%
\url{https://doi.org/10.1177/0956797620939054}
\showDOI{\tempurl}


\bibitem[\protect\citeauthoryear{Pennycook and Rand}{Pennycook and
  Rand}{2019}]%
        {pennycook_lazy_2019}
\bibfield{author}{\bibinfo{person}{Gordon Pennycook} {and}
  \bibinfo{person}{David~G. Rand}.} \bibinfo{year}{2019}\natexlab{}.
\newblock \showarticletitle{Lazy, not biased: {Susceptibility} to partisan fake
  news is better explained by lack of reasoning than by motivated reasoning}.
\newblock \bibinfo{journal}{\emph{Cognition}}  \bibinfo{volume}{188}
  (\bibinfo{date}{July} \bibinfo{year}{2019}), \bibinfo{pages}{39--50}.
\newblock
\showISSN{00100277}
\urldef\tempurl%
\url{https://doi.org/10.1016/j.cognition.2018.06.011}
\showDOI{\tempurl}


\bibitem[\protect\citeauthoryear{Pennycook and Rand}{Pennycook and
  Rand}{2021}]%
        {pennycook_psychology_2021}
\bibfield{author}{\bibinfo{person}{Gordon Pennycook} {and}
  \bibinfo{person}{David~G. Rand}.} \bibinfo{year}{2021}\natexlab{}.
\newblock \showarticletitle{The {Psychology} of {Fake} {News}}.
\newblock \bibinfo{journal}{\emph{Trends in Cognitive Sciences}}
  (\bibinfo{date}{March} \bibinfo{year}{2021}),
  \bibinfo{pages}{S1364661321000516}.
\newblock
\showISSN{13646613}
\urldef\tempurl%
\url{https://doi.org/10.1016/j.tics.2021.02.007}
\showDOI{\tempurl}


\bibitem[\protect\citeauthoryear{Pica, Pierro, Bélanger, and Kruglanski}{Pica
  et~al\mbox{.}}{2014}]%
        {pica_role_2014}
\bibfield{author}{\bibinfo{person}{Gennaro Pica}, \bibinfo{person}{Antonio
  Pierro}, \bibinfo{person}{Jocelyn~J. Bélanger}, {and}
  \bibinfo{person}{Arie~W. Kruglanski}.} \bibinfo{year}{2014}\natexlab{}.
\newblock \showarticletitle{The {Role} of {Need} for {Cognitive} {Closure} in
  {Retrieval}-{Induced} {Forgetting} and {Misinformation} {Effects} in
  {Eyewitness} {Memory}}.
\newblock \bibinfo{journal}{\emph{Social Cognition}} \bibinfo{volume}{32},
  \bibinfo{number}{4} (\bibinfo{date}{July} \bibinfo{year}{2014}),
  \bibinfo{pages}{337--359}.
\newblock
\showISSN{0278-016X}
\urldef\tempurl%
\url{https://doi.org/10.1521/soco.2014.32.4.337}
\showDOI{\tempurl}


\bibitem[\protect\citeauthoryear{Pickles, Cvejic, Nickel, Copp, Bonner, Leask,
  Ayre, Batcup, Cornell, Dakin, Dodd, Isautier, and McCaffery}{Pickles
  et~al\mbox{.}}{2021}]%
        {pickles_covid-19_2021}
\bibfield{author}{\bibinfo{person}{Kristen Pickles}, \bibinfo{person}{Erin
  Cvejic}, \bibinfo{person}{Brooke Nickel}, \bibinfo{person}{Tessa Copp},
  \bibinfo{person}{Carissa Bonner}, \bibinfo{person}{Julie Leask},
  \bibinfo{person}{Julie Ayre}, \bibinfo{person}{Carys Batcup},
  \bibinfo{person}{Samuel Cornell}, \bibinfo{person}{Thomas Dakin},
  \bibinfo{person}{Rachael~H. Dodd}, \bibinfo{person}{Jennifer M.~J. Isautier},
  {and} \bibinfo{person}{Kirsten~J. McCaffery}.}
  \bibinfo{year}{2021}\natexlab{}.
\newblock \showarticletitle{{COVID}-19 {Misinformation} {Trends} in
  {Australia}: {Prospective} {Longitudinal} {National} {Survey}}.
\newblock \bibinfo{journal}{\emph{Journal of Medical Internet Research}}
  \bibinfo{volume}{23}, \bibinfo{number}{1} (\bibinfo{date}{Jan.}
  \bibinfo{year}{2021}), \bibinfo{pages}{e23805}.
\newblock
\urldef\tempurl%
\url{https://doi.org/10.2196/23805}
\showDOI{\tempurl}


\bibitem[\protect\citeauthoryear{Pomerantsev and Weiss}{Pomerantsev and
  Weiss}{2014}]%
        {pomerantsev_how_2014}
\bibfield{author}{\bibinfo{person}{Peter Pomerantsev} {and}
  \bibinfo{person}{Michael Weiss}.} \bibinfo{year}{2014}\natexlab{}.
\newblock \showarticletitle{How the {Kremlin} {Weaponizes} {Information},
  {Culture} and {Money}}.
\newblock \bibinfo{journal}{\emph{Institute of Modern Russia}}
  (\bibinfo{year}{2014}), \bibinfo{pages}{44}.
\newblock


\bibitem[\protect\citeauthoryear{Rajadesingan, Budak, and Resnick}{Rajadesingan
  et~al\mbox{.}}{2021}]%
        {rajadesingan_political_2021}
\bibfield{author}{\bibinfo{person}{Ashwin Rajadesingan}, \bibinfo{person}{Ceren
  Budak}, {and} \bibinfo{person}{Paul Resnick}.}
  \bibinfo{year}{2021}\natexlab{}.
\newblock \showarticletitle{Political {Discussion} is {Abundant} in
  {Non}-political {Subreddits} (and {Less} {Toxic})}.
\newblock \bibinfo{journal}{\emph{Proceedings of the International AAAI
  Conference on Web and Social Media}}  \bibinfo{volume}{15}
  (\bibinfo{date}{May} \bibinfo{year}{2021}), \bibinfo{pages}{525--536}.
\newblock
\showISSN{2334-0770}
\urldef\tempurl%
\url{https://ojs.aaai.org/index.php/ICWSM/article/view/18081}
\showURL{%
\tempurl}


\bibitem[\protect\citeauthoryear{Redlawsk, Civettini, and Emmerson}{Redlawsk
  et~al\mbox{.}}{2010}]%
        {redlawsk_affective_2010}
\bibfield{author}{\bibinfo{person}{David~P. Redlawsk}, \bibinfo{person}{Andrew
  J.~W. Civettini}, {and} \bibinfo{person}{Karen~M. Emmerson}.}
  \bibinfo{year}{2010}\natexlab{}.
\newblock \showarticletitle{The {Affective} {Tipping} {Point}: {Do} {Motivated}
  {Reasoners} {Ever} “{Get} {It}”?: {The} {Affective} {Tipping} {Point}}.
\newblock \bibinfo{journal}{\emph{Political Psychology}} \bibinfo{volume}{31},
  \bibinfo{number}{4} (\bibinfo{date}{July} \bibinfo{year}{2010}),
  \bibinfo{pages}{563--593}.
\newblock
\showISSN{0162895X, 14679221}
\urldef\tempurl%
\url{https://doi.org/10.1111/j.1467-9221.2010.00772.x}
\showDOI{\tempurl}


\bibitem[\protect\citeauthoryear{Reicher, Haslam, and Rath}{Reicher
  et~al\mbox{.}}{2008}]%
        {reicher_making_2008}
\bibfield{author}{\bibinfo{person}{Stephen Reicher},
  \bibinfo{person}{S.~Alexander Haslam}, {and} \bibinfo{person}{Rakshi Rath}.}
  \bibinfo{year}{2008}\natexlab{}.
\newblock \showarticletitle{Making a {Virtue} of {Evil}: {A} {Five}-{Step}
  {Social} {Identity} {Model} of the {Development} of {Collective} {Hate}}.
\newblock \bibinfo{journal}{\emph{Social and Personality Psychology Compass}}
  \bibinfo{volume}{2}, \bibinfo{number}{3} (\bibinfo{year}{2008}),
  \bibinfo{pages}{1313--1344}.
\newblock
\showISSN{1751-9004}
\urldef\tempurl%
\url{https://doi.org/10.1111/j.1751-9004.2008.00113.x}
\showDOI{\tempurl}


\bibitem[\protect\citeauthoryear{Ribeiro, Ottoni, West, Almeida, and
  Meira}{Ribeiro et~al\mbox{.}}{2020}]%
        {ribeiro_auditing_2020}
\bibfield{author}{\bibinfo{person}{Manoel~Horta Ribeiro},
  \bibinfo{person}{Raphael Ottoni}, \bibinfo{person}{Robert West},
  \bibinfo{person}{Virgílio A.~F. Almeida}, {and} \bibinfo{person}{Wagner
  Meira}.} \bibinfo{year}{2020}\natexlab{}.
\newblock \showarticletitle{Auditing radicalization pathways on {YouTube}}. In
  \bibinfo{booktitle}{\emph{Proceedings of the 2020 {Conference} on {Fairness},
  {Accountability}, and {Transparency}}}. \bibinfo{publisher}{ACM},
  \bibinfo{address}{Barcelona Spain}, \bibinfo{pages}{131--141}.
\newblock
\showISBNx{978-1-4503-6936-7}
\urldef\tempurl%
\url{https://doi.org/10.1145/3351095.3372879}
\showDOI{\tempurl}


\bibitem[\protect\citeauthoryear{Rogers and Bhowmik}{Rogers and
  Bhowmik}{1970}]%
        {rogers_homophily-heterophily_1970}
\bibfield{author}{\bibinfo{person}{Everett~M. Rogers} {and}
  \bibinfo{person}{Dilip~K. Bhowmik}.} \bibinfo{year}{1970}\natexlab{}.
\newblock \showarticletitle{{HOMOPHILY}-{HETEROPHILY}: {RELATIONAL} {CONCEPTS}
  {FOR} {COMMUNICATION} {RESEARCH}}.
\newblock \bibinfo{journal}{\emph{Public Opinion Quarterly}}
  \bibinfo{volume}{34}, \bibinfo{number}{4} (\bibinfo{date}{Jan.}
  \bibinfo{year}{1970}), \bibinfo{pages}{523--538}.
\newblock
\showISSN{0033-362X}
\urldef\tempurl%
\url{https://doi.org/10.1086/267838}
\showDOI{\tempurl}


\bibitem[\protect\citeauthoryear{Roozenbeek, Schneider, Dryhurst, Kerr,
  Freeman, Recchia, van~der Bles, and van~der Linden}{Roozenbeek
  et~al\mbox{.}}{2020}]%
        {roozenbeek_susceptibility_2020}
\bibfield{author}{\bibinfo{person}{Jon Roozenbeek}, \bibinfo{person}{Claudia~R.
  Schneider}, \bibinfo{person}{Sarah Dryhurst}, \bibinfo{person}{John Kerr},
  \bibinfo{person}{Alexandra L.~J. Freeman}, \bibinfo{person}{Gabriel Recchia},
  \bibinfo{person}{Anne~Marthe van~der Bles}, {and} \bibinfo{person}{Sander
  van~der Linden}.} \bibinfo{year}{2020}\natexlab{}.
\newblock \showarticletitle{Susceptibility to misinformation about {COVID}-19
  around the world}.
\newblock \bibinfo{journal}{\emph{Royal Society Open Science}}
  \bibinfo{volume}{7}, \bibinfo{number}{10} (\bibinfo{year}{2020}),
  \bibinfo{pages}{201199}.
\newblock
\urldef\tempurl%
\url{https://doi.org/10.1098/rsos.201199}
\showDOI{\tempurl}


\bibitem[\protect\citeauthoryear{Ross, Greene, and House}{Ross
  et~al\mbox{.}}{1977}]%
        {ross_false_1977}
\bibfield{author}{\bibinfo{person}{Lee Ross}, \bibinfo{person}{David Greene},
  {and} \bibinfo{person}{Pamela House}.} \bibinfo{year}{1977}\natexlab{}.
\newblock \showarticletitle{The “false consensus effect”: {An} egocentric
  bias in social perception and attribution processes}.
\newblock \bibinfo{journal}{\emph{Journal of Experimental Social Psychology}}
  \bibinfo{volume}{13}, \bibinfo{number}{3} (\bibinfo{date}{May}
  \bibinfo{year}{1977}), \bibinfo{pages}{279--301}.
\newblock
\showISSN{00221031}
\urldef\tempurl%
\url{https://doi.org/10.1016/0022-1031(77)90049-X}
\showDOI{\tempurl}


\bibitem[\protect\citeauthoryear{Rothschild, Keefer, and Hauri}{Rothschild
  et~al\mbox{.}}{2020}]%
        {rothschild_defensive_2020}
\bibfield{author}{\bibinfo{person}{Zachary~K. Rothschild},
  \bibinfo{person}{Lucas~A. Keefer}, {and} \bibinfo{person}{Julianna Hauri}.}
  \bibinfo{year}{2020}\natexlab{}.
\newblock \showarticletitle{Defensive {Partisanship}? {Evidence} that
  {In}‐{Party} {Scandals} {Increase} {Out}‐{Party} {Hostility}}.
\newblock \bibinfo{journal}{\emph{Political Psychology}} (\bibinfo{date}{June}
  \bibinfo{year}{2020}), \bibinfo{pages}{pops.12680}.
\newblock
\showISSN{0162-895X, 1467-9221}
\urldef\tempurl%
\url{https://doi.org/10.1111/pops.12680}
\showDOI{\tempurl}


\bibitem[\protect\citeauthoryear{Rozenblit and Keil}{Rozenblit and
  Keil}{2002}]%
        {rozenblit_misunderstood_2002}
\bibfield{author}{\bibinfo{person}{Leonid Rozenblit} {and}
  \bibinfo{person}{Frank Keil}.} \bibinfo{year}{2002}\natexlab{}.
\newblock \showarticletitle{The misunderstood limits of folk science: an
  illusion of explanatory depth}.
\newblock \bibinfo{journal}{\emph{Cognitive Science}} \bibinfo{volume}{26},
  \bibinfo{number}{5} (\bibinfo{year}{2002}), \bibinfo{pages}{521--562}.
\newblock
\showISSN{1551-6709}
\urldef\tempurl%
\url{https://doi.org/10.1207/s15516709cog2605_1}
\showDOI{\tempurl}


\bibitem[\protect\citeauthoryear{Scherer, McPhetres, Pennycook, Kempe, Allen,
  Knoepke, Tate, and Matlock}{Scherer et~al\mbox{.}}{2021}]%
        {scherer_who_2021}
\bibfield{author}{\bibinfo{person}{Laura~D. Scherer}, \bibinfo{person}{Jon
  McPhetres}, \bibinfo{person}{Gordon Pennycook}, \bibinfo{person}{Allison
  Kempe}, \bibinfo{person}{Larry~A. Allen}, \bibinfo{person}{Christopher~E.
  Knoepke}, \bibinfo{person}{Channing~E. Tate}, {and}
  \bibinfo{person}{Daniel~D. Matlock}.} \bibinfo{year}{2021}\natexlab{}.
\newblock \showarticletitle{Who is susceptible to online health misinformation?
  {A} test of four psychosocial hypotheses}.
\newblock \bibinfo{journal}{\emph{Health Psychology}} (\bibinfo{year}{2021}),
  \bibinfo{pages}{No Pagination Specified--No Pagination Specified}.
\newblock
\showISSN{1930-7810(Electronic),0278-6133(Print)}
\urldef\tempurl%
\url{https://doi.org/10.1037/hea0000978}
\showDOI{\tempurl}


\bibitem[\protect\citeauthoryear{Sherif and Sherif}{Sherif and Sherif}{1953}]%
        {sherif_groups_1953}
\bibfield{author}{\bibinfo{person}{Muzafer Sherif} {and}
  \bibinfo{person}{Carolyn~W. Sherif}.} \bibinfo{year}{1953}\natexlab{}.
\newblock \bibinfo{booktitle}{\emph{Groups in harmony and tension; an
  integration of studies of intergroup relations}}.
\newblock \bibinfo{publisher}{Harper \& Brothers}, \bibinfo{address}{Oxford,
  England}.
\newblock


\bibitem[\protect\citeauthoryear{Shin and Thorson}{Shin and Thorson}{2017}]%
        {shin_partisan_2017}
\bibfield{author}{\bibinfo{person}{Jieun Shin} {and} \bibinfo{person}{Kjerstin
  Thorson}.} \bibinfo{year}{2017}\natexlab{}.
\newblock \showarticletitle{Partisan {Selective} {Sharing}: {The} {Biased}
  {Diffusion} of {Fact}-{Checking} {Messages} on {Social} {Media}}.
\newblock \bibinfo{journal}{\emph{Journal of Communication}}
  \bibinfo{volume}{67}, \bibinfo{number}{2} (\bibinfo{date}{April}
  \bibinfo{year}{2017}), \bibinfo{pages}{233--255}.
\newblock
\showISSN{0021-9916}
\urldef\tempurl%
\url{https://doi.org/10.1111/jcom.12284}
\showDOI{\tempurl}


\bibitem[\protect\citeauthoryear{Silvia}{Silvia}{2005}]%
        {silvia_deflecting_2005}
\bibfield{author}{\bibinfo{person}{Paul~J. Silvia}.}
  \bibinfo{year}{2005}\natexlab{}.
\newblock \showarticletitle{Deflecting {Reactance}: {The} {Role} of
  {Similarity} in {Increasing} {Compliance} and {Reducing} {Resistance}}.
\newblock \bibinfo{journal}{\emph{Basic and Applied Social Psychology}}
  \bibinfo{volume}{27}, \bibinfo{number}{3} (\bibinfo{date}{Sept.}
  \bibinfo{year}{2005}), \bibinfo{pages}{277--284}.
\newblock
\showISSN{0197-3533}
\urldef\tempurl%
\url{https://doi.org/10.1207/s15324834basp2703_9}
\showDOI{\tempurl}


\bibitem[\protect\citeauthoryear{Slater}{Slater}{2007}]%
        {slater_reinforcing_2007}
\bibfield{author}{\bibinfo{person}{Michael~D. Slater}.}
  \bibinfo{year}{2007}\natexlab{}.
\newblock \showarticletitle{Reinforcing {Spirals}: {The} {Mutual} {Influence}
  of {Media} {Selectivity} and {Media} {Effects} and {Their} {Impact} on
  {Individual} {Behavior} and {Social} {Identity}}.
\newblock \bibinfo{journal}{\emph{Communication Theory}} \bibinfo{volume}{17},
  \bibinfo{number}{3} (\bibinfo{date}{Aug.} \bibinfo{year}{2007}),
  \bibinfo{pages}{281--303}.
\newblock
\showISSN{1050-3293}
\urldef\tempurl%
\url{https://doi.org/10.1111/j.1468-2885.2007.00296.x}
\showDOI{\tempurl}


\bibitem[\protect\citeauthoryear{Smith, Wakeford, Cribbin, Barnett, and
  Hou}{Smith et~al\mbox{.}}{2020}]%
        {smith_detecting_2020}
\bibfield{author}{\bibinfo{person}{Laura~G.E. Smith}, \bibinfo{person}{Laura
  Wakeford}, \bibinfo{person}{Timothy~F. Cribbin}, \bibinfo{person}{Julie
  Barnett}, {and} \bibinfo{person}{Wai~Kai Hou}.}
  \bibinfo{year}{2020}\natexlab{}.
\newblock \showarticletitle{Detecting psychological change through mobilizing
  interactions and changes in extremist linguistic style}.
\newblock \bibinfo{journal}{\emph{Computers in Human Behavior}}
  \bibinfo{volume}{108} (\bibinfo{date}{July} \bibinfo{year}{2020}),
  \bibinfo{pages}{106298}.
\newblock
\showISSN{07475632}
\urldef\tempurl%
\url{https://doi.org/10.1016/j.chb.2020.106298}
\showDOI{\tempurl}


\bibitem[\protect\citeauthoryear{Starbird, Arif, and Wilson}{Starbird
  et~al\mbox{.}}{2019}]%
        {starbird_disinformation_2019}
\bibfield{author}{\bibinfo{person}{Kate Starbird}, \bibinfo{person}{Ahmer
  Arif}, {and} \bibinfo{person}{Tom Wilson}.} \bibinfo{year}{2019}\natexlab{}.
\newblock \showarticletitle{Disinformation as {Collaborative} {Work}:
  {Surfacing} the {Participatory} {Nature} of {Strategic} {Information}
  {Operations}}.
\newblock \bibinfo{journal}{\emph{Proceedings of the ACM on Human-Computer
  Interaction}} \bibinfo{volume}{3}, \bibinfo{number}{CSCW}
  (\bibinfo{date}{Nov.} \bibinfo{year}{2019}), \bibinfo{pages}{1--26}.
\newblock
\showISSN{2573-0142}
\urldef\tempurl%
\url{https://doi.org/10.1145/3359229}
\showDOI{\tempurl}


\bibitem[\protect\citeauthoryear{Stephan and Stephan}{Stephan and
  Stephan}{2017}]%
        {stephan_intergroup_2017}
\bibfield{author}{\bibinfo{person}{Walter~G. Stephan} {and}
  \bibinfo{person}{Cookie~White Stephan}.} \bibinfo{year}{2017}\natexlab{}.
\newblock \showarticletitle{Intergroup {Threat} {Theory}}.
\newblock In \bibinfo{booktitle}{\emph{The {International} {Encyclopedia} of
  {Intercultural} {Communication}}}. \bibinfo{publisher}{American Cancer
  Society}, \bibinfo{pages}{1--12}.
\newblock
\showISBNx{978-1-118-78366-5}
\urldef\tempurl%
\url{https://doi.org/10.1002/9781118783665.ieicc0162}
\showDOI{\tempurl}


\bibitem[\protect\citeauthoryear{Stewart, Arif, Nied, Spiro, and
  Starbird}{Stewart et~al\mbox{.}}{2017}]%
        {stewart_drawing_2017}
\bibfield{author}{\bibinfo{person}{Leo~Graiden Stewart}, \bibinfo{person}{Ahmer
  Arif}, \bibinfo{person}{A.~Conrad Nied}, \bibinfo{person}{Emma~S. Spiro},
  {and} \bibinfo{person}{Kate Starbird}.} \bibinfo{year}{2017}\natexlab{}.
\newblock \showarticletitle{Drawing the {Lines} of {Contention}: {Networked}
  {Frame} {Contests} {Within} \#{BlackLivesMatter} {Discourse}}.
\newblock \bibinfo{journal}{\emph{Proceedings of the ACM on Human-Computer
  Interaction}} \bibinfo{volume}{1}, \bibinfo{number}{CSCW}
  (\bibinfo{date}{Dec.} \bibinfo{year}{2017}), \bibinfo{pages}{96:1--96:23}.
\newblock
\urldef\tempurl%
\url{https://doi.org/10.1145/3134920}
\showDOI{\tempurl}


\bibitem[\protect\citeauthoryear{Stojanov and Halberstadt}{Stojanov and
  Halberstadt}{2019}]%
        {stojanov_conspiracy_2019}
\bibfield{author}{\bibinfo{person}{Ana Stojanov} {and} \bibinfo{person}{Jamin
  Halberstadt}.} \bibinfo{year}{2019}\natexlab{}.
\newblock \showarticletitle{The {Conspiracy} {Mentality} {Scale}}.
\newblock \bibinfo{journal}{\emph{Social Psychology}} \bibinfo{volume}{50},
  \bibinfo{number}{4} (\bibinfo{date}{July} \bibinfo{year}{2019}),
  \bibinfo{pages}{215--232}.
\newblock
\showISSN{1864-9335}
\urldef\tempurl%
\url{https://doi.org/10.1027/1864-9335/a000381}
\showDOI{\tempurl}


\bibitem[\protect\citeauthoryear{Suhay, Bello-Pardo, and Maurer}{Suhay
  et~al\mbox{.}}{2018}]%
        {suhay_polarizing_2018}
\bibfield{author}{\bibinfo{person}{Elizabeth Suhay}, \bibinfo{person}{Emily
  Bello-Pardo}, {and} \bibinfo{person}{Brianna Maurer}.}
  \bibinfo{year}{2018}\natexlab{}.
\newblock \showarticletitle{The {Polarizing} {Effects} of {Online} {Partisan}
  {Criticism}: {Evidence} from {Two} {Experiments}}.
\newblock \bibinfo{journal}{\emph{The International Journal of Press/Politics}}
  \bibinfo{volume}{23}, \bibinfo{number}{1} (\bibinfo{year}{2018}),
  \bibinfo{pages}{95--115}.
\newblock


\bibitem[\protect\citeauthoryear{Sultana and Fussell}{Sultana and
  Fussell}{2021}]%
        {sultana_dissemination_2021}
\bibfield{author}{\bibinfo{person}{Sharifa Sultana} {and}
  \bibinfo{person}{Susan~R. Fussell}.} \bibinfo{year}{2021}\natexlab{}.
\newblock \showarticletitle{Dissemination, {Situated} {Fact}-checking, and
  {Social} {Effects} of {Misinformation} among {Rural} {Bangladeshi}
  {Villagers} {During} the {COVID}-19 {Pandemic}}.
\newblock \bibinfo{journal}{\emph{Proceedings of the ACM on Human-Computer
  Interaction}} \bibinfo{volume}{5}, \bibinfo{number}{CSCW2}
  (\bibinfo{date}{Oct.} \bibinfo{year}{2021}), \bibinfo{pages}{436:1--436:34}.
\newblock
\urldef\tempurl%
\url{https://doi.org/10.1145/3479580}
\showDOI{\tempurl}


\bibitem[\protect\citeauthoryear{Swire, Berinsky, Lewandowsky, and Ecker}{Swire
  et~al\mbox{.}}{2017}]%
        {swire_processing_2017}
\bibfield{author}{\bibinfo{person}{Briony Swire}, \bibinfo{person}{Adam~J.
  Berinsky}, \bibinfo{person}{Stephan Lewandowsky}, {and}
  \bibinfo{person}{Ullrich K.~H. Ecker}.} \bibinfo{year}{2017}\natexlab{}.
\newblock \showarticletitle{Processing political misinformation: comprehending
  the {Trump} phenomenon}.
\newblock \bibinfo{journal}{\emph{Royal Society Open Science}}
  \bibinfo{volume}{4}, \bibinfo{number}{3} (\bibinfo{year}{2017}),
  \bibinfo{pages}{160802}.
\newblock
\urldef\tempurl%
\url{https://doi.org/10.1098/rsos.160802}
\showDOI{\tempurl}


\bibitem[\protect\citeauthoryear{Swire-Thompson, DeGutis, and
  Lazer}{Swire-Thompson et~al\mbox{.}}{2020}]%
        {swire-thompson_searching_2020}
\bibfield{author}{\bibinfo{person}{Briony Swire-Thompson},
  \bibinfo{person}{Joseph DeGutis}, {and} \bibinfo{person}{David Lazer}.}
  \bibinfo{year}{2020}\natexlab{}.
\newblock \showarticletitle{Searching for the {Backfire} {Effect}:
  {Measurement} and {Design} {Considerations}}.
\newblock \bibinfo{journal}{\emph{Journal of Applied Research in Memory and
  Cognition}} \bibinfo{volume}{9}, \bibinfo{number}{3} (\bibinfo{date}{Sept.}
  \bibinfo{year}{2020}), \bibinfo{pages}{286--299}.
\newblock
\showISSN{2211-3681}
\urldef\tempurl%
\url{https://doi.org/10.1016/j.jarmac.2020.06.006}
\showDOI{\tempurl}


\bibitem[\protect\citeauthoryear{Taber and Lodge}{Taber and Lodge}{2006}]%
        {taber_motivated_2006}
\bibfield{author}{\bibinfo{person}{Charles~S. Taber} {and}
  \bibinfo{person}{Milton Lodge}.} \bibinfo{year}{2006}\natexlab{}.
\newblock \showarticletitle{Motivated {Skepticism} in the {Evaluation} of
  {Political} {Beliefs}}.
\newblock \bibinfo{journal}{\emph{American Journal of Political Science}}
  \bibinfo{volume}{50}, \bibinfo{number}{3} (\bibinfo{year}{2006}),
  \bibinfo{pages}{755--769}.
\newblock
\showISSN{1540-5907}
\urldef\tempurl%
\url{https://doi.org/10.1111/j.1540-5907.2006.00214.x}
\showDOI{\tempurl}


\bibitem[\protect\citeauthoryear{Tajfel, Billig, Bundy, and Flament}{Tajfel
  et~al\mbox{.}}{1971}]%
        {tajfel_social_1971}
\bibfield{author}{\bibinfo{person}{Henri Tajfel}, \bibinfo{person}{M.~G.
  Billig}, \bibinfo{person}{R.~P. Bundy}, {and} \bibinfo{person}{Claude
  Flament}.} \bibinfo{year}{1971}\natexlab{}.
\newblock \showarticletitle{Social categorization and intergroup behaviour}.
\newblock \bibinfo{journal}{\emph{European Journal of Social Psychology}}
  \bibinfo{volume}{1}, \bibinfo{number}{2} (\bibinfo{year}{1971}),
  \bibinfo{pages}{149--178}.
\newblock
\showISSN{1099-0992}
\urldef\tempurl%
\url{https://doi.org/10.1002/ejsp.2420010202}
\showDOI{\tempurl}


\bibitem[\protect\citeauthoryear{Tajfel and Turner}{Tajfel and Turner}{2004}]%
        {jost_social_2004}
\bibfield{author}{\bibinfo{person}{Henri Tajfel} {and} \bibinfo{person}{John~C.
  Turner}.} \bibinfo{year}{2004}\natexlab{}.
\newblock \showarticletitle{The {Social} {Identity} {Theory} of {Intergroup}
  {Behavior}}.
\newblock In \bibinfo{booktitle}{\emph{Political {Psychology}}
  (\bibinfo{edition}{0} ed.)}, \bibfield{editor}{\bibinfo{person}{John~T. Jost}
  {and} \bibinfo{person}{Jim Sidanius}} (Eds.). \bibinfo{publisher}{Psychology
  Press}, \bibinfo{pages}{276--293}.
\newblock
\showISBNx{978-0-203-50598-4}
\urldef\tempurl%
\url{https://doi.org/10.4324/9780203505984-16}
\showDOI{\tempurl}


\bibitem[\protect\citeauthoryear{Terren and Borge-Bravo}{Terren and
  Borge-Bravo}{2021}]%
        {terren_echo_2021}
\bibfield{author}{\bibinfo{person}{Ludovic Terren} {and} \bibinfo{person}{Rosa
  Borge-Bravo}.} \bibinfo{year}{2021}\natexlab{}.
\newblock \showarticletitle{Echo {Chambers} on {Social} {Media}: {A}
  {Systematic} {Review} of the {Literature}}.
\newblock \bibinfo{journal}{\emph{Review of Communication Research}}
  \bibinfo{volume}{9} (\bibinfo{date}{March} \bibinfo{year}{2021}),
  \bibinfo{pages}{99--118}.
\newblock
\showISSN{2255-4165}
\urldef\tempurl%
\url{https://rcommunicationr.org/index.php/rcr/article/view/94}
\showURL{%
\tempurl}


\bibitem[\protect\citeauthoryear{Tversky and Kahneman}{Tversky and
  Kahneman}{1973}]%
        {tversky_availability_1973}
\bibfield{author}{\bibinfo{person}{Amos Tversky} {and} \bibinfo{person}{Daniel
  Kahneman}.} \bibinfo{year}{1973}\natexlab{}.
\newblock \showarticletitle{Availability: {A} heuristic for judging frequency
  and probability}.
\newblock \bibinfo{journal}{\emph{Cognitive Psychology}} \bibinfo{volume}{5},
  \bibinfo{number}{2} (\bibinfo{date}{Sept.} \bibinfo{year}{1973}),
  \bibinfo{pages}{207--232}.
\newblock
\showISSN{0010-0285}
\urldef\tempurl%
\url{https://doi.org/10.1016/0010-0285(73)90033-9}
\showDOI{\tempurl}


\bibitem[\protect\citeauthoryear{Urbanska and Guimond}{Urbanska and
  Guimond}{2018}]%
        {urbanska_swaying_2018}
\bibfield{author}{\bibinfo{person}{Karolina Urbanska} {and}
  \bibinfo{person}{Serge Guimond}.} \bibinfo{year}{2018}\natexlab{}.
\newblock \showarticletitle{Swaying to the {Extreme}: {Group} {Relative}
  {Deprivation} {Predicts} {Voting} for an {Extreme} {Right} {Party} in the
  {French} {Presidential} {Election}}.
\newblock \bibinfo{journal}{\emph{International Review of Social Psychology}}
  \bibinfo{volume}{31}, \bibinfo{number}{1} (\bibinfo{date}{Oct.}
  \bibinfo{year}{2018}), \bibinfo{pages}{26}.
\newblock
\showISSN{2119-4130}
\urldef\tempurl%
\url{https://doi.org/10.5334/irsp.201}
\showDOI{\tempurl}


\bibitem[\protect\citeauthoryear{Vaccari, Valeriani, Barberá, Jost, Nagler,
  and Tucker}{Vaccari et~al\mbox{.}}{2016}]%
        {vaccari_echo_2016}
\bibfield{author}{\bibinfo{person}{Cristian Vaccari}, \bibinfo{person}{Augusto
  Valeriani}, \bibinfo{person}{Pablo Barberá}, \bibinfo{person}{John~T. Jost},
  \bibinfo{person}{Jonathan Nagler}, {and} \bibinfo{person}{Joshua~A. Tucker}.}
  \bibinfo{year}{2016}\natexlab{}.
\newblock \showarticletitle{Of {Echo} {Chambers} and {Contrarian} {Clubs}:
  {Exposure} to {Political} {Disagreement} {Among} {German} and {Italian}
  {Users} of {Twitter}}.
\newblock \bibinfo{journal}{\emph{Social Media + Society}} \bibinfo{volume}{2},
  \bibinfo{number}{3} (\bibinfo{date}{Sept.} \bibinfo{year}{2016}),
  \bibinfo{pages}{205630511666422}.
\newblock
\showISSN{2056-3051, 2056-3051}
\urldef\tempurl%
\url{https://doi.org/10.1177/2056305116664221}
\showDOI{\tempurl}


\bibitem[\protect\citeauthoryear{Van~Bavel and Pereira}{Van~Bavel and
  Pereira}{2018}]%
        {van_bavel_partisan_2018}
\bibfield{author}{\bibinfo{person}{Jay~J. Van~Bavel} {and}
  \bibinfo{person}{Andrea Pereira}.} \bibinfo{year}{2018}\natexlab{}.
\newblock \showarticletitle{The {Partisan} {Brain}: {An} {Identity}-{Based}
  {Model} of {Political} {Belief}}.
\newblock \bibinfo{journal}{\emph{Trends in Cognitive Sciences}}
  \bibinfo{volume}{22}, \bibinfo{number}{3} (\bibinfo{date}{March}
  \bibinfo{year}{2018}), \bibinfo{pages}{213--224}.
\newblock
\showISSN{13646613}
\urldef\tempurl%
\url{https://doi.org/10.1016/j.tics.2018.01.004}
\showDOI{\tempurl}


\bibitem[\protect\citeauthoryear{Vicario, Quattrociocchi, Scala, and
  Zollo}{Vicario et~al\mbox{.}}{2019}]%
        {vicario_polarization_2019}
\bibfield{author}{\bibinfo{person}{Michela~Del Vicario},
  \bibinfo{person}{Walter Quattrociocchi}, \bibinfo{person}{Antonio Scala},
  {and} \bibinfo{person}{Fabiana Zollo}.} \bibinfo{year}{2019}\natexlab{}.
\newblock \showarticletitle{Polarization and {Fake} {News}: {Early} {Warning}
  of {Potential} {Misinformation} {Targets}}.
\newblock \bibinfo{journal}{\emph{ACM Transactions on the Web}}
  \bibinfo{volume}{13}, \bibinfo{number}{2} (\bibinfo{date}{April}
  \bibinfo{year}{2019}), \bibinfo{pages}{1--22}.
\newblock
\showISSN{1559-1131, 1559-114X}
\urldef\tempurl%
\url{https://doi.org/10.1145/3316809}
\showDOI{\tempurl}


\bibitem[\protect\citeauthoryear{Vitriol and Marsh}{Vitriol and Marsh}{2018}]%
        {vitriol_illusion_2018}
\bibfield{author}{\bibinfo{person}{Joseph~A. Vitriol} {and}
  \bibinfo{person}{Jessecae~K. Marsh}.} \bibinfo{year}{2018}\natexlab{}.
\newblock \showarticletitle{The illusion of explanatory depth and endorsement
  of conspiracy beliefs}.
\newblock \bibinfo{journal}{\emph{European Journal of Social Psychology}}
  \bibinfo{volume}{48}, \bibinfo{number}{7} (\bibinfo{year}{2018}),
  \bibinfo{pages}{955--969}.
\newblock
\showISSN{1099-0992}
\urldef\tempurl%
\url{https://doi.org/10.1002/ejsp.2504}
\showDOI{\tempurl}


\bibitem[\protect\citeauthoryear{Vosoughi, Roy, and Aral}{Vosoughi
  et~al\mbox{.}}{2018}]%
        {vosoughi_spread_2018}
\bibfield{author}{\bibinfo{person}{Soroush Vosoughi}, \bibinfo{person}{Deb
  Roy}, {and} \bibinfo{person}{Sinan Aral}.} \bibinfo{year}{2018}\natexlab{}.
\newblock \showarticletitle{The spread of true and false news online}.
\newblock \bibinfo{journal}{\emph{Science}} \bibinfo{volume}{359},
  \bibinfo{number}{6380} (\bibinfo{date}{March} \bibinfo{year}{2018}),
  \bibinfo{pages}{1146--1151}.
\newblock
\showISSN{0036-8075, 1095-9203}
\urldef\tempurl%
\url{https://doi.org/10.1126/science.aap9559}
\showDOI{\tempurl}


\bibitem[\protect\citeauthoryear{Waller and Anderson}{Waller and
  Anderson}{2021}]%
        {waller_quantifying_2021}
\bibfield{author}{\bibinfo{person}{Isaac Waller} {and} \bibinfo{person}{Ashton
  Anderson}.} \bibinfo{year}{2021}\natexlab{}.
\newblock \showarticletitle{Quantifying social organization and political
  polarization in online platforms}.
\newblock \bibinfo{journal}{\emph{Nature}} (\bibinfo{date}{Dec.}
  \bibinfo{year}{2021}), \bibinfo{pages}{1--5}.
\newblock
\showISSN{1476-4687}
\urldef\tempurl%
\url{https://doi.org/10.1038/s41586-021-04167-x}
\showDOI{\tempurl}


\bibitem[\protect\citeauthoryear{Walter, Cohen, Holbert, and Morag}{Walter
  et~al\mbox{.}}{2020}]%
        {walter_fact-checking_2020}
\bibfield{author}{\bibinfo{person}{Nathan Walter}, \bibinfo{person}{Jonathan
  Cohen}, \bibinfo{person}{R.~Lance Holbert}, {and} \bibinfo{person}{Yasmin
  Morag}.} \bibinfo{year}{2020}\natexlab{}.
\newblock \showarticletitle{Fact-{Checking}: {A} {Meta}-{Analysis} of {What}
  {Works} and for {Whom}}.
\newblock \bibinfo{journal}{\emph{Political Communication}}
  \bibinfo{volume}{37}, \bibinfo{number}{3} (\bibinfo{date}{May}
  \bibinfo{year}{2020}), \bibinfo{pages}{350--375}.
\newblock
\showISSN{1058-4609, 1091-7675}
\urldef\tempurl%
\url{https://doi.org/10.1080/10584609.2019.1668894}
\showDOI{\tempurl}


\bibitem[\protect\citeauthoryear{Walter and Tukachinsky}{Walter and
  Tukachinsky}{2020}]%
        {walter_meta-analytic_2020}
\bibfield{author}{\bibinfo{person}{Nathan Walter} {and} \bibinfo{person}{Riva
  Tukachinsky}.} \bibinfo{year}{2020}\natexlab{}.
\newblock \showarticletitle{A {Meta}-{Analytic} {Examination} of the
  {Continued} {Influence} of {Misinformation} in the {Face} of {Correction}:
  {How} {Powerful} {Is} {It}, {Why} {Does} {It} {Happen}, and {How} to {Stop}
  {It}?}
\newblock \bibinfo{journal}{\emph{Communication Research}}
  \bibinfo{volume}{47}, \bibinfo{number}{2} (\bibinfo{date}{March}
  \bibinfo{year}{2020}), \bibinfo{pages}{155--177}.
\newblock
\showISSN{0093-6502}
\urldef\tempurl%
\url{https://doi.org/10.1177/0093650219854600}
\showDOI{\tempurl}


\bibitem[\protect\citeauthoryear{Wang and Fussell}{Wang and Fussell}{2020}]%
        {wang_more_2020}
\bibfield{author}{\bibinfo{person}{Luping Wang} {and} \bibinfo{person}{Susan~R.
  Fussell}.} \bibinfo{year}{2020}\natexlab{}.
\newblock \showarticletitle{More {Than} a {Click}: {Exploring} {College}
  {Students}' {Decision}-{Making} {Processes} in {Online} {News} {Sharing}}.
\newblock \bibinfo{journal}{\emph{Proceedings of the ACM on Human-Computer
  Interaction}} \bibinfo{volume}{4}, \bibinfo{number}{GROUP}
  (\bibinfo{date}{Jan.} \bibinfo{year}{2020}), \bibinfo{pages}{09:1--09:20}.
\newblock
\urldef\tempurl%
\url{https://doi.org/10.1145/3375189}
\showDOI{\tempurl}


\bibitem[\protect\citeauthoryear{Wason}{Wason}{1960}]%
        {wason_failure_1960}
\bibfield{author}{\bibinfo{person}{P.~C. Wason}.}
  \bibinfo{year}{1960}\natexlab{}.
\newblock \showarticletitle{On the failure to eliminate hypotheses in a
  conceptual task}.
\newblock \bibinfo{journal}{\emph{Quarterly Journal of Experimental
  Psychology}} \bibinfo{volume}{12}, \bibinfo{number}{3} (\bibinfo{date}{July}
  \bibinfo{year}{1960}), \bibinfo{pages}{129--140}.
\newblock
\showISSN{0033-555X}
\urldef\tempurl%
\url{https://doi.org/10.1080/17470216008416717}
\showDOI{\tempurl}


\bibitem[\protect\citeauthoryear{Weeks, Lane, Kim, Lee, and Kwak}{Weeks
  et~al\mbox{.}}{2017}]%
        {weeks_incidental_2017}
\bibfield{author}{\bibinfo{person}{Brian~E. Weeks}, \bibinfo{person}{Daniel~S.
  Lane}, \bibinfo{person}{Dam~Hee Kim}, \bibinfo{person}{Slgi~S. Lee}, {and}
  \bibinfo{person}{Nojin Kwak}.} \bibinfo{year}{2017}\natexlab{}.
\newblock \showarticletitle{Incidental {Exposure}, {Selective} {Exposure}, and
  {Political} {Information} {Sharing}: {Integrating} {Online} {Exposure}
  {Patterns} and {Expression} on {Social} {Media}}.
\newblock \bibinfo{journal}{\emph{Journal of Computer-Mediated Communication}}
  \bibinfo{volume}{22}, \bibinfo{number}{6} (\bibinfo{date}{Nov.}
  \bibinfo{year}{2017}), \bibinfo{pages}{363--379}.
\newblock
\showISSN{1083-6101}
\urldef\tempurl%
\url{https://doi.org/10.1111/jcc4.12199}
\showDOI{\tempurl}


\bibitem[\protect\citeauthoryear{Wilkins, Livingstone, and Levine}{Wilkins
  et~al\mbox{.}}{2020}]%
        {wilkins_one_2020}
\bibfield{author}{\bibinfo{person}{Denise~J. Wilkins},
  \bibinfo{person}{Andrew~G. Livingstone}, {and} \bibinfo{person}{Mark
  Levine}.} \bibinfo{year}{2020}\natexlab{}.
\newblock \showarticletitle{One of us or one of them? {How} “peripheral”
  adverts on social media affect the social categorization of sociopolitical
  message givers.}
\newblock \bibinfo{journal}{\emph{Psychology of Popular Media}}
  (\bibinfo{date}{Nov.} \bibinfo{year}{2020}).
\newblock
\showISSN{2689-6575, 2689-6567}
\urldef\tempurl%
\url{https://doi.org/10.1037/ppm0000322}
\showDOI{\tempurl}


\bibitem[\protect\citeauthoryear{Wood and Porter}{Wood and Porter}{2019}]%
        {wood_elusive_2019}
\bibfield{author}{\bibinfo{person}{Thomas Wood} {and} \bibinfo{person}{Ethan
  Porter}.} \bibinfo{year}{2019}\natexlab{}.
\newblock \showarticletitle{The {Elusive} {Backfire} {Effect}: {Mass}
  {Attitudes}’ {Steadfast} {Factual} {Adherence}}.
\newblock \bibinfo{journal}{\emph{Political Behavior}} \bibinfo{volume}{41},
  \bibinfo{number}{1} (\bibinfo{date}{March} \bibinfo{year}{2019}),
  \bibinfo{pages}{135--163}.
\newblock
\showISSN{1573-6687}
\urldef\tempurl%
\url{https://doi.org/10.1007/s11109-018-9443-y}
\showDOI{\tempurl}


\bibitem[\protect\citeauthoryear{Yeager, Krosnick, Visser, Holbrook, and
  Tahk}{Yeager et~al\mbox{.}}{2019}]%
        {yeager_moderation_2019}
\bibfield{author}{\bibinfo{person}{David~S. Yeager}, \bibinfo{person}{Jon~A.
  Krosnick}, \bibinfo{person}{Penny~S. Visser}, \bibinfo{person}{Allyson~L.
  Holbrook}, {and} \bibinfo{person}{Alex~M. Tahk}.}
  \bibinfo{year}{2019}\natexlab{}.
\newblock \showarticletitle{Moderation of classic social psychological effects
  by demographics in the {U}.{S}. adult population: {New} opportunities for
  theoretical advancement.}
\newblock \bibinfo{journal}{\emph{Journal of Personality and Social
  Psychology}} \bibinfo{volume}{117}, \bibinfo{number}{6}
  (\bibinfo{year}{2019}), \bibinfo{pages}{e84}.
\newblock
\showISSN{1939-1315}
\urldef\tempurl%
\url{https://doi.org/10.1037/pspa0000171}
\showDOI{\tempurl}


\bibitem[\protect\citeauthoryear{Yousif, Aboody, and Keil}{Yousif
  et~al\mbox{.}}{2019}]%
        {yousif_illusion_2019}
\bibfield{author}{\bibinfo{person}{Sami~R. Yousif}, \bibinfo{person}{Rosie
  Aboody}, {and} \bibinfo{person}{Frank~C. Keil}.}
  \bibinfo{year}{2019}\natexlab{}.
\newblock \showarticletitle{The {Illusion} of {Consensus}: {A} {Failure} to
  {Distinguish} {Between} {True} and {False} {Consensus}}.
\newblock \bibinfo{journal}{\emph{Psychological Science}} \bibinfo{volume}{30},
  \bibinfo{number}{8} (\bibinfo{date}{Aug.} \bibinfo{year}{2019}),
  \bibinfo{pages}{1195--1204}.
\newblock
\showISSN{0956-7976, 1467-9280}
\urldef\tempurl%
\url{https://doi.org/10.1177/0956797619856844}
\showDOI{\tempurl}


\bibitem[\protect\citeauthoryear{Zannettou, Bradlyn, De~Cristofaro, Kwak,
  Sirivianos, Stringini, and Blackburn}{Zannettou et~al\mbox{.}}{2018}]%
        {zannettou_what_2018}
\bibfield{author}{\bibinfo{person}{Savvas Zannettou}, \bibinfo{person}{Barry
  Bradlyn}, \bibinfo{person}{Emiliano De~Cristofaro}, \bibinfo{person}{Haewoon
  Kwak}, \bibinfo{person}{Michael Sirivianos}, \bibinfo{person}{Gianluca
  Stringini}, {and} \bibinfo{person}{Jeremy Blackburn}.}
  \bibinfo{year}{2018}\natexlab{}.
\newblock \showarticletitle{What is {Gab}: {A} {Bastion} of {Free} {Speech} or
  an {Alt}-{Right} {Echo} {Chamber}}. In \bibinfo{booktitle}{\emph{Companion
  {Proceedings} of the {The} {Web} {Conference} 2018}}
  \emph{(\bibinfo{series}{{WWW} '18})}. \bibinfo{publisher}{International World
  Wide Web Conferences Steering Committee}, \bibinfo{address}{Republic and
  Canton of Geneva, CHE}, \bibinfo{pages}{1007--1014}.
\newblock
\showISBNx{978-1-4503-5640-4}
\urldef\tempurl%
\url{https://doi.org/10.1145/3184558.3191531}
\showDOI{\tempurl}


\bibitem[\protect\citeauthoryear{Zannettou, Sirivianos, Blackburn, and
  Kourtellis}{Zannettou et~al\mbox{.}}{2019}]%
        {zannettou_web_2019}
\bibfield{author}{\bibinfo{person}{Savvas Zannettou}, \bibinfo{person}{Michael
  Sirivianos}, \bibinfo{person}{Jeremy Blackburn}, {and}
  \bibinfo{person}{Nicolas Kourtellis}.} \bibinfo{year}{2019}\natexlab{}.
\newblock \showarticletitle{The {Web} of {False} {Information}: {Rumors},
  {Fake} {News}, {Hoaxes}, {Clickbait}, and {Various} {Other} {Shenanigans}}.
\newblock \bibinfo{journal}{\emph{Journal of Data and Information Quality}}
  \bibinfo{volume}{11}, \bibinfo{number}{3} (\bibinfo{date}{July}
  \bibinfo{year}{2019}), \bibinfo{pages}{1--37}.
\newblock
\showISSN{1936-1955, 1936-1963}
\urldef\tempurl%
\url{https://doi.org/10.1145/3309699}
\showDOI{\tempurl}


\bibitem[\protect\citeauthoryear{Zhang}{Zhang}{2019}]%
        {zhang_encountering_2019}
\bibfield{author}{\bibinfo{person}{Kaiping Zhang}.}
  \bibinfo{year}{2019}\natexlab{}.
\newblock \showarticletitle{Encountering {Dissimilar} {Views} in
  {Deliberation}: {Political} {Knowledge}, {Attitude} {Strength}, and {Opinion}
  {Change}}.
\newblock \bibinfo{journal}{\emph{Political Psychology}} \bibinfo{volume}{40},
  \bibinfo{number}{2} (\bibinfo{date}{April} \bibinfo{year}{2019}),
  \bibinfo{pages}{315--333}.
\newblock
\showISSN{0162-895X, 1467-9221}
\urldef\tempurl%
\url{https://doi.org/10.1111/pops.12514}
\showDOI{\tempurl}


\bibitem[\protect\citeauthoryear{Zollo}{Zollo}{2019}]%
        {zollo_dealing_2019}
\bibfield{author}{\bibinfo{person}{Fabiana Zollo}.}
  \bibinfo{year}{2019}\natexlab{}.
\newblock \showarticletitle{Dealing with digital misinformation: a polarised
  context of narratives and tribes}.
\newblock \bibinfo{journal}{\emph{EFSA Journal}} \bibinfo{volume}{17},
  \bibinfo{number}{Proceedings of the Third EFSA Scientific Conference:
  Science, Food and Society Guest Editors: Devos Y, Elliott KC and Hardy A}
  (\bibinfo{date}{July} \bibinfo{year}{2019}).
\newblock
\showISSN{18314732, 18314732}
\urldef\tempurl%
\url{https://doi.org/10.2903/j.efsa.2019.e170720}
\showDOI{\tempurl}


\bibitem[\protect\citeauthoryear{Zollo, Bessi, Del~Vicario, Scala, Caldarelli,
  Shekhtman, Havlin, and Quattrociocchi}{Zollo et~al\mbox{.}}{2017}]%
        {zollo_debunking_2017}
\bibfield{author}{\bibinfo{person}{Fabiana Zollo}, \bibinfo{person}{Alessandro
  Bessi}, \bibinfo{person}{Michela Del~Vicario}, \bibinfo{person}{Antonio
  Scala}, \bibinfo{person}{Guido Caldarelli}, \bibinfo{person}{Louis
  Shekhtman}, \bibinfo{person}{Shlomo Havlin}, {and} \bibinfo{person}{Walter
  Quattrociocchi}.} \bibinfo{year}{2017}\natexlab{}.
\newblock \showarticletitle{Debunking in a world of tribes}.
\newblock \bibinfo{journal}{\emph{PLOS ONE}} \bibinfo{volume}{12},
  \bibinfo{number}{7} (\bibinfo{date}{July} \bibinfo{year}{2017}),
  \bibinfo{pages}{e0181821}.
\newblock
\showISSN{1932-6203}
\urldef\tempurl%
\url{https://doi.org/10.1371/journal.pone.0181821}
\showDOI{\tempurl}


\bibitem[\protect\citeauthoryear{Özdemir and Özkan}{Özdemir and
  Özkan}{2020}]%
        {ozdemir_deprivation_2020}
\bibfield{author}{\bibinfo{person}{Fatih Özdemir} {and}
  \bibinfo{person}{Türker Özkan}.} \bibinfo{year}{2020}\natexlab{}.
\newblock \showarticletitle{Deprivation, identification, and extreme pro-group
  behaviors: {The} political environment in {Turkey}.}
\newblock \bibinfo{journal}{\emph{Peace and Conflict: Journal of Peace
  Psychology}} \bibinfo{volume}{26}, \bibinfo{number}{2} (\bibinfo{date}{May}
  \bibinfo{year}{2020}), \bibinfo{pages}{224--226}.
\newblock
\showISSN{1532-7949, 1078-1919}
\urldef\tempurl%
\url{https://doi.org/10.1037/pac0000392}
\showDOI{\tempurl}


\end{thebibliography}

\end{document}